\documentclass[a4paper,11pt]{article}
\pdfoutput=1

\usepackage{jheppub} 

\usepackage[T1]{fontenc} 

\usepackage{graphicx}
\usepackage{amsmath}
\usepackage{subfigure}
\usepackage{url}
\usepackage{multirow}
\usepackage{threeparttable}
\usepackage{paralist}
\usepackage{slashed}
\usepackage{color}
\usepackage{arydshln,leftidx,mathtools}

\newcommand{\GeV}{\text{GeV}}
\newcommand{\TeV}{\text{TeV}}
\newcommand{\vev}[1]{\left\langle #1 \right\rangle}
\newcommand{\bs}[1]{\boldsymbol #1}
\newcommand\scalemath[2]{\scalebox{#1}{\mbox{\ensuremath{\displaystyle #2}}}}

\DeclareRobustCommand{\Sec}[1]{Sec.~\ref{#1}}

\DeclareRobustCommand{\App}[1]{App.~\ref{#1}}
\DeclareRobustCommand{\Tab}[1]{Table~\ref{#1}}

\DeclareRobustCommand{\Fig}[1]{Fig.~\ref{#1}}
\DeclareRobustCommand{\Figs}[2]{Figs.~\ref{#1} and \ref{#2}}
\DeclareRobustCommand{\Eq}[1]{Eq.~(\ref{#1})}
\DeclareRobustCommand{\Eqs}[2]{Eqs.~(\ref{#1}) and (\ref{#2})}
\DeclareRobustCommand{\Ref}[1]{Ref.~\cite{#1}}

\newcommand{\be}{\begin{equation}}
\newcommand{\ee}{\end{equation}}

\title{\boldmath Connecting Dark Matter UV Complete Models to Direct Detection Rates via Effective Field Theory}

\author[a,b]{Francesco D'Eramo}
\author[c,d]{and Massimiliano Procura}

\affiliation[a]{Department of Physics, University of California, Berkeley, CA 94720, USA}
\affiliation[b]{Theoretical Physics Group, Lawrence Berkeley National Laboratory, Berkeley, CA 94720,
USA}
\affiliation[c]{Albert Einstein Center for Fundamental Physics, Institute for Theoretical Physics, \\
University of Bern, CH-3012 Bern, Switzerland}
\affiliation[d]{Fakult\"at f\"ur Physik, Universit\"at Wien, Boltzmanngasse 5, 1090 Vienna, Austria}

\emailAdd{fraderamo@berkeley.edu}
\emailAdd{mprocura@univie.ac.at}

\abstract{Direct searches for WIMPs are sensitive to physics well below the weak scale. In the absence of light mediators, it is fruitful to apply an Effective Field Theory (EFT) approach accounting only for dark matter (DM) interactions with Standard Model (SM) fields. We consider a singlet fermion WIMP and effective operators up to dimension 6 which are generated at the mass scale of particles mediating DM interactions with the SM. We perform a one-loop Renormalization Group Evolution (RGE) analysis, evolving these effective operators from the mediators mass scale to the nuclear scales probed by direct searches. We apply our results to models with DM velocity-suppressed interactions, DM couplings only to heavy quarks, leptophilic DM and Higgs portal, which without our analysis would not get constrained from direct detection bounds. Remarkably, a large parameter space region for these models is found to be excluded as a consequence of spin-independent couplings induced by SM loops. In addition to these examples, we stress that more general renormalizable models for singlet fermion WIMP can be matched onto our EFT framework, and the subsequent model-independent RGE can be used to compute direct detection rates. Our results allow us to properly connect the different energy scales involved in constraining WIMP models, and to combine information from direct detection with other complementary searches, such as collider and indirect detection.}

\begin{document} 

\hfill UCB-PTH-14/38, UWThPh-2014-27

\maketitle
\flushbottom


\section{Introduction}
\label{sec:intro}

The nature of dark matter (DM) is one of the main open questions in particle physics. Among many candidates~\cite{Jungman:1995df,Bertone:2004pz,Feng:2010gw}, a Weakly Interacting Massive Particle (WIMP) with relic abundance obtained through thermal freeze-out~\cite{Lee:1977ua,Scherrer:1985zt,Srednicki:1988ce,Gondolo:1990dk} is quite appealing. Motivated frameworks for physics beyond the Standard Model (SM) naturally have a WIMP candidate~\cite{Goldberg:1983nd,Ellis:1983ew,Falk:1994es,Servant:2002aq,Cheng:2002ej,Birkedal:2006fz}, and it is suggestive that the same theory addressing the hierarchy problem also provides us with a DM candidate. Another exciting feature of the WIMPs is the fact that their typical couplings to SM particles are in the correct ballpark to give signals at colliders, direct and indirect detection experiments. Each of these searches is more sensitive to a certain parameter space region, so the WIMP paradigm can be tested with multiple and complementary methods.

Direct detection experiments play a peculiar role among these searches, since they probe energy scales much lower than the weak scale. For example, a $1 \, {\rm TeV}$ DM particle with a typical velocity $v / c \sim 10^{-3}$ scattering off a Xenon target cannot lead to a nuclear recoil energy larger than about $200 \, {\rm keV}$. The relevant physics at such small scales can be described by a non-relativistic Effective Field Theory (EFT), by integrating out short-distance effects and keeping only the relevant low-energy degrees of freedom~\cite{Fan:2010gt,Fitzpatrick:2012ix,Fitzpatrick:2012ib,Anand:2013yka,Gresham:2014vja,Alves:2014yha,Anand:2014kea,Catena:2014epa,Hill:2014yxa}. The analysis of other searches requires to go beyond this non-relativistic EFT. A simplifying hypothesis is to assume that the DM is the {\emph{only}} non-SM particle experimentally accessible~\cite{Beltran:2010ww,Goodman:2010yf,Goodman:2010ku,Rajaraman:2011wf,Bai:2010hh,Fox:2011fx,Fox:2011pm,Cheung:2012gi,Bai:2012xg,Buckley:2013jwa,Vecchi:2013iza,Fedderke:2014wda,Matsumoto:2014rxa}, with interactions parameterized by non-renormalizable operators originated from the exchange of heavy mediator particles. The validity of this approach does not extend all the way up to the LHC center of mass energy~\cite{Shoemaker:2011vi,Busoni:2013lha,Profumo:2013hqa,Busoni:2014sya,Endo:2014mja,Busoni:2014haa,Racco:2015dxa}, motivating the recent effort towards simplified models with mediator fields in the spectrum~\cite{Chang:2013oia,An:2013xka,DiFranzo:2013vra,Buchmueller:2013dya,Papucci:2014iwa,deSimone:2014pda,Bai:2014osa,Chang:2014tea,Agrawal:2014ufa,Hamaguchi:2014pja,Garny:2014waa,Askew:2014kqa,Buchmueller:2014yoa,Malik:2014ggr,Abdallah:2014hon,Buckley:2014fba,Harris:2014hga}. 

In this work we develop a formalism to connect DM models to nuclear scales probed by direct detection. As shown in Refs.~\cite{Kopp:2009et,Hill:2011be,Frandsen:2012db,Haisch:2012kf,Haisch:2013uaa,Hill:2013hoa,Hill:2014yka,Kopp:2014tsa,Crivellin:2014qxa,Crivellin:2014gpa}, there are examples where this large separation of scales has remarkable implications when a comparison with experiments is attempted. We focus on models for fermion DM with no SM gauge charge, and we assume that all the non-SM particles (with the possible exception of the DM itself) are above the weak scale. Fermion singlets cannot communicate with the SM at a renormalizable level, thus DM interactions must be necessarily mediated by these heavy particles. Once they are integrated out, it is possible to make a connection with nuclear scales that does not depend on the specific model we started from. This general analysis is the goal of our paper, where we connect physics at the mediator mass scales with direct detection observables. A generic model of singlet fermion DM, Dirac or Majorana, can be matched onto our EFT framework at the mass scale of the mediator particles.

\begin{figure}
\centering
\includegraphics[width=5.8in]{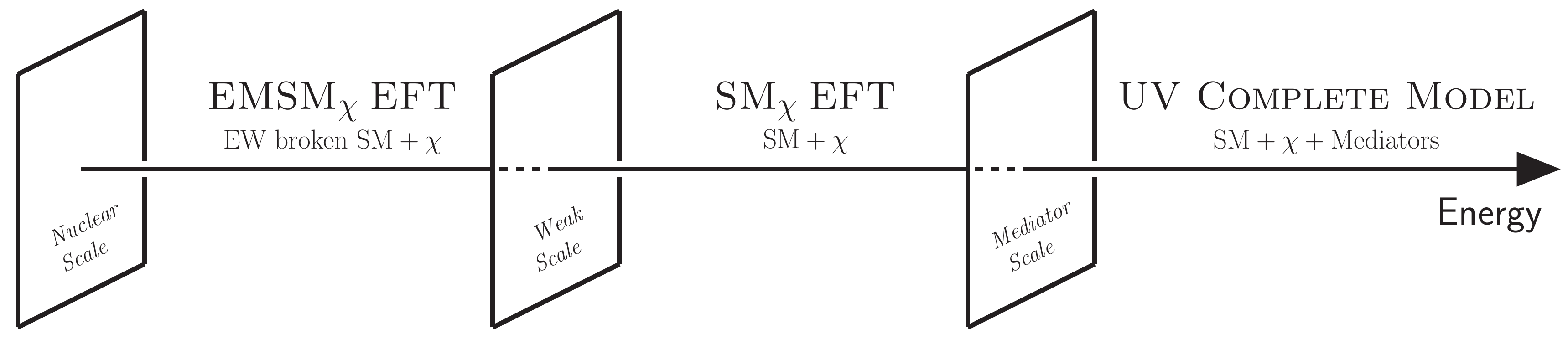}
\caption{Effective Field Theories used in this work. The fields mediating DM interactions with the SM are integrated out at the scale $\Lambda$. The operators of the ${\rm SM}_\chi$ EFT are evolved down to the EWSB scale, where electroweak states are integrated out. There a matching onto the ${\rm EMSM}_\chi$ EFT is performed. Finally, the operators are evolved down to the nuclear scale probed by direct searches.
}
\label{fig:EFTs}
\end{figure}

Our setup is sketched in \Fig{fig:EFTs}. At high energy scales we imagine the DM field $\chi$ embedded in an ultraviolet (UV) complete model. In our model-independent analysis we can neglect the UV details and consider the low-energy EFT with only $\chi$ and the SM fields in the spectrum. As emphasized in \Ref{Crivellin:2014qxa}, a systematic study allows us to identify mixing among operators and bound interactions that are poorly constrained otherwise. For this reason we start from the most general basis of operators up to dimension 6, defined at the EFT cutoff $\Lambda$, which corresponds to the mediators mass scale. The operators are evolved via a proper one-loop Renormalization Group (RG) analysis down to the ElectroWeak Symmetry Breaking (EWSB) scale, where the $W$ and $Z$ gauge bosons, the Higgs boson and the top quark are integrated out. This procedure defines a different EFT, with only strong and electromagnetic gauge interactions and $5$ quark flavors. We perform the one-loop RG analysis in this EFT as well, taking into account threshold corrections to the Wilson coefficients from integrating out the $b$ and $c$ quarks and the $\tau$ lepton, and evolving down to the nuclear scale $\mu \sim 1 - 2 \, {\rm GeV}$ at which hadronic matrix elements are evaluated~(see e.g. \cite{Gondolo:2004sc,Crivellin:2013ipa,Hill:2014yxa}). 

Why is this study relevant? After all, RG corrections are of the order of $\log (\Lambda / 1 \, {\rm GeV})$ multiplied by a loop factor, and pushing $\Lambda$ to $10 \, \TeV$ barely changes the order of magnitude for the rate. One may question the usefulness of a precise evaluation of the cross section in a pre-discovery era. We are certainly not after such a precision, and our focus is rather on models where loop effects are the dominant contribution. 

The only DM interactions at the nuclear scale relevant for direct detection involve the $u, d, s$ quarks, gluons and photons. However, many motivated models have mediator fields coupling the DM particle to heavy SM states and/or leptons. In these cases the main contribution to direct detection rates comes from loop effects. Furthermore, different  light quarks couplings yield direct detection cross sections which could differ by orders of magnitude, as Goodman and Witten showed in their seminal paper~\cite{Goodman:1984dc}. If the mediator fields induce suppressed couplings to light quarks (e.g. DM velocity-suppressed and/or spin-dependent interactions), loop-induced couplings to non-suppressed operators are again the dominant contribution. The best current experimental limits come from XENON100~\cite{Aprile:2012nq} and LUX~\cite{Akerib:2013tjd}, and will be significantly improved soon by SCDMS, XENON1T, DARKSIDE G2 and LZ (see for example \Ref{Cushman:2013zza}). They rule out electroweak processes with $Z$ boson exchange by orders of magnitude, and are therefore powerful enough to put constraints even on loop-induced processes.

The paper is structured as follows. The bases of independent operators for both the EFTs in \Fig{fig:EFTs} as well as matching conditions at the EWSB scale are discussed in \Sec{sec:EFTs}. The RGE equations in both EFTs are presented in \Sec{sec:RG}, with details on loop calculations contained in \App{app:Loops}. The reader only interested in our results, not in their derivation, can safely jump from \Sec{sec:EFTs} to \Sec{sec:App}, where we present the applications of our results to spin-independent searches. Consistently with the spirit of this work, we focus on examples where the DM has either suppressed couplings to light quarks or couplings only to heavy SM states. In these cases our loop effects are the main contribution to spin-independent direct detection rates. In \App{app:recipe} we give a straightforward recipe that allows one to apply our results and constrain UV complete fermion WIMP models that give rise to dimension 6 effective operators. \Sec{sec:discussion} contains our conclusions. 


\section{The Effective Theories for Singlet Fermion Dark Matter}
\label{sec:EFTs}

Our conceptual starting point is a renormalizable model for a fermion DM field $\chi$ that is a SM gauge singlet. Interactions between $\chi$ and the SM degrees of freedom $\psi_{\rm SM}$ are due to the exchange of mediator fields $\Phi$. The typical mass of the $\Phi$'s is assumed to be greater than the Fermi scale, and at such scales the full SM gauge symmetry $SU(3)_c \times SU(2)_L \times U(1)_Y$ is unbroken. The Lagrangian of the UV complete model schematically reads
\be
\mathcal{L}_{\rm UV} = \mathcal{L}_{\rm SM} + \overline{\chi} (i \slashed{\partial} - m_\chi ) \chi + \mathcal{L}_{\rm med}(\psi_{\rm SM}, \chi, \Phi) \ .
\label{eq:LagA}
\ee
Integrating out the mediators at the scale $\Lambda$ generates what we call ${\rm SM}_\chi$ EFT, containing only $\chi$ and the whole SM field content as its degrees of freedom. Many explicit realizations for $\mathcal{L}_{\rm UV}$ exist in the literature, and they can all be matched onto the ${\rm SM}_\chi$ EFT at the cutoff scale $\Lambda$. The regime of validity of this EFT extends all the way down to the EWSB scale, where the heavy EW states ($W$, $Z$, $h$ and $t$-quark) have to be integrated out and the residual gauge symmetry is $SU(3)_c \times U(1)_{\rm em}$. For this reason we employ a different EFT below the EWSB scale, with only a $SU(3)_c \times U(1)_{\rm em}$ gauge symmetry and $5$ quark flavors, which we call ${\rm EMSM}_\chi$ EFT (where EMSM stands for SM with only electromagnetic interactions). In the remaining part of this Section we give a basis of independent operators for both EFTs up to mass dimension~6, as well as a prescription for how to match ${\rm SM}_\chi$ EFT onto ${\rm EMSM}_\chi$ EFT at the EWSB scale.


\subsection{${\rm SM}_\chi$ Effective Theory}
\label{eq:SMchibasis}

Right below the mediator scale $\Lambda$ all the SM degrees of freedom are in the spectrum, and the $SU(3)_c \times SU(2)_L \times U(1)_Y$ SM gauge group is unbroken. Integrating out the mediators $\Phi$ in \Eq{eq:LagA} generates an infinite tower of higher dimensional operators
\be
\mathcal{L}_{{\rm SM}_\chi} = \mathcal{L}_{\rm SM} + \overline{\chi} \left( i \slashed{\partial} - m_\chi \right) \chi + 
\sum_{d>4} \sum_\alpha \frac{c^{(d)}_\alpha}{\Lambda^{d-4}} \mathcal{O}_\alpha^{(d)} \ .
\label{eq:LagSMchi}
\ee
Our conventions for the SM Lagrangian $\mathcal{L}_{\rm SM}$ are summarized in \App{app:SM}. In particular, since SM fermions are in a chiral representation of the gauge group, we use the matter fields
\be
\mathcal{F}_{{\rm SM}} = \left\{q_L^{(i)}, u_R^{(i)}, d_R^{(i)}, l_L^{(i)}, e_R^{(i)}, H \right\} \ .
\ee
The index $i$ runs over the three different SM fermion generations, and the gauge quantum numbers are assigned as in \Tab{tab:SMfields}. The index $\alpha$ runs over all gauge invariant operators of a given dimension $d$, with the dimensionless Wilson coefficients $c^{(d)}_\alpha$ encoding unresolved dynamics. These coefficients are renormalization-scale dependent, and we will quantify this dependence in the next Section.

\begin{table} 
\begin{center}
\begin{tabular}{ c || c c c c c | c}
 &  $q_L^i$ &  $u_R^i$ & $d_R^i$ & $l_L^i$ &$e_R^i$ & $H$ \\ 
\hline  \hline 
$SU(3)_c$ & $\bs 3$ & $ \bs 3$ & $ \bs 3$ & $\bs 1$ & $\bs 1$ & $\bs 1$ \\
$SU(2)_L$ & $\bs 2$ & $\bs 1$ & $\bs 1$ & $\bs 2$ & $\bs 1$ & $\bs 2$ \\
$U(1)_Y$  & $+1/6$ & $+2/3$ & $-1/3$ & $-1/2$ & $-1$ & $+1/2$ \\
\end{tabular}
\end{center}
\caption{SM matter fields and gauge charges in the unbroken phase. $q^i_L$ and $l^i_L$ are left-handed fermions, $u^i_R$, $d^i_R$ and $e^i_R$ are right-handed fermions. The index $i$ runs over the three generations.}
\label{tab:SMfields}
\end{table}

Without the need of specifying the responsible symmetry, we make sure the DM field is stable by requiring that every operator contains at least two $\chi$ fields. As an example, if DM is stabilized by a $Z_2$ symmetry, only operators with an even number of $\chi$ fields are allowed. Furthermore, our focus is on DM elastic scattering off target nuclei, thus we only need to consider operators with two DM fields. In our study we adopt the following basis of DM bilinears $\mathcal{O}_{\alpha \chi}$~\footnote{For effective operators up to dimension 6 and neglecting velocity suppressed effects this is a complete basis.} 
\be
\mathcal{O}_{\alpha \chi} = \left\{ \overline{\chi} \chi \ , \; \overline{\chi} \gamma^5 \chi  \ , \; \overline{\chi} \gamma^\mu \chi \ ,  \; 
\overline{\chi} \gamma^\mu \gamma^5 \chi \ , \;  \overline{\chi} \sigma^{\mu\nu} \chi \right\} \ .
\label{eq:DMbilinears}
\ee

Upon applying Fierz identities if necessary, each higher dimensional operator $\mathcal{O}_\alpha^{(d)}$ appearing in \Eq{eq:LagSMchi} with $d \leq 6$ and relevant to our analysis can be written as a product of a DM bilinear and SM fields
\be 
\mathcal{O}_\alpha^{(d)} = \mathcal{O}^{(3)}_{\alpha \chi} \; \times \;  \mathcal{O}_{\alpha {\rm SM}}^{(d-3)} \ .
\label{eq:EffOper}
\ee
The part involving only SM fields $\mathcal{O}_{\alpha {\rm SM}}^{(d-3)}$ has mass dimension $d-3$, is a SM gauge singlet but not necessarily a Lorentz singlet. We derive a complete basis of operators for the ${\rm SM}_\chi$ EFT up to dimension~6 following this strategy: we first identify all possible gauge singlets $\mathcal{O}_{\alpha {\rm SM}}^{(d-3)}$ by employing the same procedure described in Refs.~\cite{Buchmuller:1985jz,Grzadkowski:2010es}, then we take all allowed Lorentz invariant contractions with DM bilinears in \Eq{eq:DMbilinears}.

The first operators to look for are at dimension~5, which implies that we need gauge invariant SM operators $\mathcal{O}_{\alpha {\rm SM}}^{(2)}$ with mass dimension~2. The following options are available
\be
\mathcal{O}_{\alpha {\rm SM}}^{(2)} =  \left\{ H^\dag H \ , \; B_{\mu\nu} \ , \; \epsilon_{\mu\nu\rho\sigma}B^{\rho\sigma} \right\} \ ,
\ee
where $H$ and $B_{\mu\nu}$ are the Higgs doublet and the hypercharge field strength, respectively. The Lorentz invariant combinations with DM bilinears are listed in \Tab{tab:TheoryBdim5}. Since they are the lowest dimensional non-renormalizable operators, they do not mix onto other ones. As is well known, the resulting long-range interaction arising from the dipole operators severely constrain their Wilson coefficients~\cite{Barger:2010gv,Banks:2010eh,Fortin:2011hv}. The dimension 5 Higgs portal operator induces interactions with the gluon field strength once heavy quarks are integrated out~\cite{Shifman}, and current experiments are probing cross sections in its typical range~\cite{Kim:2006af,Kim:2008pp,MarchRussell:2008yu,Kanemura:2010sh,LopezHonorez:2012kv}. DM interactions to the Higgs also yields mono-Higgs events at colliders~\cite{Petrov:2013nia,Carpenter:2013xra,Berlin:2014cfa}, and for light enough DM ($m_\chi < m_{{h}}/2$) they contribute to the invisible Higgs decay width~\cite{D'Eramo:2007ga,Pospelov:2011yp,Bai:2011wz,Greljo:2013wja}. No interesting mixing takes place in this dimension 5 sector~\cite{Crivellin:2014qxa}, and for this reason our RG analysis will focus on dimension 6 operators, which we now identify.

\begin{table} 
\begin{center}
\renewcommand{\arraystretch}{1.2}
\begin{tabular}{ c | c || c | c }
Symbol & Operator & Symbol & Operator  \\ \hline 
$\mathcal{O}_{S}$ & $\overline{\chi} \chi \, H^\dag H$ & $\mathcal{O}_{M_B}$ & $\overline{\chi} \sigma^{\mu\nu} \chi \, B_{\mu\nu}$  \\
$\mathcal{O}_{P}$ & $\overline{\chi} \gamma^5 \chi \, H^\dag H$ & $\mathcal{O}_{E_B}$ & $\overline{\chi} \sigma^{\mu\nu} \chi \,  \epsilon_{\mu\nu\rho\sigma}B^{\rho\sigma}   $
\end{tabular}
\end{center}
\caption{Basis of dimension~5 operators for the ${\rm SM}_\chi$ Effective Theory.}
\label{tab:TheoryBdim5}
\end{table}
For dimension~6 operators, the relevant SM structures of dimension~3 are the currents
\be
\mathcal{O}_{\alpha {\rm SM}}^{(3)} =  \left\{ \overline{q^i_L} \gamma_\mu q^i_L, \; \overline{u^i_R} \gamma_\mu u^i_R,  
\; \overline{d^i_R} \gamma_\mu d^i_R, \; \overline{l^i_L} \gamma_\mu l^i_L, \; \overline{e^i_R} \gamma_\mu e^i_R, \; 
H^\dag \, i \overleftrightarrow{D}_\mu H \right\} \ ,
\label{eq:SMcurrents}
\ee
where we do not assume any flavor violation The index $i$ runs over the three different fermion generations, thus the above vector has $5 \times 3 + 1= 16$ components. The double-arrow derivative entering the Higgs current reads
\be
H^\dag \overleftrightarrow{D}_\mu H \equiv H^\dag ( D_\mu H ) -  ( D_\mu H^\dag ) H \ ,
\label{eq:Higgsdouble}
\ee
with the covariant derivative defined as in \Eq{eq:Dcov} of \App{app:SM}.

Lorentz invariant operators can be obtained by contracting the currents in \Eq{eq:SMcurrents} with a DM current $\overline{\chi} \,\Gamma^\mu \chi$, where both vector $\Gamma^\mu = \gamma^\mu$ and axial $\Gamma^\mu = \gamma^\mu \gamma^5$ currents are possible. This gives a total of $16 \times 2 = 32$ independent operators. However, since $\chi$ is a singlet, the DM current $\overline{\chi} \,\Gamma^\mu \chi$ is invariant under RG evolution, thus we can study two $16$-dimensional sectors separately. The basis for dimension~6 operators with a specific DM current $\overline{\chi} \,\Gamma^\mu \chi$ is shown in \Tab{tab:TheoryBdim6}. 
For future convenience, we introduce a 16-dimensional vector of Wilson coefficients
\be 
\mathcal{C}_{{\rm SM}_\chi}^T \!\equiv\! \left(\!\begin{array}{ccc:cc|ccc:cc|ccc:cc||c}
c_{\Gamma q}^{(1)} & c_{\Gamma u}^{(1)} & c_{\Gamma d}^{(1)}\,&\, c_{\Gamma l}^{(1)} & c_{\Gamma e}^{(1)} \,&\, 
c_{\Gamma q}^{(2)} & c_{\Gamma u}^{(2)} & c_{\Gamma d}^{(2)} \,&\, c_{\Gamma l}^{(2)} & c_{\Gamma e}^{(2)} \,&\,
c_{\Gamma q}^{(3)} & c_{\Gamma u}^{(3)} & c_{\Gamma d}^{(3)} \,&\, c_{\Gamma l}^{(3)} & c_{\Gamma e}^{(3)} \,&\,
c_{\Gamma \!H} \end{array}\!\right),
\label{ed:cdef}
\ee
where $c_\alpha$ is associated with the operator $\mathcal{O}_\alpha$ in \Tab{tab:TheoryBdim6}. The solid double line divides DM interactions with the Higgs from the ones with SM fermions. The solid single lines divide different SM generations and within each generation quarks and leptons are divided by a dashed line. 

\begin{table} 
\begin{center}
\renewcommand{\arraystretch}{1.3}
\begin{tabular}{c | c || c | c || c | c }
Symbol & Operator & Symbol & Operator & Symbol & Operator  \\ \hline 
$\mathcal{O}_{\Gamma q}^{(i)}$ & $\overline{\chi} \,\Gamma^\mu \chi \, \overline{q^i_L} \gamma_\mu q^i_L$ & 
$\mathcal{O}_{\Gamma l}^{(i)}$ & $\overline{\chi} \,\Gamma^\mu \chi \, \overline{l^i_L} \gamma_\mu l^i_L$ & 
$\mathcal{O}_{\Gamma H}^{(i)}$ & $\overline{\chi} \, \Gamma^\mu \chi \, H^\dag i\overleftrightarrow{D}_\mu H $ \\
$\mathcal{O}_{\Gamma u}^{(i)}$ & $\overline{\chi} \,\Gamma^\mu \chi \, \overline{u^i_R} \gamma_\mu u^i_R$ & 
$\mathcal{O}_{\Gamma e}^{(i)}$ & $\overline{\chi} \,\Gamma^\mu \chi \, \overline{e^i_R} \gamma_\mu e^i_R$ & 
$$ & $$  \\
$\mathcal{O}_{\Gamma d}^{(i)}$ & $\overline{\chi} \,\Gamma^\mu \chi \, \overline{d^i_R} \gamma_\mu d^i_R$ & 
$$ & $$ & 
$$ & $$ \\
\end{tabular}
\end{center}
\caption{Basis of dimension~6 operators for the ${\rm SM}_\chi$ EFT. The first two columns have three different replicas, corresponding to the SM generations. We consider a generic $\overline{\chi} \,\Gamma^\mu \chi$, which can be either a vector ($\Gamma^\mu = \gamma^\mu$) or an axial ($\Gamma^\mu = \gamma^\mu \gamma^5$) DM current or any linear combination of them.}
\label{tab:TheoryBdim6}
\end{table}

We stress that the dimension 6 operator
\be
\mathcal{O}_{\Gamma B} = g^\prime \frac{c_B}{\Lambda^2}  \; \overline{\chi} \, \Gamma^\mu \chi \; \partial^\nu B_{\nu\mu}
\label{eq:redundantSection2}
\ee
does not need to be included in our list since it can be expressed as a linear combination of the ones listed in \Tab{tab:TheoryBdim6} by using classical equation of motion~\cite{Arzt:1993gz} for the hypercharge field strength (see \Eq{eq:Beom}). More specifically, the effect of this operator can be absorbed into the following shifts of the Wilson coefficients
\begin{align}
\label{eq:replacementB1} c^{(i)}_{\Gamma q} \rightarrow & \, c^{(i)}_{\Gamma q}  - g^{\prime\,2} y_q c_B \ , \\
c^{(i)}_{\Gamma u} \rightarrow & \, c^{(i)}_{\Gamma u}  - g^{\prime\,2} y_u c_B \ , \\
c^{(i)}_{\Gamma d} \rightarrow & \, c^{(i)}_{\Gamma d}  - g^{\prime\,2} y_d c_B \ , \\
c^{(i)}_{\Gamma l} \rightarrow & \, c^{(i)}_{\Gamma l}  - g^{\prime\,2} y_l c_B \ , \\
c^{(i)}_{\Gamma e} \rightarrow & \, c^{(i)}_{\Gamma e}  - g^{\prime\,2} y_e c_B \ , \\ 
\label{eq:replacementB6} c_H \rightarrow & \, c_H - g^{\prime\,2} y_H c_B \ .
\end{align}

\subsection{${\rm EMSM}_\chi$ Effective Theory}
The construction of the operator basis for the ${\rm EMSM}_\chi$ EFT is analogous. The Lagrangian reads
\be
\mathcal{L}_{{\rm EMSM}_\chi} = \mathcal{L}_{\rm EMSM} + \overline{\chi} \left( i \slashed{\partial} - m_\chi \right) \chi + 
\sum_{d>4} \sum_\alpha \frac{c^{(d)}_\alpha}{\Lambda^{d-4}} \mathcal{O}_\alpha^{(d)} \ .
\label{eq:LagEMSMchi}
\ee
Details and conventions for the renormalizable EMSM part can be found in \App{app:EMSM}. Since matter fields fill vector-like representations of the gauge group in this phase, we employ the set of Dirac fermions
\be
\mathcal{F}_{{\rm EMSM}} = \left\{u^{i}, d^{i}, e^{i} \right\} \ .
\ee
The index $i$ runs again over the three different SM generations, but this time without the top quark (i.e. without $u^{(3)}$). Gauge quantum numbers are assigned as in \Tab{tab:SMfields2}. The effective operators are still of the form
\be 
\mathcal{O}_\alpha^{(d)} = \mathcal{O}^{(3)}_{\alpha \chi} \; \times \;  \mathcal{O}_{\alpha {\rm EMSM}}^{(d-3)} \ ,
\label{eq:EffOper2}
\ee
with DM bilinears  listed in \Eq{eq:DMbilinears}. 

\begin{table} 
\begin{center}
\begin{tabular}{ c || c c c}
 &  $u^i$ &  $d^i$ & $e^i$ \\ 
\hline  \hline 
$SU(3)_c$ & $\bs 3$ & $ \bs 3$ & $ \bs 1$  \\
$U(1)_{\rm em}$  & $+2/3$ & $-1/3$ & $-1$ 
\end{tabular}
\end{center}
\caption{SM matter fields and their gauge quantum numbers in the broken phase. In this case $q^i$, $u^i$ and $e^i$ are Dirac fermions. The index $i = 1,2,3$ runs over the threes different generations. The top quark (i.e. $u^3$) is not included.}
\label{tab:SMfields2}
\end{table}

For dimension~5 operators we only have interactions with the electromagnetic field, i.e.
\be
\mathcal{O}_{\alpha {\rm EMSM}}^{(2)} =  \left\{ F_{\mu\nu} \ , \; \epsilon_{\mu\nu\rho\sigma}F^{\rho\sigma} \right\} \ ,
\ee
and the Lorentz invariant contractions with DM bilinears are listed in \Tab{tab:TheoryCdim5}.
\begin{table}
\begin{center}
\renewcommand{\arraystretch}{1.2}
\begin{tabular}{ c | c}
Symbol & Operator \\ \hline 
$\mathcal{O}_{M_F}$ & $\overline{\chi} \sigma^{\mu\nu} \chi \, F_{\mu\nu}$  \\
$\mathcal{O}_{E_F}$ & $\overline{\chi} \sigma^{\mu\nu} \chi \,  \epsilon_{\mu\nu\rho\sigma}F^{\rho\sigma}   $
\end{tabular}
\end{center}
\caption{Basis of dimension~5 operators for the ${\rm EMSM}_\chi$ Effective Theory.}
\label{tab:TheoryCdim5}
\end{table}
For dimension~6 operators we have the SM currents
\be
\mathcal{O}_{\alpha {\rm EMSM}}^{(3)} =  
\left\{ \overline{u^i} \gamma_\mu u^i \ , \;  \overline{u^i} \gamma_\mu \gamma_5 u^i \ ,   \; 
\overline{d^i} \gamma_\mu d^i \ , \;  \overline{d^i} \gamma_\mu \gamma_5 d^i \ ,   \; 
\overline{e^i} \gamma_\mu e^i \ , \;  \overline{e^i} \gamma_\mu \gamma_5 e^i  \right\} \ .
\ee
The top quark is not in the spectrum, thus we count $6 \times 3 - 2 = 16$ independent currents. Also in this case they can be contracted with either a vector or an axial DM current, giving a total of $32$ independent operators. Each $16$ dimensional sector shown in \Tab{tab:TheoryCdim6} can be studied separately. In analogy to \Eq{ed:cdef}, we define the vector
\be
\scalemath{0.9}{\mathcal{C}_{{\rm EMSM}_\chi}^T \!=\! \left(\!\begin{array}{ccccc|ccc||ccccc|ccc}
c_{\Gamma Vu}^{(1)} & c_{\Gamma Vd}^{(1)} & c_{\Gamma Vu}^{(2)} & c_{\Gamma Vd}^{(2)} & c_{\Gamma Vd}^{(3)} & 
c_{\Gamma Ve}^{(1)} &c_{\Gamma Ve}^{(2)}  & c_{\Gamma Ve}^{(3)}  & c_{\Gamma Au}^{(1)} & c_{\Gamma Ad}^{(1)} & c_{\Gamma Au}^{(2)} & c_{\Gamma Ad}^{(2)} & c_{\Gamma Ad}^{(3)} & 
c_{\Gamma Ae}^{(1)} &c_{\Gamma Ae}^{(2)}  & c_{\Gamma Ae}^{(3)} \end{array}\!\right). }
\label{ed:cdef2}
\ee
Here, the solid double line is used to divide DM couplings to a vector or an axial SM current, whereas single solid lines divide quarks from leptons.

\begin{table}
\begin{center}
\renewcommand{\arraystretch}{1.3}
\begin{tabular}{c | c || c | c || c | c }
Symbol & Operator & Symbol & Operator & Symbol & Operator  \\ \hline 
$\mathcal{O}_{\Gamma Vu}^{(i)}$ & $\overline{\chi} \,\Gamma^\mu \chi \, \overline{u^i} \gamma_\mu u^i$ & 
$\mathcal{O}_{\Gamma Vd}^{(i)}$ & $\overline{\chi} \,\Gamma^\mu \chi \, \overline{d^i} \gamma_\mu d^i$ & 
$\mathcal{O}_{\Gamma Ve}^{(i)}$ & $\overline{\chi} \,\Gamma^\mu \chi \, \overline{e^i} \gamma_\mu e^i$ \\
$\mathcal{O}_{\Gamma Au}^{(i)}$ & $\overline{\chi} \,\Gamma^\mu \chi \, \overline{u^i} \gamma_\mu \gamma _5 u^i$ & 
$\mathcal{O}_{\Gamma Ad}^{(i)}$ & $\overline{\chi} \,\Gamma^\mu \chi \, \overline{d^i} \gamma_\mu \gamma _5 d^i$ & 
$\mathcal{O}_{\Gamma Ae}^{(i)}$ & $\overline{\chi} \,\Gamma^\mu \chi \, \overline{e^i} \gamma_\mu \gamma _5 e^i$
\end{tabular}
\end{center}
\caption{Basis of dimension~6 operators for the ${\rm EMSM}_\chi$ Effective Theory. Each operator has three different replicas, corresponding to the three SM generations. The DM bilinear can have both vector or axial currents, namely $\Gamma = \{V, A\}$, where $V^\mu = \gamma^\mu$ and $A^\mu = \gamma^\mu\gamma^5$.}
\label{tab:TheoryCdim6}
\end{table}

The redundant dimension~6 operator in this case is
\be
\mathcal{O}_{\Gamma F} = e \frac{c_F}{\Lambda^2}  \; \overline{\chi} \, \Gamma^\mu \chi \; \partial^\nu F_{\nu\mu} \ .
\label{eq:redundantphotonSection2}
\ee
Equations of motion for the electromagnetic field strength (see \Eq{eq:Feom}) translates this operator into a linear combination of the ones listed in \Tab{tab:TheoryCdim6}, which equivalently amounts to this shift of the Wilson coefficients for the operators with SM vector currents
\begin{align}
c^{(i)}_{\Gamma V u} \rightarrow & \, c^{(i)}_{\Gamma V u}  - e^2 Q_u c_F \ , \\
c^{(i)}_{\Gamma V d} \rightarrow & \, c^{(i)}_{\Gamma V d}  - e^2 Q_d c_F \ , \\
c^{(i)}_{\Gamma V e} \rightarrow & \, c^{(i)}_{\Gamma V e}  - e^2 Q_e c_F \ .
\end{align}
The operators with SM axial currents are not affected, since the photon only couples to vector currents.

\subsection{Matching the two EFTs at the EWSB scale}
\label{sec:matching}

We conclude this Section by giving matching conditions between the two theories, namely the relations between the Wilson coefficients in \Eq{ed:cdef2} and those in \Eq{ed:cdef}, both evaluated at the EWSB scale, which is smaller than $\Lambda$ in this setup. As we will see shortly, the leading contribution arises already at tree level, therefore we do not need to consider the subleading one-loop contributions.

When performing the tree-level matching, going from left- and right-handed currents to vector and axial currents is straightforward. But this is not the full story, since the operator coupling the DM to the Higgs current leads to the following contribution obtained by giving the Higgs doublets an EWSB VEV. The result is to induce an effective tree-level coupling between the DM and the $Z$ boson
\be
\mathcal{L}_{\chi\chi Z} = \frac{c_H}{\Lambda^2} \,  \overline{\chi} \, \Gamma^\mu \chi \,  \vev{H^\dag} \, 
i \overleftrightarrow{D}_\mu \vev{H} =  -  \frac{c_H}{\Lambda^2} \, v^2 \, \sqrt{g^2 + g^{\prime \, 2}} \; \overline{\chi} \, \Gamma^\mu \chi \, Z_\mu   \ .
\label{eq:chichiZ}
\ee
The $Z$ boson also couples to SM fermions
\be
\mathcal{L}_{\rm N.C.}^Z = \frac{g}{2 c_w}  \, Z_\mu J^\mu_0 \ , 
\ee
where the neutral current $J^\mu_0$ is defined as follows
\begin{align}
J^\mu_0 = & \,  \sum_f \left[g_{V f} \, \overline{f} \gamma^\mu f + g_{A f}  \, \overline{f} \gamma^\mu \gamma^5 f \right] \ , \\
g_{V f} = & \, T_f^3 - 2 s_w^2 Q_f \ , \\ 
g_{A f} = & \, - T_f^3  \ .
\end{align}
Here, $T^3_f$ is the third component of the weak isospin, $s_w$ the sine of the weak mixing angle and $Q_f$ the fermion electromagnetic charge. The coefficients for the SM fermions explicitly read
\be
\renewcommand{\arraystretch}{1.3}
\begin{array}{lccccccclcccccccl}
g_{Vu} =  \frac{1}{2} - \frac{4}{3} s_w^2  \ , & & & & & & & & g_{Vd} = - \frac{1}{2} + \frac{2}{3} s_w^2 \ , & & & & & & & & 
 g_{Ve} = - \frac{1}{2} + 2 s_w^2   \ ,  \\ 
 g_{Au} =  - \frac{1}{2} \ , & & & & & & & & g_{Ad} = \frac{1}{2}  \ ,  & & & & & & & &  g_{Ae} = \frac{1}{2}  \ .
\end{array} 
\label{eq:SMfermionstoZ}
\ee
Integrating out the $Z$ boson gives rise to the Fermi Lagrangian for SM neutral currents
\be
\mathcal{L}_{\rm Fermi} = - \frac{G_F}{\sqrt{2}} \, J^\mu_0 J_{0\mu} \ .
\ee
Analogously, tree-level $Z$ exchange gives a finite threshold corrections to the Wilson coefficients of the ${\rm EMSM}_\chi$ EFT. The complete matching conditions read 
\begin{align}
\label{eq:matching1} c_{\Gamma Vu}^{(i)} = & \, \frac{c_{\Gamma q}^{(i)} + c_{\Gamma u}^{(i)}}{2} +  c_H \, g_{Vu} \ ,  \\
\label{eq:matching2} c_{\Gamma Vd}^{(i)} = & \, \frac{c_{\Gamma q}^{(i)} + c_{\Gamma d}^{(i)}}{2} +  c_H \, g_{Vd} \ , \\
\label{eq:matching3} c_{\Gamma Ve}^{(i)} = & \, \frac{c_{\Gamma l}^{(i)} + c_{\Gamma e}^{(i)}}{2} +  c_H \, g_{Ve} \ , \\
\label{eq:matching4} c_{\Gamma Au}^{(i)} = & \, \frac{- c_{\Gamma q}^{(i)} + c_{\Gamma u}^{(i)}}{2} +  c_H \, g_{Au} \ , \\
\label{eq:matching5} c_{\Gamma Ad}^{(i)} = & \, \frac{- c_{\Gamma q}^{(i)} + c_{\Gamma d}^{(i)}}{2} +  c_H \, g_{Ad} \ , \\
\label{eq:matching6} c_{\Gamma Ae}^{(i)} = & \, \frac{- c_{\Gamma l}^{(i)} + c_{\Gamma e}^{(i)}}{2} +  c_H \, g_{Ae} \ .
\end{align}


\section{Renormalization Group Evolution}
\label{sec:RG}

We present the complete one-loop RG equations in both EFTs. Here, we only show Feynman diagrams and quote final results. Regularization and renormalization at one loop in both EFTs are detailedly discussed in App.\ref{app:Loops}. As explained in the previous Section, no interesting loop effect takes place among the dimension 5 operators, besides the well known heavy quark threshold contribution from the Higgs portal~\cite{Shifman}. Thus we focus on dimension 6 operators.

\subsection{From the messenger scale to the EWSB scale}
\label{sec:RG1}

The evolution of the Wilson coefficients in \Eq{ed:cdef} is described by the differential equation
\be
\frac{d \, \mathcal{C}_{{\rm SM}_\chi}}{d \ln \mu}  = \gamma_{{\rm SM}_\chi} \mathcal{C}_{{\rm SM}_\chi} \ ,
\ee
where $\mu$ is the renormalization scale and $\gamma_{{\rm SM}_\chi}$ is the anomalous dimension matrix. Our goal here is to fill out the $16 \times 16 = 256$ entries of the matrix $\gamma_{{\rm SM}_\chi}$.

\begin{figure}
\centering
\includegraphics[width=5.8in]{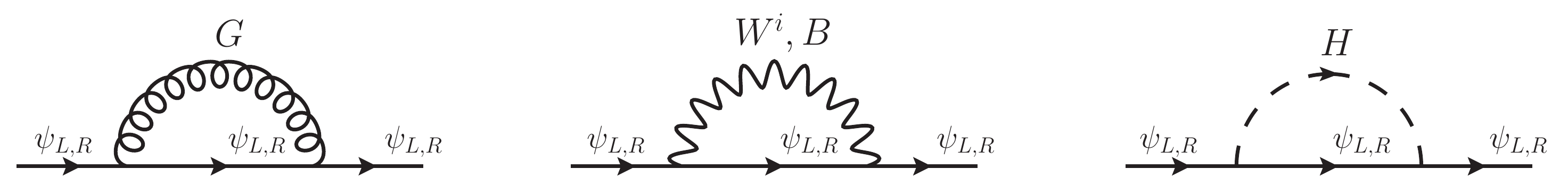}
\caption{External legs corrections for SM fermions.}
\label{fig:LegsFeynmanDiagrams1}
\end{figure}

\begin{figure}
\centering
\includegraphics[width=5.8in]{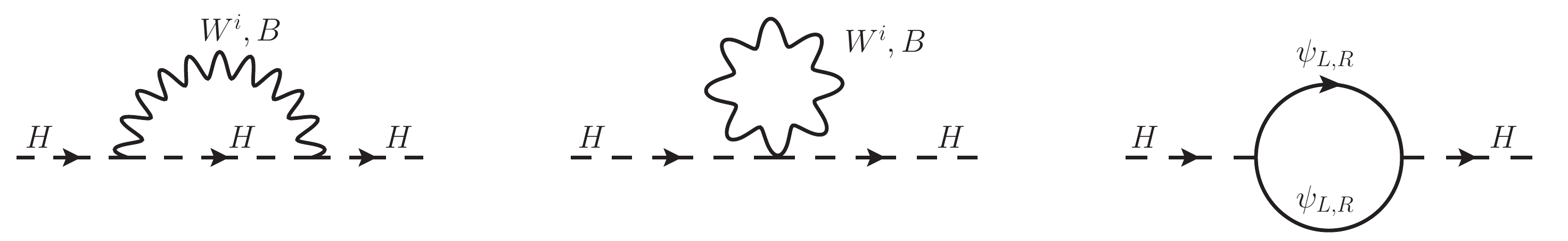}
\caption{External legs for SM Higgs.}
\label{fig:LegsFeynmanDiagrams2}
\end{figure}

We start our one-loop analysis in this theory by considering external legs corrections. Since the DM field is a gauge singlet, these contributions only involve SM fields and interactions. We perform the field renormalizations
\be
\psi_i \rightarrow Z_{\psi_i}^{1/2} \psi_i \ , \qquad \qquad H \rightarrow Z_H^{1/2} H \ ,
\ee
where $\psi_i$ is any SM fermion, and we do it in such a way to subtract the infinite part from the residue of each one-loop propagator. There are only two possible sources for this effect, which are gauge and Yukawa interactions. As is well know, the Higgs quartic coupling does not induce a one-loop contribution to the wave-function renormalization. The relevant Feynman diagrams are shown in \Figs{fig:LegsFeynmanDiagrams1}{fig:LegsFeynmanDiagrams2} for fermion and Higgs fields, respectively. 

When considering vertex corrections, one still has to deal only with these two interactions. We organize the presentation by fixing the external legs of a specific amplitude, and then identifying all the possible one-loop contributions. In other words, we fix a given effective operator from the ones in \Tab{tab:TheoryCdim6} and then look for operators mixing into it. 
\begin{figure}
\centering
\includegraphics[scale=0.45]{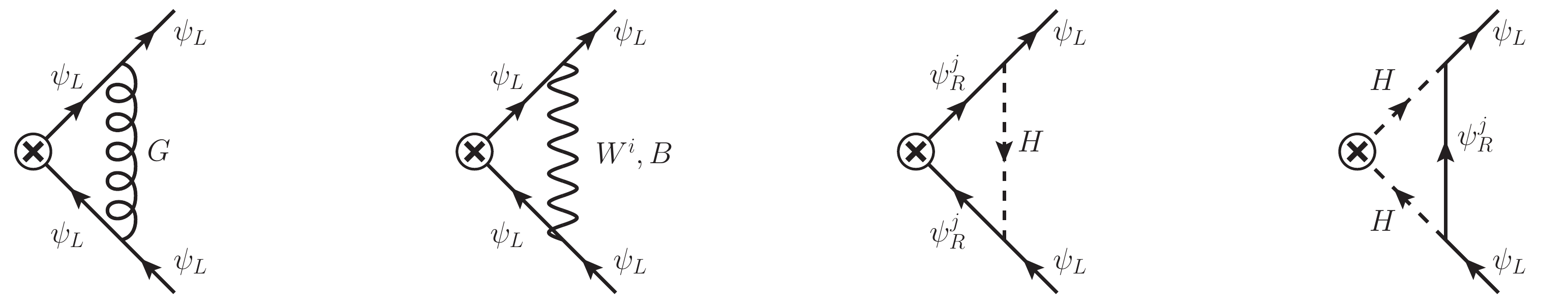}
\caption{One loop corrections to the Wilson coefficient $c_L$ in the ${\rm SM}_\chi$ EFT, where the crossed circle denotes the effective vertex between SM fields and the DM bilinear. The index $j$ for right-handed fermions can be either $u$ or $d$. The diagrams for the one-loop corrections to $c_R^u$ and $c_R^d$ are analogous, but without the $W^i_\mu$ gauge bosons in the loop.}
\label{fig:VertexFeynmanDiagrams1}
\end{figure}
\begin{figure}
\centering
\includegraphics[scale=0.45]{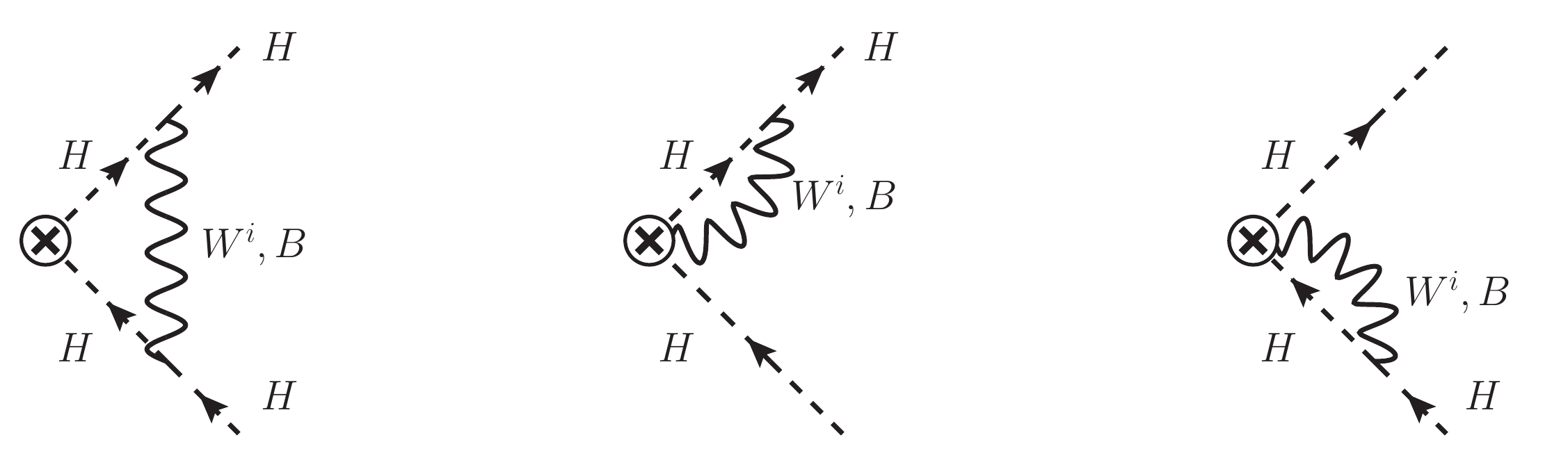} \\ 
\includegraphics[scale=0.45]{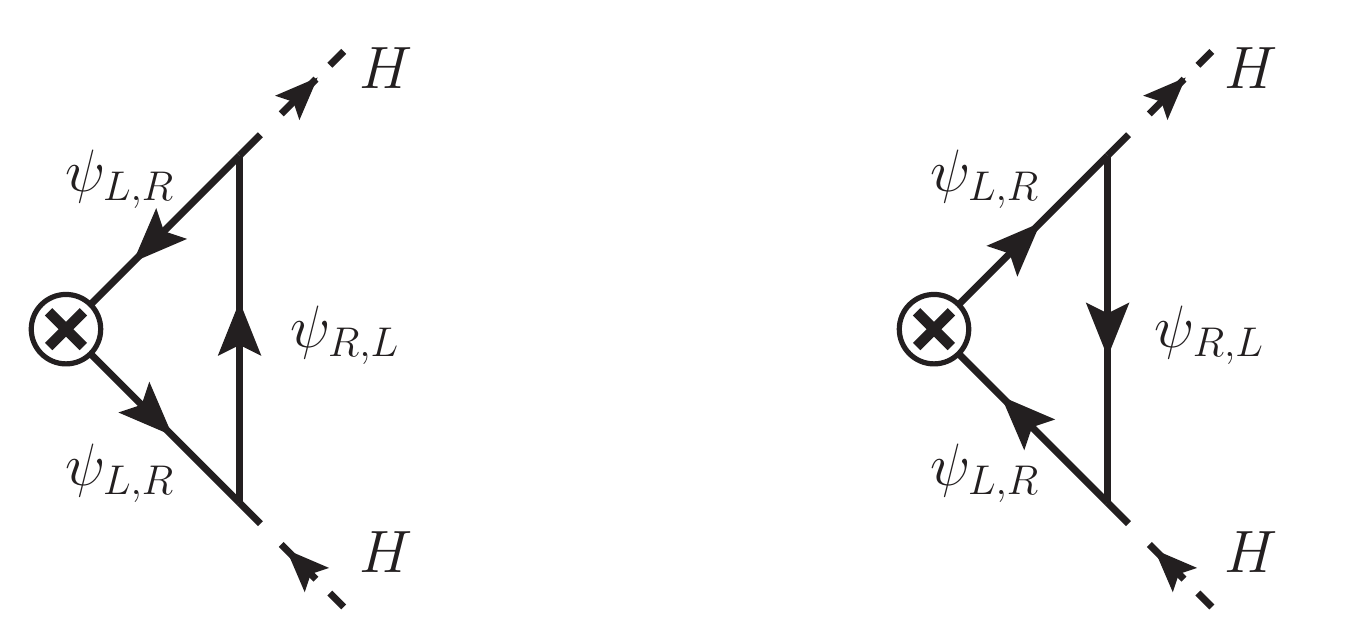} 
\caption{One loop corrections to the Wilson coefficient $c_H$ in the ${\rm SM}_\chi$ EFT.  The crossed circle notation is the same as \Fig{fig:VertexFeynmanDiagrams1}. In the first row we have corrections from gauge interactions, in the second row from Yukawa interactions.}
\label{fig:VertexFeynmanDiagrams2}
\end{figure}

\begin{figure}
\centering
\includegraphics[scale=0.45]{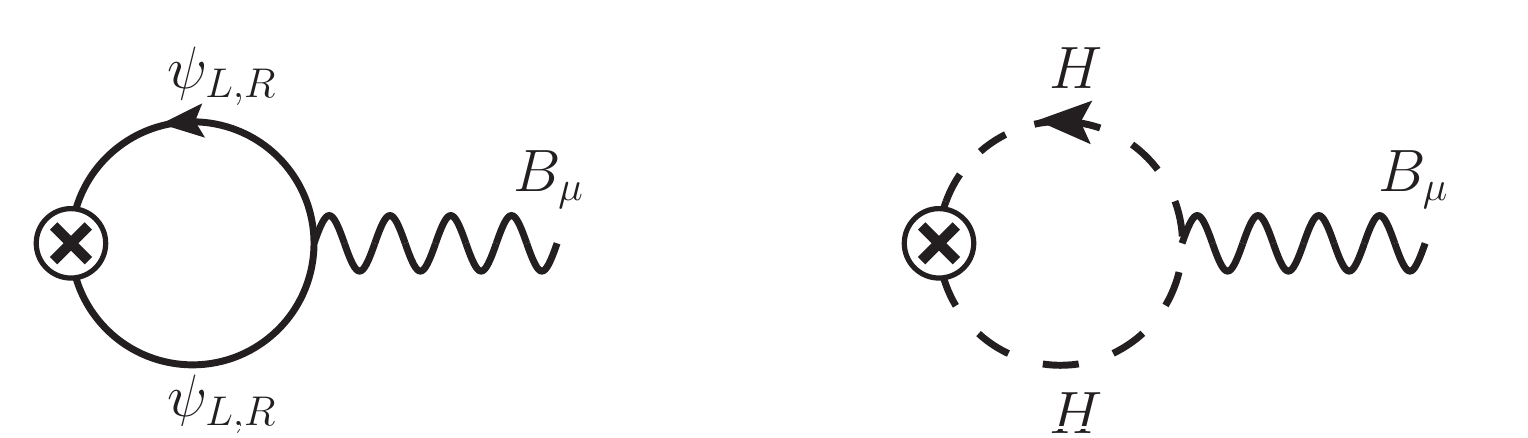}
\caption{One loop corrections to the Wilson coefficient $c_B$ of the redundant operator in the ${\rm SM}_\chi$ EFT. The crossed circle notation is the same as \Fig{fig:VertexFeynmanDiagrams1}.}
\label{fig:VertexFeynmanDiagrams3}
\end{figure}

We start from the loop corrections to the Wilson coefficient $c_L$, which can be induced by gauge interactions (diagonal renormalization) and by Yukawa interactions (off-diagonal renormalization). The associated Feynman diagrams are shown in \Fig{fig:VertexFeynmanDiagrams1}. Loop effects for $c_R^u$ and $c_R^d$ are analogous, with the important difference that right-handed fermions have no $SU(2)_L$ interactions, and therefore there are no diagrams with $W_\mu^i$ in the loop.

The analysis of loop corrections to $c_H$ involves many more Feynman diagrams. The associated operator describes the DM interaction with two Higgs bosons, and we expect by gauge invariance also diagrams with two Higgses and one electroweak gauge boson. We have computed all the possible one-loop diagrams, both the ones with only two Higgs fields on the external legs and the ones with an additional $W_\mu^i$ or a $B_\mu$ gauge boson on the external legs, and checked that they combine in a gauge invariant way to give the Higgs current defined in \Eq{eq:Higgsdouble}. We show in \Fig{fig:VertexFeynmanDiagrams2} only the diagrams with two external Higgs bosons, which get contributions from gauge and Yukawa interactions. For the latter, we checked that the associated diagrams with an external $B_\mu$ combine in a gauge invariant way with the others only after using the SM hypercharge values, as it should be. 

Finally, loop diagrams in \Fig{fig:VertexFeynmanDiagrams3} radiatively induce a contribution to the Wilson coefficient $c_B$ of the redundant operator in \Eq{eq:redundantSection2}. By consistently using the equation of motion, this translates into a shift for the independent operators as in Eqs.~(\ref{eq:replacementB1})-(\ref{eq:replacementB6}).\footnote{Alternatively, instead of dealing with a redundant operator, we can restrict ourselves to a minimal basis. In this case one has to compute one-loop corrections to Wilson coefficients coming from one-particle-reducible (penguin-type) diagrams. We explicitly checked that the two procedures lead to the same results, as it should be.}.

The explicit expression for the one-loop amplitudes, as well as the consequent derivation of the RG equations, are presented in \App{app:SMchi}. As shown there, the gauge interactions contribution to wave-function renormalization exactly cancels with the associated vertex corrections. This is consistent with the Ward identities in abelian gauge theories, and with the fact that non-abelian gauge interactions renormalize only axial currents starting at two loops. The final anomalous dimension matrix has two main pieces
\be
\gamma_{{\rm SM}_\chi} = \left. \gamma_{{\rm SM}_\chi} \right|_\lambda + \left. \gamma_{{\rm SM}_\chi} \right|_Y \ ,
\label{eq:gammatotalSMchi}
\ee
where the contribution proportional to the hypercharge gauge couplings is a consequence of the diagrams in \Fig{fig:VertexFeynmanDiagrams3}. The explicit expressions read
\be
\left. \!\!\!\!\!\! \gamma_{{\rm SM}_\chi} \right|_\lambda\! =\! \scalemath{0.58}{\frac{1}{8\pi^2}  
\left(\renewcommand{\arraystretch}{1.6}
\begin{array}{ccc:cc|ccc:cc|ccc:cc||c}
 \left(\lambda_u^2 + \lambda_d^2\right) / 2  & - \lambda_u^2 / 2 & - \lambda_d^2 / 2 & 0 & 0 & 0 & 0 & 0 & 0 & 0 & 0 & 0 & 0 & 0 & 0 & \left(\lambda_u^2 + \lambda_d^2\right) / 2 \\
- \lambda_u^2 & \lambda_u^2 & 0 & 0 & 0 & 0 & 0 & 0 & 0 & 0 & 0 & 0 & 0 & 0 & 0 & \lambda_u^2 \\
- \lambda_d^2 & 0 & \lambda_d^2 & 0 & 0 & 0 & 0 & 0 & 0 & 0 & 0 & 0 & 0 & 0 & 0 & \lambda_d^2 \\ \hdashline
0 & 0 & 0 & \lambda_e^2 / 2  &  - \lambda_e^2 / 2 & 0 & 0 & 0 & 0 & 0 & 0 & 0 & 0 & 0 & 0 & \lambda_e^2 / 2 \\
0 & 0 & 0 & - \lambda_e^2 & \lambda_e^2 & 0 & 0 & 0 & 0 & 0 & 0 & 0 & 0 & 0 & 0 & \lambda_e^2 \\ \hline
0 & 0 & 0 & 0 & 0 &\left(\lambda_c^2 + \lambda_s^2\right) / 2  & - \lambda_c^2 / 2 & - \lambda_s^2 / 2 & 0 & 0 & 0 & 0 & 0 & 0 & 0 & \left(\lambda_c^2 + \lambda_s^2\right) / 2  \\
0 & 0 & 0 & 0 & 0  & - \lambda_c^2 & \lambda_c^2 & 0 & 0 & 0 & 0 & 0 & 0 & 0 & 0  &  \lambda_c^2 \\
0 & 0 & 0 & 0 & 0  & - \lambda_s^2 & 0 & \lambda_s^2 & 0 & 0 & 0 & 0 & 0 & 0 & 0  & \lambda_s^2  \\ \hdashline
0 & 0 & 0 & 0 & 0 & 0 & 0 & 0 & \lambda_\mu^2 / 2  &  - \lambda_\mu^2 / 2 & 0 & 0 & 0 & 0 & 0 &  \lambda_\mu^2 / 2 \\
0 & 0 & 0 & 0 & 0 & 0 & 0 & 0 & - \lambda_\mu^2   &  \lambda_\mu^2  & 0 & 0 & 0 & 0 & 0 &  \lambda_\mu^2  \\ \hline
0 & 0 & 0 & 0 & 0 & 0 & 0 & 0 & 0 & 0 & \left(\lambda_t^2 + \lambda_b^2\right) / 2  & - \lambda_t^2 / 2 & - \lambda_b^2 / 2  & 0 & 0 &  \left(\lambda_t^2 + \lambda_b^2\right) / 2 \\
0 & 0 & 0 & 0 & 0 & 0 & 0 & 0 & 0 & 0 & - \lambda_t^2 & \lambda_t^2 & 0 & 0 & 0 &  \lambda_t^2 \\
0 & 0 & 0 & 0 & 0 & 0 & 0 & 0 & 0 & 0 & - \lambda_b^2 & 0 & \lambda_b^2 & 0 & 0 &  \lambda_b^2 \\ \hdashline
0 & 0 & 0 & 0 & 0 & 0 & 0 & 0 & 0 & 0 & 0 & 0 & 0 & \lambda_\tau^2 / 2  & - \lambda_\tau^2 / 2 & \lambda_\tau^2 / 2 \\
0 & 0 & 0 & 0 & 0 & 0 & 0 & 0 & 0 & 0 & 0 & 0 & 0 & - \lambda_\tau^2 & \lambda_\tau^2 & \lambda_\tau^2 \\  \hline\hline
3 \left(\lambda_u^2 - \lambda_d^2\right) &  - 3 \lambda_u^2 &  3 \lambda_d^2  &  - \lambda_e^2 & \lambda_e^2  &  
3 \left(\lambda_c^2 - \lambda_s^2\right) &  - 3 \lambda_c^2 &  3 \lambda_s^2  &  - \lambda_\mu^2 & \lambda_\mu^2 & 3 \left(\lambda_t^2 - \lambda_b^2\right) &  - 3 \lambda_t^2 &  3 \lambda_b^2 & - \lambda_\tau^2 &  \lambda_\tau^2  & 3 \sum_q \lambda_q^2 + \sum_l \lambda_l^2
\end{array}\right)}, 
\label{eq:gammalambda}
\ee
for the Yukawa contribution, and 
\be
\left. \!\!\!\!\!\!\gamma_{{\rm SM}_\chi} \right|_Y \!=\! \scalemath{0.88}{\frac{4}{3} \frac{g^{\prime\,2}}{16 \pi^2}  
 \tiny  \left(\renewcommand{\arraystretch}{1.9}
 \begin{array}{ccc:cc|ccc:cc|ccc:cc||c}
6 y_q^2 & 3 y_q y_u  & 3 y_q y_d & 2 y_q y_l & y_q y_e & 6 y_q^2 & 3 y_q y_u  & 3 y_q y_d & 2 y_q y_l & y_q y_e & 
6 y_q^2 & 3 y_q y_u  & 3 y_q y_d & 2 y_q y_l & y_q y_e & y_q y_H \\
6 y_u y_q & 3 y^2_u  & 3 y_u  y_d & 2 y_u  y_l & y_u  y_e & 6 y_u y_q & 3 y_u^2  & 3 y_u y_d & 2 y_u  y_l & y_u y_e & 
6 y_u y_q & 3 y^2_u  & 3 y_u  y_d & 2 y_u  y_l & y_u  y_e & y_u y_H \\
6 y_d y_q & 3 y_d y_u  & 3 y^2_d & 2 y_d  y_l & y_d  y_e & 6 y_d y_q & 3 y_d y_u  & 3 y^2_d & 2 y_d  y_l & y_d  y_e & 
6 y_d y_q & 3 y_d y_u  & 3 y^2_d & 2 y_d  y_l & y_d  y_e & y_d y_H \\ \hdashline
6 y_l y_q & 3 y_l y_u  & 3 y_l y_d & 2 y^2_l & y_l  y_e & 6 y_l y_q & 3 y_l y_u  & 3 y_l y_d & 2 y^2_l & y_l  y_e & 
6 y_l y_q & 3 y_l y_u  & 3 y_l y_d & 2 y^2_l & y_l  y_e & y_l y_H \\
6 y_e y_q & 3 y_e y_u  & 3 y_e y_d & 2 y_e y_l & y^2_e & 6 y_e y_q & 3 y_e y_u  & 3 y_e y_d & 2 y_e y_l & y^2_e & 
6 y_e y_q & 3 y_e y_u  & 3 y_e y_d & 2 y_e y_l & y^2_e & y_e y_H \\ \hline
6 y_q^2 & 3 y_q y_u  & 3 y_q y_d & 2 y_q y_l & y_q y_e & 6 y_q^2 & 3 y_q y_u  & 3 y_q y_d & 2 y_q y_l & y_q y_e & 
6 y_q^2 & 3 y_q y_u  & 3 y_q y_d & 2 y_q y_l & y_q y_e & y_q y_H \\
6 y_u y_q & 3 y^2_u  & 3 y_u  y_d & 2 y_u  y_l & y_u  y_e & 6 y_u y_q & 3 y_u^2  & 3 y_u y_d & 2 y_u  y_l & y_u y_e & 
6 y_u y_q & 3 y^2_u  & 3 y_u  y_d & 2 y_u  y_l & y_u  y_e & y_u y_H \\
6 y_d y_q & 3 y_d y_u  & 3 y^2_d & 2 y_d  y_l & y_d  y_e & 6 y_d y_q & 3 y_d y_u  & 3 y^2_d & 2 y_d  y_l & y_d  y_e & 
6 y_d y_q & 3 y_d y_u  & 3 y^2_d & 2 y_d  y_l & y_d  y_e & y_d y_H \\ \hdashline
6 y_l y_q & 3 y_l y_u  & 3 y_l y_d & 2 y^2_l & y_l  y_e & 6 y_l y_q & 3 y_l y_u  & 3 y_l y_d & 2 y^2_l & y_l  y_e & 
6 y_l y_q & 3 y_l y_u  & 3 y_l y_d & 2 y^2_l & y_l  y_e & y_l y_H \\
6 y_e y_q & 3 y_e y_u  & 3 y_e y_d & 2 y_e y_l & y^2_e & 6 y_e y_q & 3 y_e y_u  & 3 y_e y_d & 2 y_e y_l & y^2_e & 
6 y_e y_q & 3 y_e y_u  & 3 y_e y_d & 2 y_e y_l & y^2_e & y_e y_H \\ \hline
6 y_q^2 & 3 y_q y_u  & 3 y_q y_d & 2 y_q y_l & y_q y_e & 6 y_q^2 & 3 y_q y_u  & 3 y_q y_d & 2 y_q y_l & y_q y_e & 
6 y_q^2 & 3 y_q y_u  & 3 y_q y_d & 2 y_q y_l & y_q y_e & y_q y_H \\
6 y_u y_q & 3 y^2_u  & 3 y_u  y_d & 2 y_u  y_l & y_u  y_e & 6 y_u y_q & 3 y_u^2  & 3 y_u y_d & 2 y_u  y_l & y_u y_e & 
6 y_u y_q & 3 y^2_u  & 3 y_u  y_d & 2 y_u  y_l & y_u  y_e & y_u y_H \\
6 y_d y_q & 3 y_d y_u  & 3 y^2_d & 2 y_d  y_l & y_d  y_e & 6 y_d y_q & 3 y_d y_u  & 3 y^2_d & 2 y_d  y_l & y_d  y_e & 
6 y_d y_q & 3 y_d y_u  & 3 y^2_d & 2 y_d  y_l & y_d  y_e & y_d y_H \\ \hdashline
6 y_l y_q & 3 y_l y_u  & 3 y_l y_d & 2 y^2_l & y_l  y_e & 6 y_l y_q & 3 y_l y_u  & 3 y_l y_d & 2 y^2_l & y_l  y_e & 
6 y_l y_q & 3 y_l y_u  & 3 y_l y_d & 2 y^2_l & y_l  y_e & y_l y_H \\
6 y_e y_q & 3 y_e y_u  & 3 y_e y_d & 2 y_e y_l & y^2_e & 6 y_e y_q & 3 y_e y_u  & 3 y_e y_d & 2 y_e y_l & y^2_e & 
6 y_e y_q & 3 y_e y_u  & 3 y_e y_d & 2 y_e y_l & y^2_e & y_e y_H \\ \hline\hline
6 y_H y_q & 3 y_H y_u  & 3 y_H y_d & 2 y_H y_l & y_H y_e & 6 y_H y_q & 3 y_H y_u  & 3 y_H y_d & 2 y_H y_l & y_H y_e &  
6 y_H y_q & 3 y_H y_u  & 3 y_H y_d & 2 y_H y_l & y_H y_e &  y^2_H
\end{array}\right)}, 
\label{eq:gammaY}
\ee
with the hypercharges as in \Eq{eq:SMhypervalues}.


\subsection{From the EWSB scale to the nuclear scale}
\label{sec:RG2}

The Wilson coefficients given in \Eq{ed:cdef2} for the EFT below the EWSB scale evolve according to
\be
\frac{d \, \mathcal{C}_{{\rm EMSM}_\chi}}{d \ln \mu}  = \gamma_{{\rm EMSM}_\chi} \mathcal{C}_{{\rm EMSM}_\chi}.
\ee
We now discuss how to obtain the $16\times 16$ anomalous dimension matrix $ \gamma_{{\rm EMSM}_\chi}$.
\begin{figure}
\centering
\includegraphics[scale=0.45]{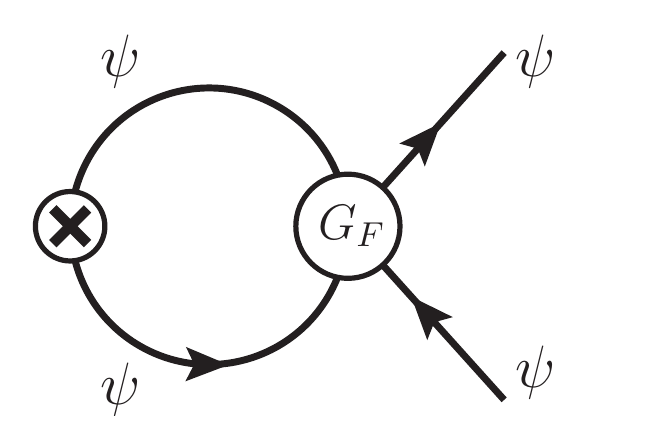} 
\caption{Vertex corrections in the ${\rm EMSM}_\chi$ EFT induced by SM four-fermion interactions.}
\label{fig:VertexFeynmanDiagrams4}
\end{figure}
The external leg corrections only come from the gauge sector. For strong interactions they are identical to the ones in the ${\rm SM}_\chi$ EFT, and for electromagnetic interactions they can be easily obtained from the analogous hypercharge diagrams. Their effect is again to cancel out against the associated vertex corrections.

Also in this case there are two classes of vertex corrections. The first ones are due to the SM four-fermion interactions, in the way we show in \Fig{fig:VertexFeynmanDiagrams4}. This diagram in the ${\rm EMSM}_\chi$ EFT is the analogous of the correction to $c_H$ discussed in the ${\rm SM}_\chi$ EFT, but the $Z$ boson is integrated out in this phase of the theory. Despite the fact that these diagrams are suppressed by the Fermi constant, we keep them to be consistent with the analysis above the EWSB scale, since their contribution is proportional to $G_F \, m_\psi^2 \propto \lambda_\psi^2$.

The second effect is the radiative correction to the Wilson coefficient $c_F$ of the redundant operator in \Eq{eq:redundantphotonSection2}. The diagrams are analogous to the fermion loop in \Fig{fig:VertexFeynmanDiagrams3}, but this time with an external photon. We evaluate them, and use the equations of motion to induce the evolution of the independent operators.\footnote{If a minimal operator basis is used, photon penguin diagrams have to be evaluated. We checked that we obtain the same results with both procedures.}
The anomalous dimension is
\be
\gamma_{{\rm EMSM}_\chi} = \left. \gamma_{{\rm EMSM}_\chi} \right|_m + \left. \gamma_{{\rm EMSM}_\chi} \right|_{\rm em} \ .
\ee
The running driven by fermion masses reads
\be
\left.\!\!\!\!\!\!\!\!\gamma_{{\rm EMSM}_\chi} \right|_m  \!=\!  \scalemath{0.62}{ \frac{\sqrt{2} G_F}{\pi^2}  
\left(\renewcommand{\arraystretch}{1.6}
\begin{array}{ccccc|ccc||ccccc|ccc}
0 & 0 & 0 & 0 & 0 & 0 & 0 & 0 & 3 m_u^2 g_{Au} \, g_{Vu} & 3 m_d^2 g_{Ad} \, g_{Vu} & 3 m_c^2 g_{Au} \, g_{Vu} & 3 m_s^2 g_{Ad} \, g_{Vu} & 3 m_b^2 g_{Ad} \, g_{Vu} & m_e^2 g_{Ae} \, g_{Vu} & m_\mu^2 g_{Ae} \, g_{Vu} & m_\tau^2 g_{Ae} \, g_{Vu} \\ 
0 & 0 & 0 & 0 & 0 & 0 & 0 & 0 & 3 m_u^2 g_{Au} \, g_{Vd} & 3 m_d^2 g_{Ad} \, g_{Vd} & 3 m_c^2 g_{Au} \, g_{Vd} & 3 m_s^2 g_{Ad} \, g_{Vd} & 3 m_b^2 g_{Ad} \, g_{Vd} & m_e^2 g_{Ae} \, g_{Vd} & m_\mu^2 g_{Ae} \, g_{Vd} & m_\tau^2 g_{Ae} \, g_{Vd} \\ 
0 & 0 & 0 & 0 & 0 & 0 & 0 & 0 & 3 m_u^2 g_{Au} \, g_{Vu} & 3 m_d^2 g_{Ad} \, g_{Vu} & 3 m_c^2 g_{Au} \, g_{Vu} & 3 m_s^2 g_{Ad} \, g_{Vu} & 3 m_b^2 g_{Ad} \, g_{Vu} & m_e^2 g_{Ae} \, g_{Vu} & m_\mu^2 g_{Ae} \, g_{Vu} & m_\tau^2 g_{Ae} \, g_{Vu} \\ 
0 & 0 & 0 & 0 & 0 & 0 & 0 & 0 & 3 m_u^2 g_{Au} \, g_{Vd} & 3 m_d^2 g_{Ad} \, g_{Vd} & 3 m_c^2 g_{Au} \, g_{Vd} & 3 m_s^2 g_{Ad} \, g_{Vd} & 3 m_b^2 g_{Ad} \, g_{Vd} & m_e^2 g_{Ae} \, g_{Vd} & m_\mu^2 g_{Ae} \, g_{Vd} & m_\tau^2 g_{Ae} \, g_{Vd} \\ 
0 & 0 & 0 & 0 & 0 & 0 & 0 & 0 & 3 m_u^2 g_{Au} \, g_{Vd} & 3 m_d^2 g_{Ad} \, g_{Vd} & 3 m_c^2 g_{Au} \, g_{Vd} & 3 m_s^2 g_{Ad} \, g_{Vd} & 3 m_b^2 g_{Ad} \, g_{Vd} & m_e^2 g_{Ae} \, g_{Vd} & m_\mu^2 g_{Ae} \, g_{Vd} & m_\tau^2 g_{Ae} \, g_{Vd} \\  \hline 
0 & 0 & 0 & 0 & 0 & 0 & 0 & 0 & 3 m_u^2 g_{Au} \, g_{Ve} & 3 m_d^2 g_{Ad} \, g_{Ve} & 3 m_c^2 g_{Au} \, g_{Ve} & 3 m_s^2 g_{Ad} \, g_{Ve} & 3 m_b^2 g_{Ad} \, g_{Ve} & m_e^2 g_{Ae} \, g_{Ve} & m_\mu^2 g_{Ae} \, g_{Ve} & m_\tau^2 g_{Ae} \, g_{Ve} \\ 
0 & 0 & 0 & 0 & 0 & 0 & 0 & 0 & 3 m_u^2 g_{Au} \, g_{Ve} & 3 m_d^2 g_{Ad} \, g_{Ve} & 3 m_c^2 g_{Au} \, g_{Ve} & 3 m_s^2 g_{Ad} \, g_{Ve} & 3 m_b^2 g_{Ad} \, g_{Ve} & m_e^2 g_{Ae} \, g_{Ve} & m_\mu^2 g_{Ae} \, g_{Ve} & m_\tau^2 g_{Ae} \, g_{Ve} \\ 
0 & 0 & 0 & 0 & 0 & 0 & 0 & 0 & 3 m_u^2 g_{Au} \, g_{Ve} & 3 m_d^2 g_{Ad} \, g_{Ve} & 3 m_c^2 g_{Au} \, g_{Ve} & 3 m_s^2 g_{Ad} \, g_{Ve} & 3 m_b^2 g_{Ad} \, g_{Ve} & m_e^2 g_{Ae} \, g_{Ve} & m_\mu^2 g_{Ae} \, g_{Ve} & m_\tau^2 g_{Ae} \, g_{Ve} \\ \hline \hline
0 & 0 & 0 & 0 & 0 & 0 & 0 & 0 & 3 m_u^2 g_{Au} \, g_{Au} & 3 m_d^2 g_{Ad} \, g_{Au} & 3 m_c^2 g_{Au} \, g_{Au} & 3 m_s^2 g_{Ad} \, g_{Au} & 3 m_b^2 g_{Ad} \, g_{Au} & m_e^2 g_{Ae} \, g_{Au} & m_\mu^2 g_{Ae} \, g_{Au} & m_\tau^2 g_{Ae} \, g_{Au} \\ 
0 & 0 & 0 & 0 & 0 & 0 & 0 & 0 & 3 m_u^2 g_{Au} \, g_{Ad} & 3 m_d^2 g_{Ad} \, g_{Ad} & 3 m_c^2 g_{Au} \, g_{Ad} & 3 m_s^2 g_{Ad} \, g_{Ad} & 3 m_b^2 g_{Ad} \, g_{Ad} & m_e^2 g_{Ae} \, g_{Ad} & m_\mu^2 g_{Ae} \, g_{Ad} & m_\tau^2 g_{Ae} \, g_{Ad} \\ 
0 & 0 & 0 & 0 & 0 & 0 & 0 & 0 & 3 m_u^2 g_{Au} \, g_{Au} & 3 m_d^2 g_{Ad} \, g_{Au} & 3 m_c^2 g_{Au} \, g_{Au} & 3 m_s^2 g_{Ad} \, g_{Au} & 3 m_b^2 g_{Ad} \, g_{Au} & m_e^2 g_{Ae} \, g_{Au} & m_\mu^2 g_{Ae} \, g_{Au} & m_\tau^2 g_{Ae} \, g_{Au} \\ 
0 & 0 & 0 & 0 & 0 & 0 & 0 & 0 & 3 m_u^2 g_{Au} \, g_{Ad} & 3 m_d^2 g_{Ad} \, g_{Ad} & 3 m_c^2 g_{Au} \, g_{Ad} & 3 m_s^2 g_{Ad} \, g_{Ad} & 3 m_b^2 g_{Ad} \, g_{Ad} & m_e^2 g_{Ae} \, g_{Ad} & m_\mu^2 g_{Ae} \, g_{Ad} & m_\tau^2 g_{Ae} \, g_{Ad} \\ 
0 & 0 & 0 & 0 & 0 & 0 & 0 & 0 & 3 m_u^2 g_{Au} \, g_{Ad} & 3 m_d^2 g_{Ad} \, g_{Ad} & 3 m_c^2 g_{Au} \, g_{Ad} & 3 m_s^2 g_{Ad} \, g_{Ad} & 3 m_b^2 g_{Ad} \, g_{Ad} & m_e^2 g_{Ae} \, g_{Ad} & m_\mu^2 g_{Ae} \, g_{Ad} & m_\tau^2 g_{Ae} \, g_{Ad} \\  \hline 
0 & 0 & 0 & 0 & 0 & 0 & 0 & 0 & 3 m_u^2 g_{Au} \, g_{Ae} & 3 m_d^2 g_{Ad} \, g_{Ae} & 3 m_c^2 g_{Au} \, g_{Ae} & 3 m_s^2 g_{Ad} \, g_{Ae} & 3 m_b^2 g_{Ad} \, g_{Ae} & m_e^2 g_{Ae} \, g_{Ae} & m_\mu^2 g_{Ae} \, g_{Ae} & m_\tau^2 g_{Ae} \, g_{Ae} \\ 
0 & 0 & 0 & 0 & 0 & 0 & 0 & 0 & 3 m_u^2 g_{Au} \, g_{Ae} & 3 m_d^2 g_{Ad} \, g_{Ae} & 3 m_c^2 g_{Au} \, g_{Ae} & 3 m_s^2 g_{Ad} \, g_{Ae} & 3 m_b^2 g_{Ad} \, g_{Ae} & m_e^2 g_{Ae} \, g_{Ae} & m_\mu^2 g_{Ae} \, g_{Ae} & m_\tau^2 g_{Ae} \, g_{Ae} \\ 
0 & 0 & 0 & 0 & 0 & 0 & 0 & 0 & 3 m_u^2 g_{Au} \, g_{Ae} & 3 m_d^2 g_{Ad} \, g_{Ae} & 3 m_c^2 g_{Au} \, g_{Ae} & 3 m_s^2 g_{Ad} \, g_{Ae} & 3 m_b^2 g_{Ad} \, g_{Ae} & m_e^2 g_{Ae} \, g_{Ae} & m_\mu^2 g_{Ae} \, g_{Ae} & m_\tau^2 g_{Ae} \, g_{Ae} 
\end{array}\right)},  
\ee
whereas the electromagnetic interactions give \footnote{We correct an overall sign typo in the $5 \times 5$ quark block given in \Ref{Crivellin:2014qxa}.}
\be
\left. \!\!\!\!\!\!\!\!\gamma_{{\rm EMSM}_\chi} \right|_{\rm em}  \!=\!  \scalemath{0.77}{\frac{8}{3} \frac{e^2}{16 \pi^2}
\left(\renewcommand{\arraystretch}{1.3}
\begin{array}{ccccc|ccc||ccccc|ccc}
3 Q_u^2 & 3 Q_u Q_d & 3 Q_u^2 & 3 Q_u Q_d & 3 Q_u Q_d & Q_u Q_e & Q_u Q_e & Q_u Q_e & 0 & 0 & 0 & 0 & 0 & 0 & 0 & 0 \\
3 Q_d Q_u & 3 Q_d^2 & 3 Q_d Q_u & 3 Q_d^2 & 3 Q_d^2 & Q_d Q_e & Q_d Q_e & Q_d Q_e & 0 & 0 & 0 & 0 & 0 & 0 & 0 & 0 \\
3 Q_u^2 & 3 Q_u Q_d & 3 Q_u^2 & 3 Q_u Q_d & 3 Q_u Q_d & Q_u Q_e & Q_u Q_e & Q_u Q_e & 0 & 0 & 0 & 0 & 0 & 0 & 0 & 0 \\
3 Q_d Q_u & 3 Q_d^2 & 3 Q_d Q_u & 3 Q_d^2 & 3 Q_d^2 & Q_d Q_e & Q_d Q_e & Q_d Q_e & 0 & 0 & 0 & 0 & 0 & 0 & 0 & 0 \\
3 Q_d Q_u & 3 Q_d^2 & 3 Q_d Q_u & 3 Q_d^2 & 3 Q_d^2 & Q_d Q_e & Q_d Q_e & Q_d Q_e & 0 & 0 & 0 & 0 & 0 & 0 & 0 & 0 \\ \hline 
3 Q_e Q_u & 3 Q_e Q_d & 3 Q_e Q_u & 3 Q_e Q_d & 3 Q_e Q_d & Q_e^2 & Q_e^2 & Q_e^2 & 0 & 0 & 0 & 0 & 0 & 0 & 0 & 0 \\
3 Q_e Q_u & 3 Q_e Q_d & 3 Q_e Q_u & 3 Q_e Q_d & 3 Q_e Q_d & Q_e^2 & Q_e^2 & Q_e^2 & 0 & 0 & 0 & 0 & 0 & 0 & 0 & 0 \\
3 Q_e Q_u & 3 Q_e Q_d & 3 Q_e Q_u & 3 Q_e Q_d & 3 Q_e Q_d & Q_e^2 & Q_e^2 & Q_e^2 & 0 & 0 & 0 & 0 & 0 & 0 & 0 & 0 \\ \hline \hline
0 & 0 & 0 & 0 & 0 & 0 & 0 & 0 & 0 & 0 & 0 & 0 & 0 & 0 & 0 & 0 \\
0 & 0 & 0 & 0 & 0 & 0 & 0 & 0 & 0 & 0 & 0 & 0 & 0 & 0 & 0 & 0 \\
0 & 0 & 0 & 0 & 0 & 0 & 0 & 0 & 0 & 0 & 0 & 0 & 0 & 0 & 0 & 0 \\
0 & 0 & 0 & 0 & 0 & 0 & 0 & 0 & 0 & 0 & 0 & 0 & 0 & 0 & 0 & 0 \\
0 & 0 & 0 & 0 & 0 & 0 & 0 & 0 & 0 & 0 & 0 & 0 & 0 & 0 & 0 & 0 \\ \hline
0 & 0 & 0 & 0 & 0 & 0 & 0 & 0 & 0 & 0 & 0 & 0 & 0 & 0 & 0 & 0 \\
0 & 0 & 0 & 0 & 0 & 0 & 0 & 0 & 0 & 0 & 0 & 0 & 0 & 0 & 0 & 0 \\
0 & 0 & 0 & 0 & 0 & 0 & 0 & 0 & 0 & 0 & 0 & 0 & 0 & 0 & 0 & 0
\end{array}\right)}.
\label{eq:gammaEM}
\ee


\section{Applications to Spin-Independent Searches}
\label{sec:App}

In \Sec{sec:RG} we presented the full $16 \times 16$ anomalous dimension matrices describing the one-loop RG evolution for dimension 6 operators above and below the EWSB scale. As promised in the introduction, these details can be skipped by a reader only interested in our final results. For the benefit of such a reader, we now briefly summarize the RG procedure.

The boundary conditions for the RG system are the ${\rm SM}_\chi$ EFT Wilson coefficients at the cutoff $\Lambda$. In a generic UV complete model with mediators heavier than the weak scale, they are obtained by integrating out the mediator fields. Then we evolve them down to the EWSB scale, which we take equal to the $Z$ boson mass. It is convenient to introduce the following dimensionless variable related to the renormalization scale $\mu$,
\be
t \equiv \ln \left[ \frac{\mu}{m_Z}\right] \ .
\ee
In this notation the matching is performed at $ t = 0$, whereas the Wilson coefficient $c_\Lambda$ are specified at the cutoff scale, $t_\Lambda = \ln\left[\Lambda / m_Z\right]$. The RG evolution in the ${\rm SM}_\chi$ EFT is obtained by solving the system of differential equations
\begin{align}
\frac{d \, \mathcal{C}_{{\rm SM}_\chi}}{d t}  = &\,  \gamma_{{\rm SM}_\chi} \mathcal{C}_{{\rm SM}_\chi} \ , \qquad \qquad \qquad 0 \leq t \leq t_\Lambda \ , \\
\mathcal{C}_{{\rm SM}_\chi}(t_\Lambda) = & \, c_\Lambda \ ,
\end{align}
with the Wilson coefficients vector $\mathcal{C}_{{\rm SM}_\chi}$ defined in \Eq{ed:cdef}, and the explicitly expression for the anomalous dimension matrix $\gamma_{{\rm SM}_\chi}$ given in \Sec{sec:RG1}. Once at $t = 0$, we perform the matching between the two theories as described in \Sec{sec:matching}. The subsequent RG evolution for the Wilson coefficients $\mathcal{C}_{{\rm EMSM}_\chi}$ defined in \Eq{ed:cdef2} is described by
\be
\frac{d \, \mathcal{C}_{{\rm EMSM}_\chi}}{d t}  = \gamma_{{\rm EMSM}_\chi} \mathcal{C}_{{\rm EMSM}_\chi} \ , \qquad \qquad \qquad t_N \leq t \leq 0 \ , 
\ee
with the explicit $\gamma_{{\rm EMSM}_\chi}$ given in \Sec{sec:RG2} and $t_N = \ln\left[1 \, {\rm GeV} / m_Z \right] \simeq - 4.51$. The outcome of this three-step procedure is the array of Wilson coefficients at the nuclear scale $c_N$. We only perform linear operations on the Wilson coefficients, therefore we have
\be
c_N = U_\Lambda c_\Lambda \ .
\label{eq:linearevolution}
\ee
The $\Lambda$-dependent evolution matrix $U_\Lambda$ is derived in \App{app:RGsol} and for a user-friendly recipe we refer to \App{app:recipe}. 

The rest of this Section is devoted to applying \Eq{eq:linearevolution} to limits from direct detection experiments. We focus on spin-independent searches, since they have much stronger bounds, and this has two implications. First, we need to consider effective operators with DM vector currents $\overline{\chi} \gamma^\mu \chi$. For pure elastic scattering this operator is non vanishing only for Dirac fermions, but our results are also valid for inelastic scattering of two splitted Majorana states~\cite{TuckerSmith:2001hy}. Second, matrix elements of SM fermion currents have only contributions from valence quarks in the target nuclei, therefore the direct detection cross section at zero momentum transfer and low DM velocities reads
 \be
\sigma_{\mathcal{N}}^{\rm SI} = \frac{m_\chi^2\, m_\mathcal{N}^2}{(m_\chi + m_\mathcal{N})^2\,\pi\,\Lambda^4}\,
\left| c_{VV u}^{(1)}  (A+Z) + c_{VV d}^{(1)} (2 A - Z)  \right|^2 \ .
\label{eq:SIsigma}
 \ee
Here, $c_{VV u}^{(1)}$ and $c_{VV d}^{(1)}$ are the first two component of the vector defined in \Eq{ed:cdef2}, whereas $A$, $Z$ and $m_\mathcal{N}$ are the mass number, atomic number and mass of the target nucleus $\mathcal{N}$, respectively.

In what follows, we consider specific choices of Wilson coefficients $c_\Lambda$ at the cutoff scale and we evolve them down to the nuclear scale as in \Eq{eq:linearevolution}. The running of the Yukawa couplings above the EWSB scale is performed according to \Ref{Buttazzo:2013uya} and of the quark masses below $m_Z$ using the results in \Ref{Xing:2011aa}. We compare the predicted rate as in \Eq{eq:SIsigma} to the experimental limits, and extract bounds on the Wilson coefficients. Our results are model independent, in the sense that every UV complete model generating that specific set $c_\Lambda$ when matched on the ${\rm SM}_\chi$ is subject to our constraints. 

\subsection{D5 and D7 operators}
\label{sec:D5D7}

The connection between different DM negative searches is often expressed in terms of limits on the coefficients for the effective operators introduced in \Ref{Goodman:2010ku}. For a vector current of a fermion WIMP, the relevant operators involving quarks are
\begin{align}
\label{eq:D5def} \mathcal{L}_{\rm D5} = & \, \frac{c_{{\rm D5}}}{\Lambda^2} \; \overline{\chi} \,\gamma^\mu \chi \;
\left[ \sum_i \overline{u^i} \gamma_\mu u^i + \sum_i \overline{d^i} \gamma_\mu d^i \right] \ ,  \\ 
\label{eq:D7def} \mathcal{L}_{\rm D7} = & \, \frac{c_{{\rm D7}}}{\Lambda^2} \; \overline{\chi} \,\gamma^\mu \chi \;
\left[ \sum_i \overline{u^i} \gamma_\mu \gamma_5 u^i + \sum_i \overline{d^i} \gamma_\mu \gamma_5 d^i  \right] \ .
\end{align}
We now connect this description to the notation used in this paper, and explore the consequences of connecting EFT scales.

Keeping the complementarity among different searches in mind (e.g. between collider and direct searches as in \Ref{Goodman:2010ku}), we take the operators in \Eqs{eq:D5def}{eq:D7def} as defined at the EFT cutoff $\Lambda$. In other words these are operators in the ${\rm SM}_\chi$ EFT. Considering flavor universal coupling to SM quarks, D5 and D7 are reproduced by this set of Wilson coefficients
\be
\left. c_\Lambda^T\right|_{\rm D5, D7} =  \left(\begin{array}{ccc:cc|ccc:cc|ccc:cc||c}
c_L & c_R & c_R & 0 & 0 & 
c_L  & c_R & c_R & 0 & 0 &
c_L & c_R & c_R & 0 & 0 & 0 \end{array}\right) \ ,
\ee
where
\begin{align}
c_{{\rm D5}} = & \, \frac{c_L + c_R}{2}  \ , \\
c_{{\rm D7}} = & \, \frac{- c_L + c_R}{2}  \ . 
\end{align}

\begin{figure}
\begin{center}
\includegraphics[scale=0.49]{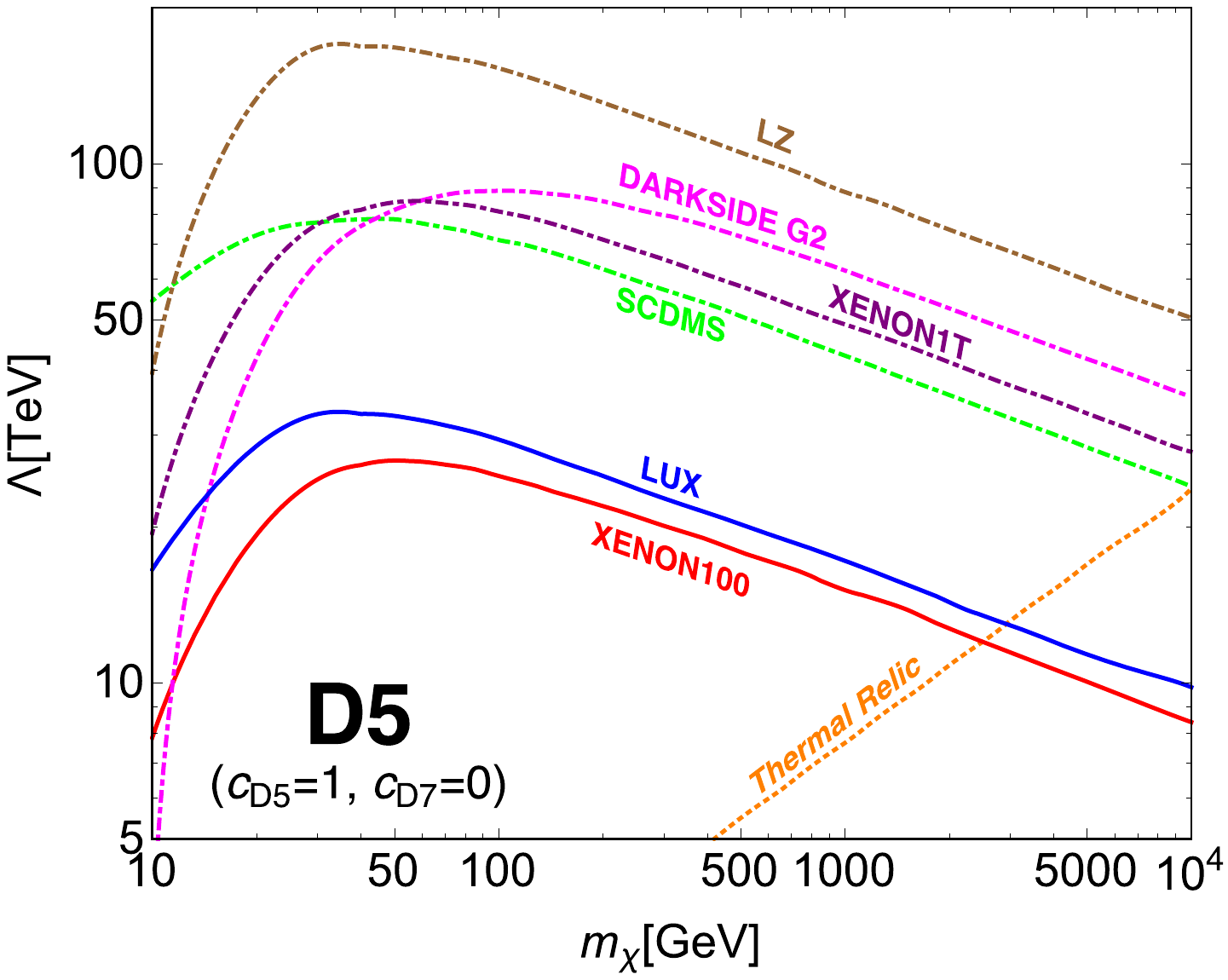} $\quad$ \includegraphics[scale=0.48]{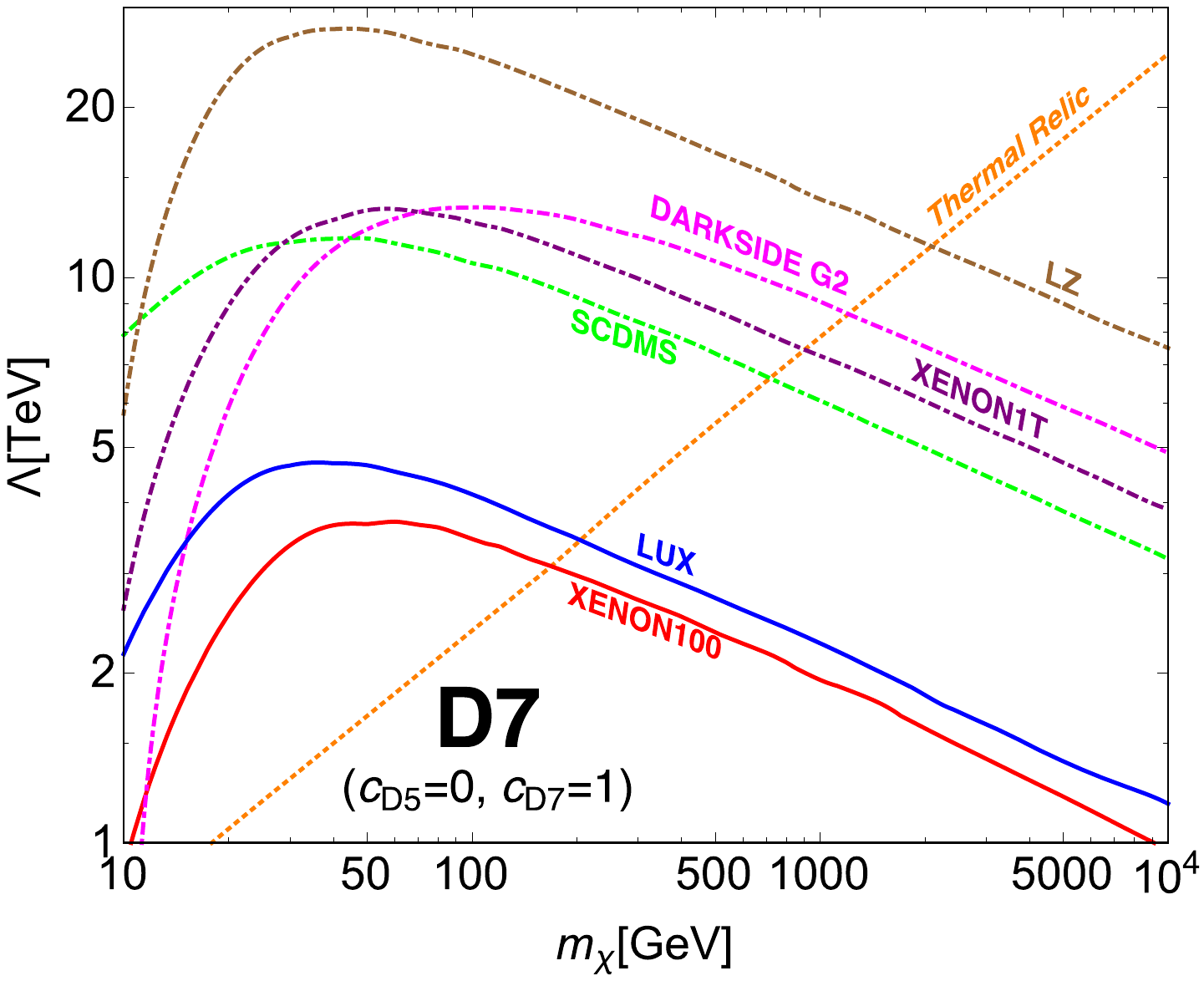}  \vspace{0.5cm} \\
\includegraphics[scale=0.49]{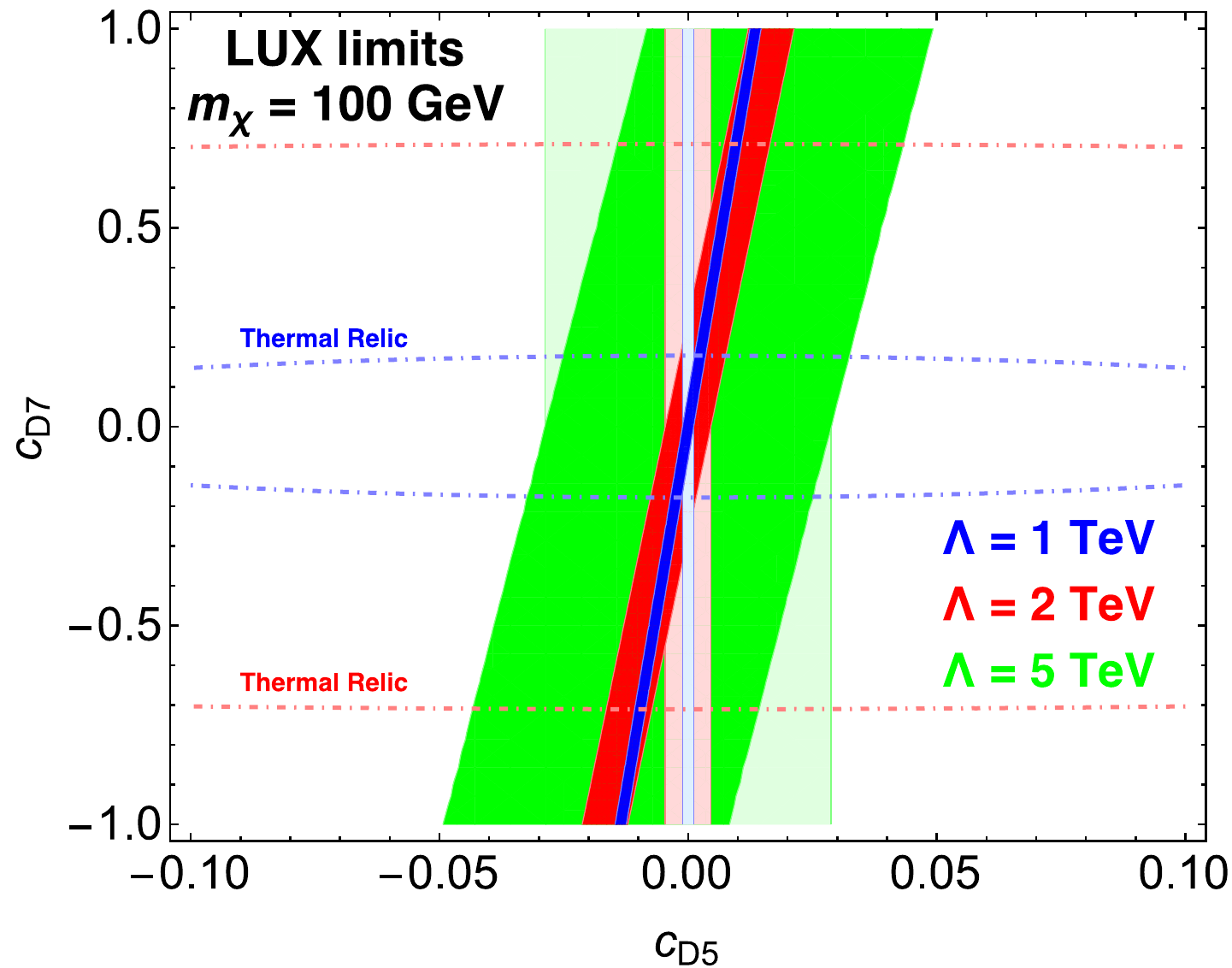} $\quad$ \includegraphics[scale=0.48]{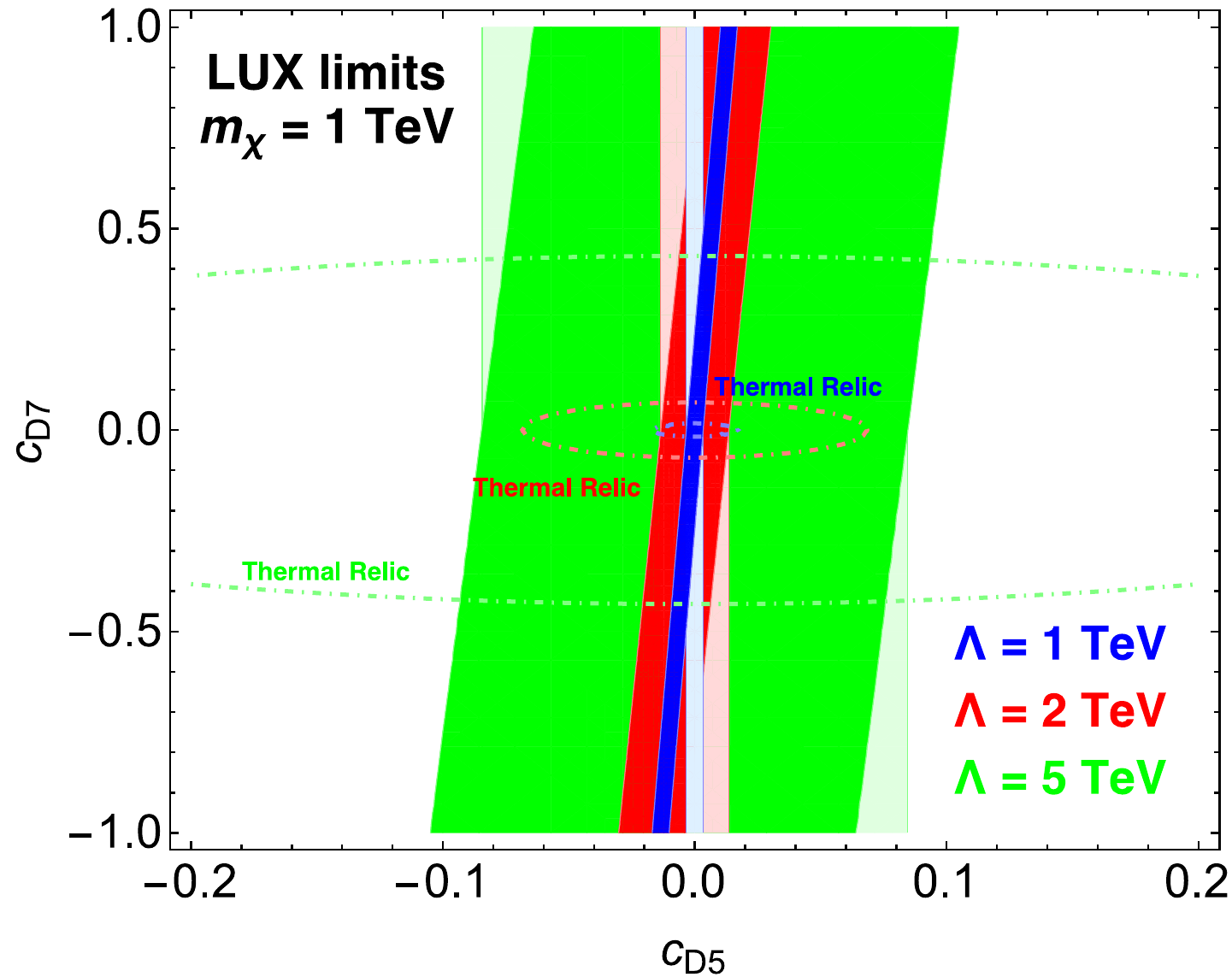}  
\end{center}
\caption{Experimental limits from direct detection for the D5 and D7 operators. In the top-left (right) panel we consider only D5 (D7) switched on at the scale $\Lambda$, and plot the lower bounds on $\Lambda$ from XENON100 (solid, red)~\cite{Aprile:2012nq} and LUX (solid, blue)~\cite{Akerib:2013tjd}, as well as projected limits from SCDMS (dot-dashed, green),  XENON1T (dot-dashed, purple), DARKSIDE G2 (dot-dashed, magenta), LZ (dot-dashed, brown)~\cite{Cushman:2013zza}. The dotted orange line gives the correct thermal relic density. In the bottom panel we fix $m_\chi$ and plot the region allowed by LUX in the $(c_{D5}, c_{D7})$ plane for three different values of $\Lambda$. The faded bands, which only constrain the vector coupling $c_{\rm D5}$, show the limits which would be obtained ignoring our analysis. We also plot the thermal relic lines whenever they are in the parameter space region under consideration.}
\label{fig:App1}
\end{figure}

Our results are shown in the four panels of \Fig{fig:App1}. In the top-left panel we consider the case where only D5 is switched on, and plot current and projected experimental limits in the $(m_\chi, \Lambda)$ plane for $c_{\rm D5} = 1$. As is well known, quite high scales for the mediator masses are necessary to be consistent with experimental exclusion bounds. We gain valuable information from this plot: given the extremely strong constraints on this operator, we are still likely to get useful limits on the scale $\Lambda$ in other cases where the dominant contribution to direct detection rates is via D5 generated by SM loop effects. We deal with these cases in the next subsections, but we first complete the discussion of the (D5, D7) set.

The top-right panel of \Fig{fig:App1} shows the analogous case where only D7 is switched on at the scale $\Lambda$, with $c_{\rm D7} = 1$. The limits on $\Lambda$ are weaker than the case of D5, but still in the multi-TeV region~\cite{Crivellin:2014qxa}. For both D5 and D7 we also plot the line that gives a correct thermal relic density, obtained using the annihilation cross section in \Ref{MarchRussell:2012hi}. Current limits exclude thermal relics with mass $m_\chi \gtrsim 3 \, \TeV$ for D5, with a potential of excluding DM masses of the order of $10 \, \TeV$ by forthcoming experiments. The weaker limits for D7 are still in the range $m_\chi \gtrsim 200 \, \GeV$, which can be improved to reach $\TeV$ masses in the future.

Both upper panels are for either $c_5$ or $c_7$  equal to $1$. To relax this assumption one cannot just rescale the vertical axis by $\Lambda \rightarrow \Lambda c_i^{-1/2}$. While it is true that the rate in \Eq{eq:SIsigma} scales as $c_i^2 \Lambda^{-4}$, there is an additional $\Lambda$ dependence in the running Wilson coefficient $c_i(\Lambda)$. For this reasons, when presenting our results we specify both $c_i$ and $\Lambda$.

We go beyond the $c_i = 1$ approximation in the left and right bottom panels of \Fig{fig:App1}. In each panel we fix the DM mass value and explore the allowed region in the $(c_5, c_7)$ plane for three different values of $\Lambda$. If we ignore the running, the allowed regions would be the faded vertical bands, which implies no restriction at all on $c_{D7}$. However, the RG evolution mixes different operators, and the actual experimental limits are the oblique bands. 

To summarize, if SM loop effects are included, it is not consistent to assume that $c_5 = 0$ or $c_7 = 0$ at all energy scales.
This is our main point in this subsection. Furthermore, any sensible UV completion is likely to generate both $c_5$ and $c_7$, at least at one loop~\cite{Freytsis:2010ne}. The constraints that such a model has to satisfy are the ones in the bottom panels of \Fig{fig:App1}.

\subsection{DM interacting with heavy quarks}
\label{sec:HQ}

\begin{figure}
\begin{center}
\includegraphics[scale=0.49]{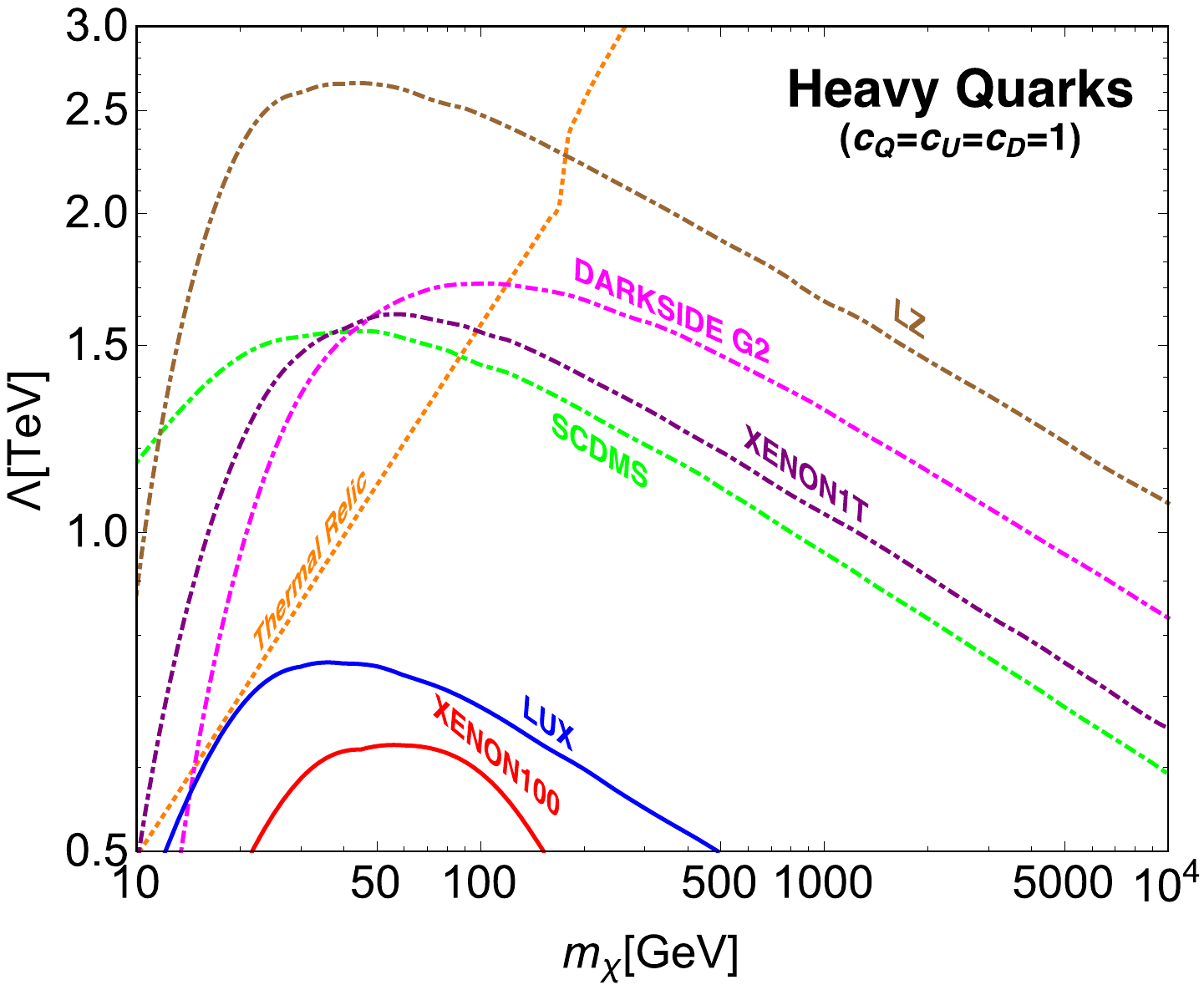} $\quad$ \includegraphics[scale=0.48]{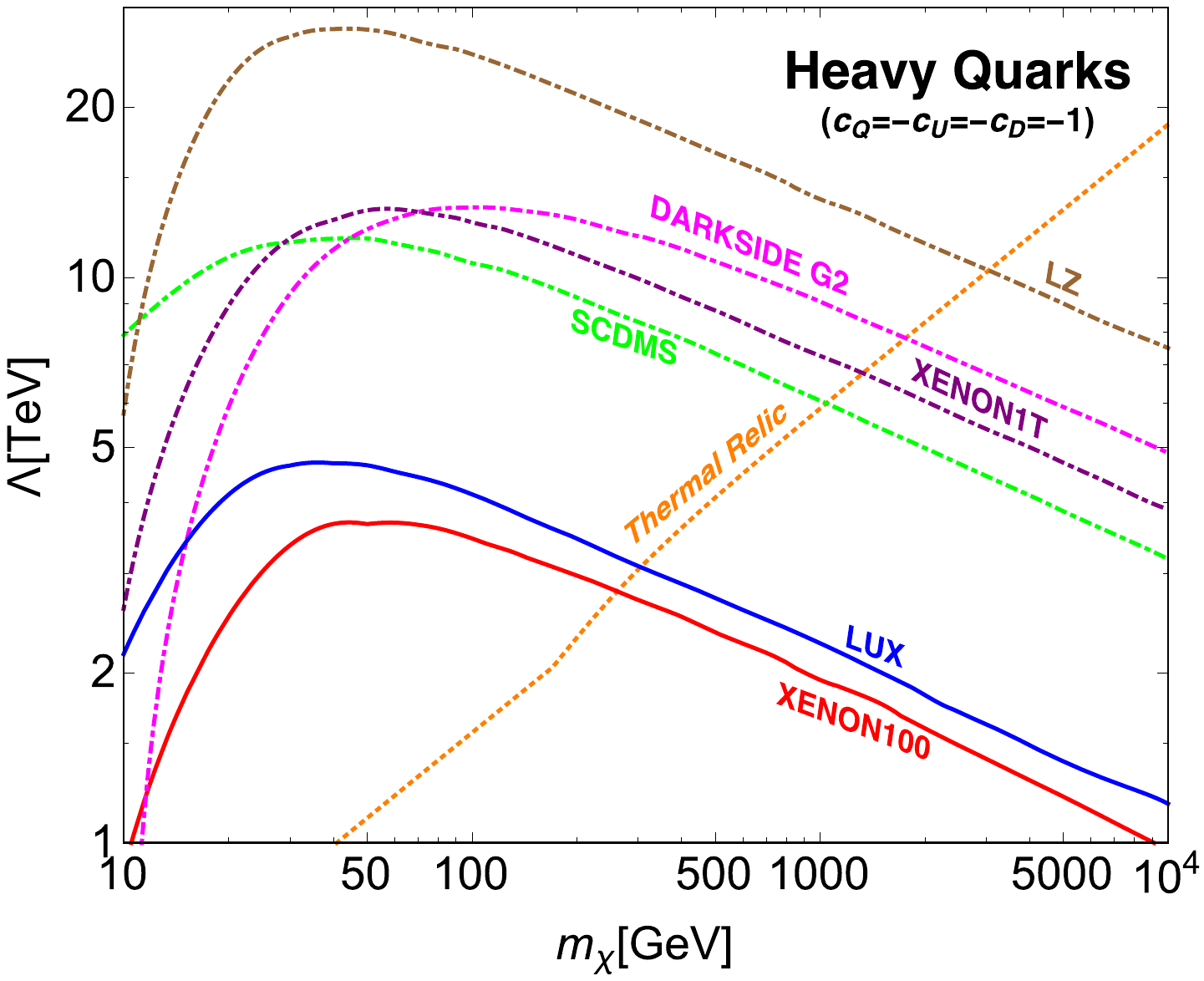}  \vspace{0.5cm} \\
\includegraphics[scale=0.49]{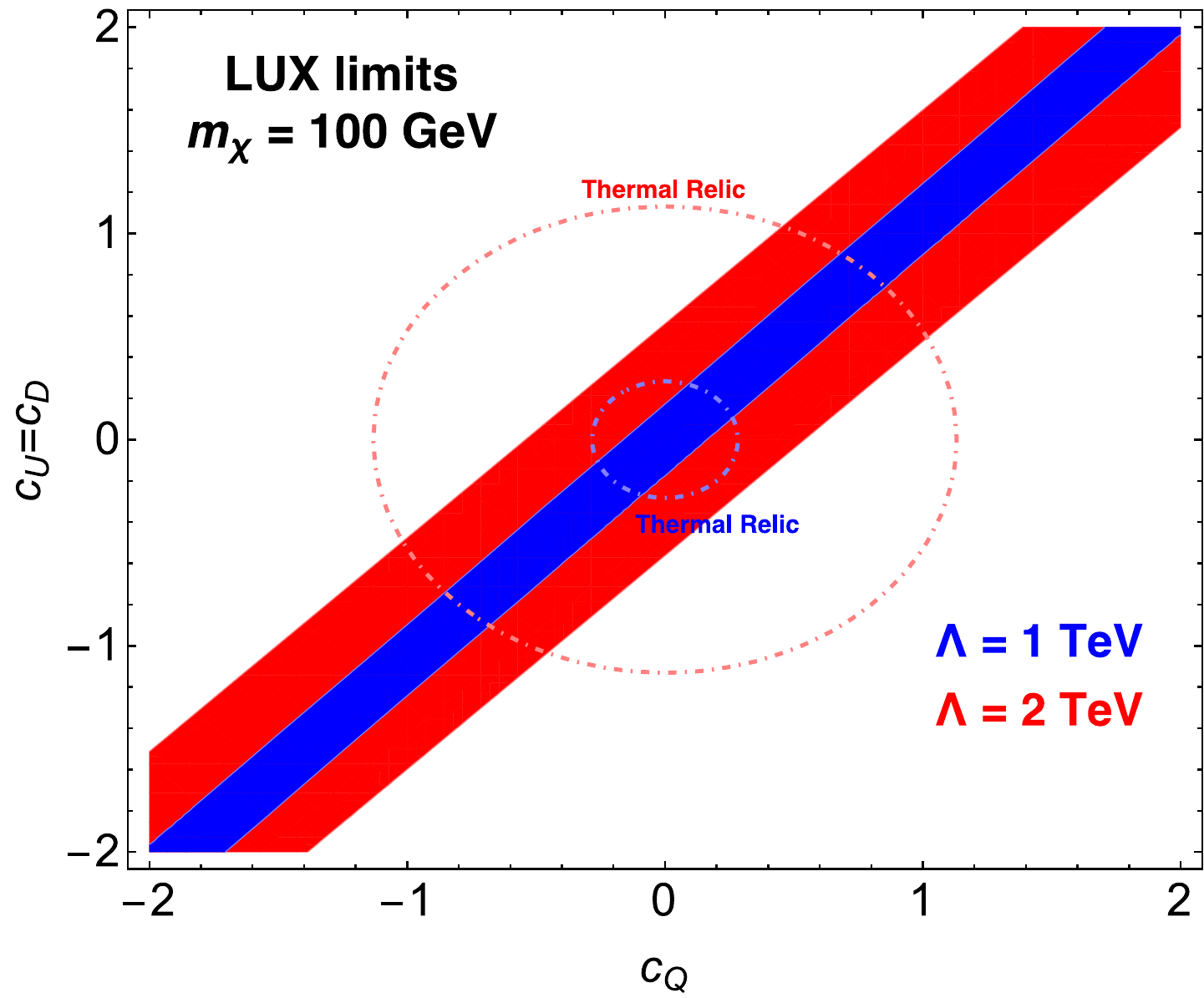} $\quad$ \includegraphics[scale=0.35]{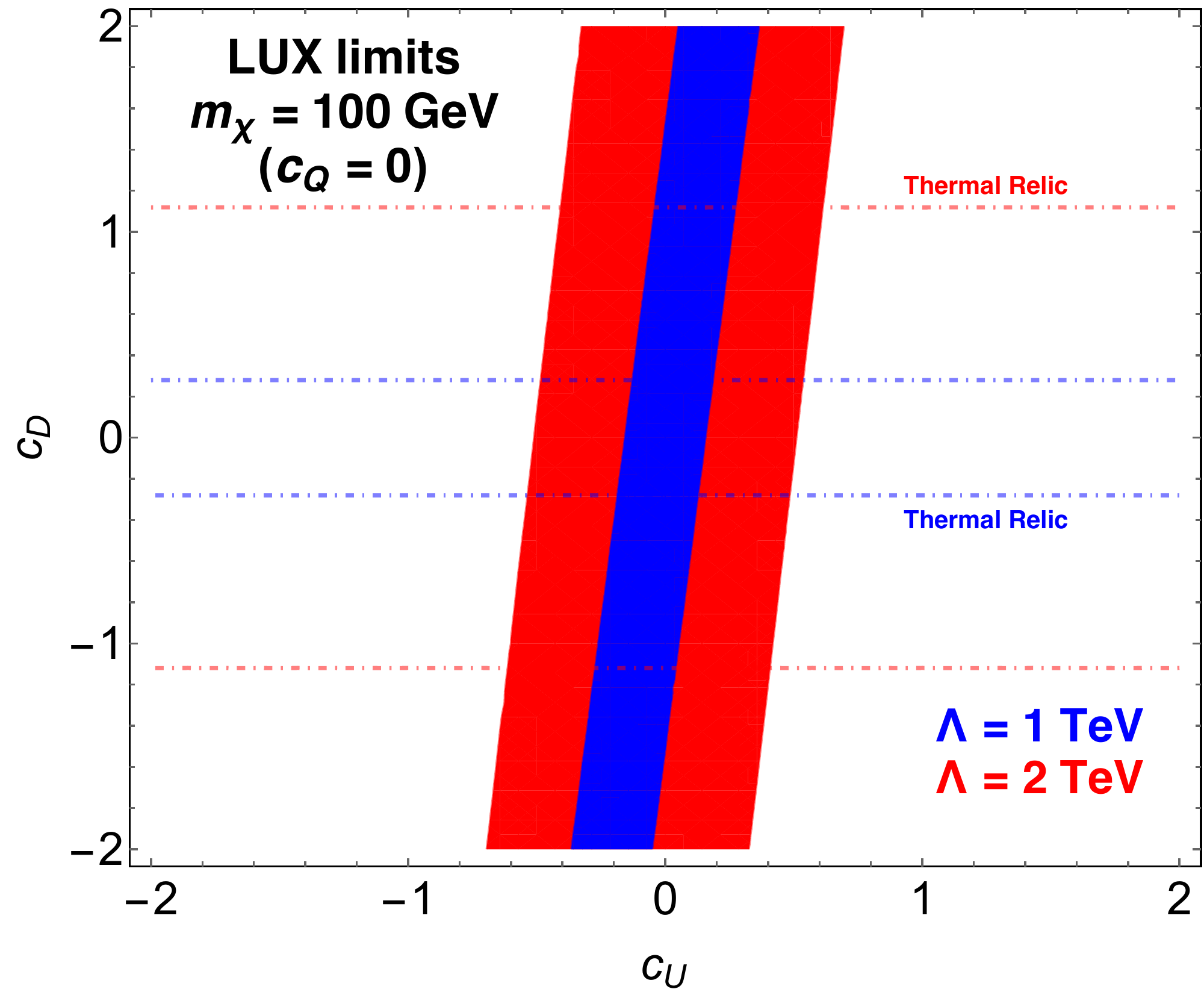} 
\end{center}
\caption{Same as \Fig{fig:App1} but for DM vector current interactions with heavy quarks.}
\label{fig:App2}
\end{figure}

Let us now focus on a different class of models, where the DM vector current only couples to heavy SM quarks
\be
\left. c_\Lambda^T\right|_{\rm HQ} = \left(\begin{array}{ccc:cc|ccc:cc|ccc:cc||c}
0 & 0 & 0 & 0 & 0 & 
0 & 0 & 0 & 0 & 0 &
c_Q & c_U & c_D & 0 & 0 & 0 \end{array}\right) \ .
\ee
The results for this case are shown in \Fig{fig:App2}. In the two top panels of the figure we fix the Wilson coefficients and identify the allowed region in the $(m_\chi, \Lambda)$ plane. We consider two opposite cases. In the top-left panel we set $c_Q = c_U = c_D = 1$, or in other words we couple the DM to heavy quark vector currents at the EFT cutoff. The limits are shown again in the $(m_\chi, \Lambda)$ plane. In the top-right panel we choose the Wilson coefficients such that the DM couples to heavy quark axial currents.

We observe a distinctive feature of couplings to heavy quarks: the limits for interactions to axial currents are much stronger than the ones for vector currents, unlike for the case of D5 and D7 (see top panels in \Fig{fig:App1}). When considering D5 in the previous case, we had couplings to vector currents of light quarks already at tree level, which explains why the limits were much stronger than the loop induced couplings when starting from D7. Here, couplings to light quarks are induced via loop effects in both cases, therefore we have to look at the RG equations and see how this is achieved. The largest coupling driving the mixing onto light quark currents is the top Yukawa, and its effect is encoded in the anomalous dimension matrix in \Eq{eq:gammalambda}. As it turns out, for coupling to vector currents of SM fermions there is no contribution to the running from Yukawa interactions, and the mixing is driven by the sub-leading hypercharge contribution (see \Eq{eq:gammaY}). On the contrary, for couplings to SM axial currents the top Yukawa contribution is maximal, and a substantial mixing onto the Higgs operator $\mathcal{O}_{VH}$ is radiatively induced, which in turns gives interactions to light quarks once the $Z$ boson is integrated out at the EWSB scale.

We also consider cases beyond the $|c_i| = 1$ limit. In the bottom-left panel of \Fig{fig:App2} we fix the DM mass to $m_\chi = 100 \, {\rm GeV}$, impose the isospin conserving condition $c_U = c_D$, and show the allowed region in the $(c_Q, c_U)$ plane for three different values of $\Lambda$. Unsurprisingly, the allowed region lies close to the diagonal line $c_Q = c_D$, since limits are weaker for coupling to vector currents. In the bottom-right panel we consider isospin violation by coupling the DM only to right handed quarks (i.e. $c_Q = 0$) and identifying the allowed region in the $(c_U, c_D)$ plane. The bands are close to the vertical line going through $c_U = 0$, since the effect is driven by the top Yukawa.

\subsection{Leptophilic Dark Matter}
\label{sec:leptons}

Another interesting possibility are leptophilic DM models 
\be
\left. c_\Lambda^T\right|_{\rm Leptoph.} = \left(\begin{array}{ccc:cc|ccc:cc|ccc:cc||c}
0 & 0 & 0 & c_l & c_e & 
0 & 0 & 0 & c_l & c_e &
0 & 0 & 0 & c_l & c_e & 0 \end{array}\right) \ ,
\ee
where for simplicity we consider flavor universal coupling to leptons. In such models there are many sources of couplings to light quarks currents. The Yukawa coupling of the $\tau$ induces a mixing into the Higgs current, which in turn leads to a coupling to light quarks when the $Z$ is integrated out. Hypercharge (electromagnetic) interactions above (below) the EWSB scale also induce mixing onto light quark currents. 

\begin{figure}
\begin{center}
\includegraphics[scale=0.49]{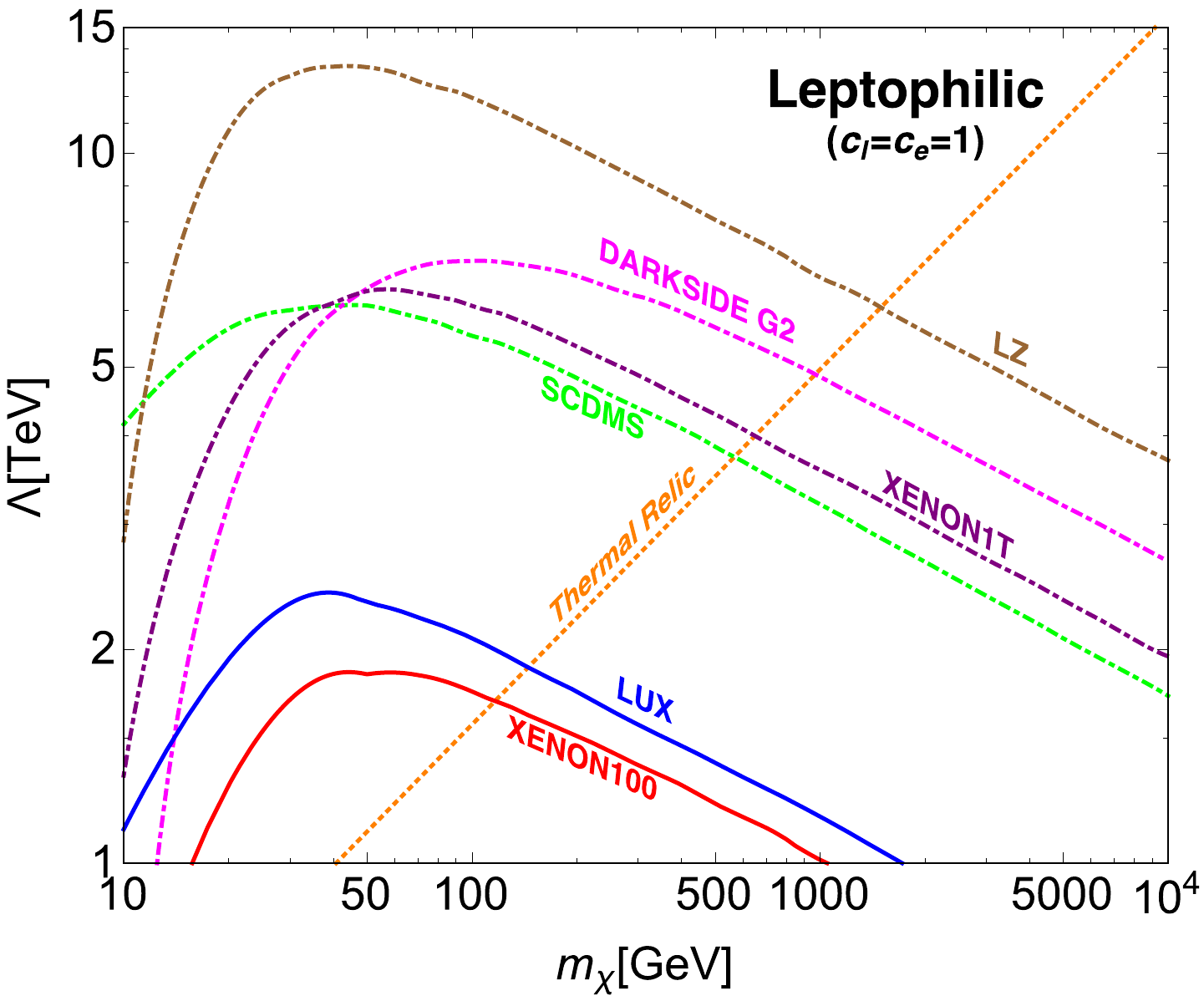} $\quad$ \includegraphics[scale=0.35]{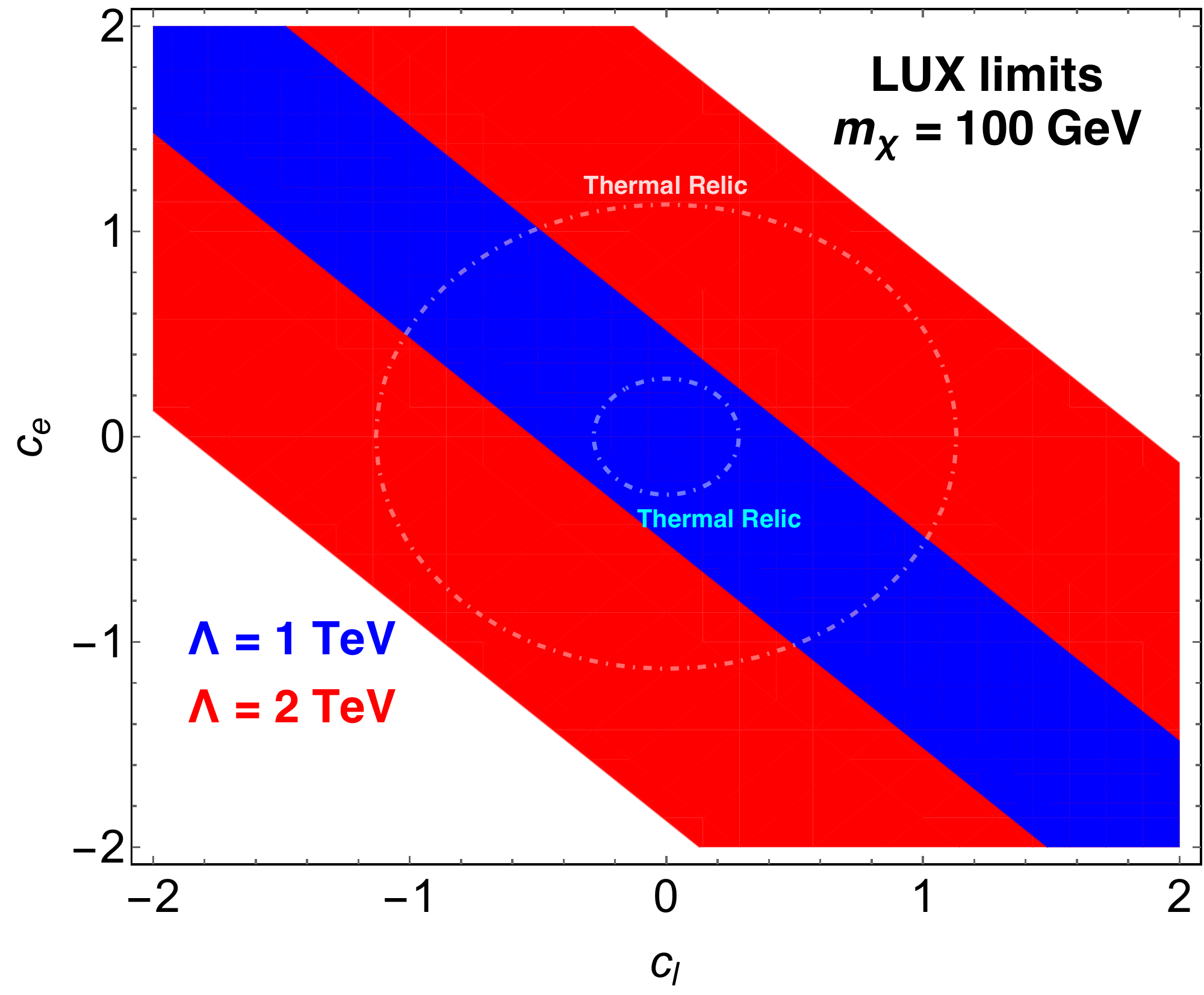}  
\end{center}
\caption{Same as \Fig{fig:App1} but for DM vector current interactions with leptons.}
\label{fig:App3}
\end{figure}

Since it turns out that the mixing driven by Yukawa couplings does not lead to any appreciable constraint, the direct detection rate is induced by hypercharge and electromagnetic interactions. As manifest from the explicit matrices in \Eqs{eq:gammaY}{eq:gammaEM}, such a mixing does not affect axial currents, therefore in this case we can only put limits on interactions with vector currents of leptons. This is in contrast with the heavy quarks case, where the limits we found were much stronger for interactions with axial currents. Our results for coupling vector lepton currents to the vector DM current are shown in the left panel of \Fig{fig:App3}. The right panel of \Fig{fig:App3} displays the allowed regions in the $(c_l, c_e)$ plane for fixed DM mass and two different values of $\Lambda$. Since we do not have here an order one coupling like $\lambda_t$, the bands are wider than in the heavy quarks case. We still have the characteristic orientation along a diagonal, although this time along the one described $c_l \sim - c_e$, for the reasons explained above.

\subsection{Dimension 6 Higgs Portal}
\label{eq:Higgsdim6}

\begin{figure}
\begin{center}
\includegraphics[scale=0.55]{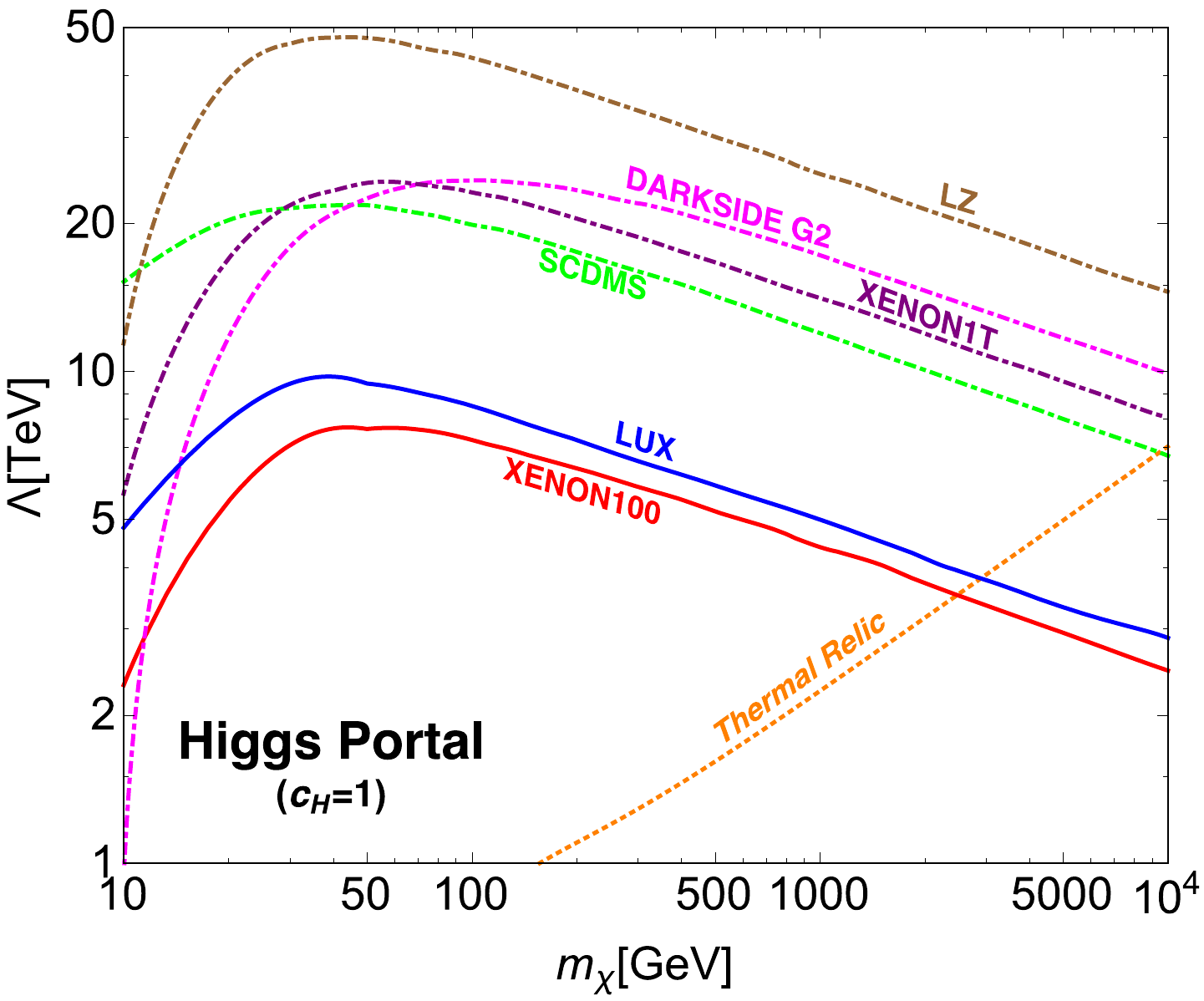} $\quad$
\end{center}
\caption{Same as \Fig{fig:App1} but for DM vector current interactions with the SM Higgs doublet.}
\label{fig:App4}
\end{figure}

The last example we discuss is DM communicating with SM fields only via Higgs couplings
\be
\left. c_\Lambda^T\right|_{\rm Higgs} =  \left(\begin{array}{ccc:cc|ccc:cc|ccc:cc||c}
0 & 0 & 0 & 0 & 0 & 
0 & 0 & 0 & 0 & 0 &
0 & 0 & 0 & 0 & 0 & c_H \end{array}\right) \ .
\label{eq:HiggsPortal}
\ee
This dimension 6 Higgs portal implies a DM vector current tree-level coupling with the $Z$ boson as a consequence of EWSB, as shown in \Eq{eq:chichiZ}.  As usual, by evolving the Wilson coefficients to the nuclear scale, we quantify the implications of negative direct searches. In \Fig{fig:App4} we fix $c_H = 1$ and plot the current and projected exclusion limits in the $(m_\chi, \Lambda)$ plane, as well as the line giving a thermal relic. The constraints are pretty severe, and future experiments can rule out thermal relics up to masses of the order of $10 \, \TeV$. Models with a $Z^\prime$ portal field generate a Wilson coefficient array as in \Eq{eq:HiggsPortal} if the spin-1 mediator only couples to the SM Higgs. The stringent limits persist in any $Z^\prime$ portal model even if the $Z^\prime$ couples to other SM fields besides the Higgs.

\section{Discussion}
\label{sec:discussion}

In the last decade we witnessed an enormous progress by direct searches, which excluded a large parameter space fraction for WIMPs. In particular, models with tree-level vector-vector interactions with light quarks (i.e. D5 in \Eq{eq:D5def}) are severely challenged, as shown in the left panels of our \Fig{fig:App1}. This is not surprising, since spin-independent cross sections arising from tree-level $Z$ boson exchange were ruled out years ago. This effort continues with the new LUX run and with forthcoming experiments, which will soon either discover DM or improve the exclusion limits by one or two orders of magnitude. At the same time, center of mass energies at the LHC will be increased up to a factor of two, and new regions of mass and couplings in DM simplified models will be probed. Last but not least, new generation ground-based and satellite experiments will look for products of DM reactions, constraining the Milky Way WIMPs annihilation rate. We will soon get access to many complementary results on WIMPs property.

When combining the negative outcome of different searches, it is crucial to properly handle the separation among the scales probed by the different experiments. This paper fills the gap between the mediator scales and the nuclear scales for singlet fermion DM models. As sketched in \Fig{fig:EFTs}, we assumed that all non-SM particles (except possibly the DM) are heavier than the weak scale, and we considered the EFT obtained from integrating out these heavy degrees of freedom. It is certainly true that assuming contact interactions with SM fields is not always justified in a collider environment. However, given the typical recoil energy for DM-nucleus scattering, this is always the case for direct searches. Every DM model with mediator fields heavier than the weak scale (possibly constrained by collider or indirect searches) can be matched onto an EFT with only DM interactions with the SM (the ${\rm SM}_\chi$ EFT in \Fig{fig:EFTs}), and the subsequent model-independent one-loop RG evolution properly connects physics at the mediator scales to direct detection observables.

We used a systematic power counting in the suppression scale of the contact effective interactions, and we considered non-renormalizable operators up to dimension 6. Mixing among operators is a standard phenomenon in the dimension 6 sector, where SM interactions mix every operator into all the others. We identified the interactions responsible for these mixings, and computed the associated anomalous dimension matrices. Despite the size of the strong coupling constant $\alpha_s$, QCD interactions do not play any significant role since SM quark currents are not renormalized at one loop. QCD contributions in general cannot induce flavor- or parity-violating mixing, and unless there is a cancellation among subamplitudes~\cite{Hill:2011be,Hill:2013hoa,Hill:2014yka,Hill:2014yxa} they can only modify the overall rate by a numerical factor of order one. The RGE in our case was instead driven by Yukawa and electroweak interactions. The former can mix SM fermion currents onto Higgs currents, which in turn generates DM interactions with SM fermion currents once the $Z$ boson is integrated out. The latter mixes every SM current onto all the others. Given this rich mixing structure, for an arbitrary choice of DM interactions we are very likely to radiatively induce the strongly constrained D5 vector-vector operator defined in \Eq{eq:D5def}, and get significant bounds from experimental results.

In \Sec{sec:App} we discussed applications of our results, focusing on examples where the operator mixing is the main source for direct detection rates. For dimensionless couplings of order one, in most of the cases we found multi-TeV constraints on the mediator mass scale. This rules out a significant region of parameter space where the DM can be a thermal relic, although one can always resort to a non-standard cosmological history~\cite{Giudice:2000ex,D'Eramo:2011ec}. For fixed DM and mediator masses, being consistent with the exclusion limits implies a peculiar alignment among the different dimensionless couplings. Besides the examples discussed here, further singlet fermion DM models generating dimension 6 operators at the mediator scale can be matched onto our framework, and exclusion bounds from direct searches can be derived.

Our work shows how to systematically account for loop effects in fermion singlet DM models. We envision several possible future directions. The next natural step for singlet fermion is to include dimension 7 operators, that are suppressed by an additional power of $1 / \Lambda$. These include DM interactions with (pseudo)scalar and tensor currents of SM fermions and with gauge boson field strengths. Interesting mixing effects involving specific operators of this type have already been studied and used to explore the complementarity of direct searches with indirect detection~\cite{Frandsen:2012db} and searches at hadron colliders~\cite{Haisch:2013uaa,Crivellin:2014gpa}. Singlet scalar WIMPs belong to a separate chapter, since scalar and vector currents have different mass dimensions, and dimension 6 operators would involve also SM field strengths. Moreover, a singlet scalar WIMP can have renormalizable interactions with SM fields~\cite{Burgess:2000yq}. 

On a different route, one can apply the results of our paper to simplified models. The operators considered in this work are generated for example by the exchange of a spin-$1$ mediator. Complementarity among different searches in this class of models have been investigated in Refs.~\cite{Alves:2013tqa,Arcadi:2013qia,Lebedev:2014bba,Alves:2015pea}, focusing on parameter space regions with a tree-level direct detection signal. It is interesting to use the setup built in this paper to extend these studies to cases where the dominant effects are given by the operators discussed in \Sec{sec:App}, such that the direct detection rate is loop induced. As we eagerly await for new data, and hopefully for a DM discovery, it is important to stress that loop effects extend the notion of complementarity among different DM searches.


\acknowledgments

We are very thankful to Mikhail Solon, Philip Tanedo and Tim Tait for useful conversations. F.D. is supported by the Miller Institute for Basic Research in Science.  M.P. acknowledges support by the Swiss National Science Foundation.

\appendix


\section{Standard Model: Conventions and Results}

In this Appendix we define our conventions and collect useful SM results. Our conventions for the space-time metric, the spinor and gauge fields are the same as in \Ref{Peskin:1995ev}. 
\subsection*{Lagrangian in the $SU(3)_c \times SU(2)_L \times U(1)_Y$ phase}
\label{app:SM}
Above the EWSB scale the SM has an unbroken $SU(3)_c \times SU(2)_L \times U(1)_Y$ gauge symmetry, and the matter fields shown in \Tab{tab:SMfields}. We divide the full Lagrangian into four contributions
\be
\mathcal{L}_{\rm SM} = \mathcal{L}_{\rm YM} + \mathcal{L}_{\rm fermions} + \mathcal{L}_{\rm Higgs} + \mathcal{L}_{\rm Yukawa} \ .
\ee

The Lagrangian for the gauge sector read
\be
\mathcal{L}_{\rm YM} = - \frac{1}{4} G^{A \, \mu\nu} G^{A}_{\mu\nu} - \frac{1}{4} W^{I \, \mu\nu} W^{I}_{\mu\nu} - 
\frac{1}{4} B^{\mu\nu} B_{\mu\nu} \ ,  
\label{eq:LagSMYM}
\ee
where the indices $A = 1, \ldots, 8$ and $I = 1, 2, 3$ run over the adjoint representations of $SU(3)_c$ and $SU(2)_L$, respectively. We define a covariant derivative acting on matter fields as follows
\be
D_\mu = \partial_\mu - i g_s \frac{\lambda^A}{2} G^A_\mu - i g \frac{\sigma^I}{2} W^I_\mu - i g^\prime Y B_\mu \ .
\label{eq:Dcov}
\ee
The Gell-Mann ($\lambda^A$) and Pauli ($\sigma^I$) matrices act on color and weak-isospin indices (if any), respectively, whereas the hypercharge $Y$ is assigned as in \Tab{tab:SMfields}. Then we have
\begin{align}
\mathcal{L}_{\rm fermions} = & \, \sum_{i=1}^3 \; \left[ \overline{q_L^i} \, i \slashed{D} q_L^i + \overline{u_R^i} \, i \slashed{D} u_R^i + \overline{d_R^i} \, i \slashed{D} d_R^i + \overline{l_L^i} \, i \slashed{D} l_L^i  + \overline{e_R^i} \, i \slashed{D} e_R^i \right] \ , \\
 \mathcal{L}_{\rm Higgs}  = & \, \left( D^\mu H \right)^\dag \left( D_\mu H \right) + \mu^2 H^\dag H - \lambda \left( H^\dag H \right)^2 \ .
\end{align}
Finally, the Yukawa couplings between fermions and the Higgs doublet are
\be
 \mathcal{L}_{\rm Yukawa} = - \lambda_{u^{(i)}} \; \overline{q^i_L} \, \tilde{H} u^i_R - \lambda_{d^{(i)}} \; \overline{q^i_L} \, H d^j_R - \lambda_{e^{(i)}}  \; \overline{l^i_L} \, H e^j_R  + {\rm h.c.} \ , \qquad \qquad \tilde{H} = \epsilon H^* \ .
\ee
For the purpose of this work we neglect flavor violating terms, therefore we take the Yukawa matrices to be diagonal. The $SU(2)_L \times U(1)_Y \rightarrow U(1)_{\rm em}$ breaking Higgs VEV is chosen as
\be
\langle H^T \rangle = ( \begin{array}{cc} 0 & v \end{array} ) \ , \qquad \qquad \qquad 
v = \frac{\mu}{\sqrt{2 \lambda}} = 174 \; {\rm GeV} \ ,
\label{eq:Higgsvev}
\ee
which gives mass terms for the electroweak gauge bosons and the fermions.

\subsection*{Lagrangian in the $SU(3)_c \times \times U(1)_{\rm em}$ phase}
\label{app:EMSM}

Below the EWSB scale, the SM model has a residual $SU(3)_c \times U(1)_{\rm em}$ gauge symmetry. Matter fermions fill vector-like representations of the gauge group, therefore it is convenient to use the Dirac fermions shown in \Tab{tab:SMfields2}. The Lagrangian has two contributions
\be
\mathcal{L}_{\rm EMSM} = \mathcal{L}_{\rm YM} + \mathcal{L}_{\rm fermions} \ .
\ee
The Yang-Mills term is analogous to the one in \Eq{eq:LagSMYM}, but this time with the photon field strength $F_{\mu\nu}$ in the electroweak sector. The fermion piece reads
\be
\mathcal{L}_{\rm fermions} = \sum_{i=1}^2 \; \overline{u^i} \, \left( i \slashed{D} - m_u^i \right) u^i + 
 \sum_{i=1}^3 \overline{d^i} \, \left( i \slashed{D} - m_d^i \right) d^i + \sum_{i=1}^3 \overline{e^i} \, \left( i \slashed{D} - m_e^i \right) e^i  \ ,
\ee
where we also include fermion masses. The fermion covariant derivative is defined in analogy to \Eq{eq:Dcov}, but with only strong and electromagnetic interactions. 

\subsection*{Useful Equations of Motion}

The only equations of motion needed to get rid of redundant operators are the ones for abelian gauge bosons. In the unbroken phase, the hypercharge field strength satisfies
\be
\partial^\nu B_{\nu\mu} + g^\prime J^{(Y)}_\mu = 0 \ ,
\label{eq:Beom}
\ee
where we define the current
\be
J^{(Y)}_\mu = \sum_{i=1}^3 \left[y_q \,  \overline{q^i_L} \gamma_\mu q^i_L + 
y_u \, \overline{u^i_R} \gamma_\mu u^i_R + y_d \, \overline{d^i_R} \gamma_\mu d^i_R + 
y_l \, \overline{l^i_L} \gamma_\mu l^i_L + y_e \, \overline{e^i_R} \gamma_\mu e^i_R \right] + 
y_H \, H^\dag i \overleftrightarrow{D}_\mu H \ .
\ee
Analogously, for the electromagnetic field strength in the broken phase
\be
\partial^\nu F_{\nu\mu} + e J^{(\rm em)}_\mu = 0 \ ,
\label{eq:Feom}
\ee
where the electromagnetic current reads
\be
J^{(\rm em)}_\mu = Q_u \, \sum_{i=1}^2   \overline{u^i} \gamma_\mu u^i + 
Q_d \, \sum_{i=1}^3 \overline{d^i} \gamma_\mu d^i + Q_e \, \sum_{i=1}^3 \overline{e^i} \gamma_\mu e^i  \ .
\ee

 
\section{Loops and RG Equations} 
\label{app:Loops}

In this Appendix we give results for one-loop diagrams, from which we derive the RG equations. We regularize UV divergences by computing loop integrals in $d = 4 - 2 \epsilon$ dimensions, and subtract infinities using a mass-independent subtraction scheme. As done throughout the paper, we divide the discussion in two cases, correspondent to the two different EFTs. 
 
\subsection*{RGE in the ${\rm SM}_\chi$ EFT}
\label{app:SMchi}

Here we discuss the renormalizion of the EFT described in \Sec{eq:SMchibasis}, with a specific focus on the dimension 6 operators listed in \Tab{tab:TheoryBdim6}. 
The wave-function renormalization factors, generated by the one-loop diagrams in Figs.~\ref{fig:LegsFeynmanDiagrams1} and \ref{fig:LegsFeynmanDiagrams2} have the following $1/\epsilon$-pole structure
\begin{align}
\label{eq:Z1} Z_{q_L^{i}} -1 = & \,  - \frac{g_s^2}{16 \pi^2 \epsilon} \, \mathcal{C}_2(\bs 3) - \frac{g^2}{16 \pi^2 \epsilon} \, \mathcal{C}_2(\bs 2) - \frac{g^{\prime\,2}}{16 \pi^2 \epsilon} \, y_q^2 - \frac{ \lambda_{u^{(i)}}^{2} + \lambda_{d^{(i)}}^{2}  }{32 \pi^2 \epsilon} \ , \\
Z_{u_R^{i}} - 1 = & \,  - \frac{g_s^2}{16 \pi^2 \epsilon} \, \mathcal{C}_2(\bs 3) 
  - \frac{g^{\prime\,2}}{16 \pi^2 \epsilon} \, y_u^2 - \frac{\lambda_{u^{(i)}}^{2} }{16 \pi^2 \epsilon} \ ,  \\
 Z_{d_R^{i}} - 1 = & \,  - \frac{g_s^2}{16 \pi^2 \epsilon} \, \mathcal{C}_2(\bs 3) 
  - \frac{g^{\prime\,2}}{16 \pi^2 \epsilon} \, y_d^2 - \frac{\lambda_{d^{(i)}}^{2}  }{16 \pi^2 \epsilon} \ , \\
  Z_{l_L^{i}} -1 = & \,  - \frac{g^2}{16 \pi^2 \epsilon} \, \mathcal{C}_2(\bs 2) - \frac{g^{\prime\,2}}{16 \pi^2 \epsilon} \, y_l^2 - \frac{ \lambda_{l^{i}}^{2}  }{32 \pi^2 \epsilon} \ , \\
\label{eq:Z5}    Z_{e_R^{i}} - 1 = & \,    - \frac{g^{\prime\,2}}{16 \pi^2 \epsilon} \, y_e^2 - \frac{\lambda_{l^{(i)}}^{2}}{16 \pi^2 \epsilon} \ , \\
Z_H - 1 = & \,  \frac{g^2}{8 \pi^2 \epsilon} \mathcal{C}_2({\bs 2}) + \frac{g^{\prime \, 2}}{8 \pi^2 \epsilon} y_H^2
 - 3 \sum_i \frac{ \lambda_{u^{(i)}}^{2} + \lambda_{d^{(i)}}^{2}}{16 \pi^2 \epsilon}
  - \sum_i \frac{ \lambda_{e^{(i)}}^{2}}{16 \pi^2 \epsilon}      \ .
\end{align}
Here $\mathcal{C}_2(\bs N)$ denotes the Casimir for the $SU(N)$ fundamental representation, and the hypercharge values are assigned as follows
\be
y_q = \frac{1}{6} \ , \qquad   y_u = \frac{2}{3} \ , \qquad  y_d = - \frac{1}{3} \ , \qquad  
y_l = - \frac{1}{2} \ , \qquad   y_e = -1 \ , \qquad \quad  y_H = \frac{1}{2}  \ . 
\label{eq:SMhypervalues}
\ee
We have also define the Yukawa coupling vectors
\begin{align}
\lambda_{u^{(i)}} = & \, \left(\lambda_u, \lambda_c, \lambda_t \right) \, , \;\;\;
\lambda_{d^{(i)}} =  \, \left(\lambda_d, \lambda_s, \lambda_b \right) \, , \;\;\;
\lambda_{e^{(i)}} =  \, \left(\lambda_e, \lambda_\mu, \lambda_\tau \right) \, .
\end{align}
The Higgs quartic coupling does not induce any field renormalization.
We now move to vertex corrections shown in Figs.~\ref{fig:VertexFeynmanDiagrams1} and \ref{fig:VertexFeynmanDiagrams2}. The sum of diagrams with gauge and Yukawa vertices gives the following one-loop corrections to the Wilson coefficients
\begin{align}
\delta c_{\Gamma q}^{(i)} = & \, \frac{g_s^2}{16\pi^2\epsilon} \mathcal{C}_2(\bs 3) c_{\Gamma q}^{(i)} + \frac{g^2}{16\pi^2\epsilon} \mathcal{C}_2(\bs 2) c_{\Gamma q}^{(i)} + \frac{g^{\prime \,2}}{16\pi^2\epsilon} y_q^2 c_{\Gamma q}^{(i)} + \\ \nonumber & 
\frac{\lambda_{u^{(i)}}^{2}}{32 \pi^2\epsilon} c_{\Gamma u}^{(i)} +  \frac{\lambda_{d^{(i)}}^{2}}{32 \pi^2\epsilon} c_{\Gamma d}^{(i)} - \frac{\lambda_{u^{(i)}}^{2} + \lambda_{d^{(i)}}^{2}}{32 \pi^2 \epsilon} c_{\Gamma H} \ , \\
\delta c_{\Gamma u}^{(i)} = & \, \frac{g_s^2}{16\pi^2\epsilon} \mathcal{C}_2(\bs 3) c_{\Gamma u}^{(i)} +  \frac{g^{\prime \,2}}{16\pi^2\epsilon} y_u^2  c_{\Gamma u}^{(i)} + \frac{\lambda_{u^{(i)}}^{2}}{16 \pi^2\epsilon}  c_{\Gamma q}^{(i)} - \frac{\lambda_{u^{(i)}}^{2}}{16 \pi^2 \epsilon} c_{\Gamma H} \ , \\
\delta c_{\Gamma d}^{(i)} = & \, \frac{g_s^2}{16\pi^2\epsilon} \mathcal{C}_2(\bs 3) c_{\Gamma d}^{(i)} +  \frac{g^{\prime \,2}}{16\pi^2\epsilon} y_d^2 c_{\Gamma d}^{(i)} + \frac{\lambda_{d^{(i)}}^{2}}{16 \pi^2\epsilon} c_{\Gamma q}^{(i)} - \frac{\lambda_{d^{(i)}}^{2}}{16 \pi^2 \epsilon} c_{\Gamma H} \ , \\
\delta c_{\Gamma l}^{(i)} = & \, \frac{g^2}{16\pi^2\epsilon} \mathcal{C}_2(\bs 2) c_{\Gamma l}^{(i)} + \frac{g^{\prime \,2}}{16\pi^2\epsilon} y_l^2 c_{\Gamma l}^{(i)} + \frac{\lambda_{l^{(i)}}^{2}}{32 \pi^2\epsilon} c_{\Gamma e}^{(i)}  - \frac{\lambda_{l^{(i)}}^{2}}{32 \pi^2 \epsilon} c_{\Gamma H} \ , \\
\delta c_{\Gamma e}^{(i)} = & \,  \frac{g^{\prime \,2}}{16\pi^2\epsilon} y_e^2 c_{\Gamma e}^{(i)}  + \frac{\lambda_{l^{(i)}}^{2}}{16 \pi^2\epsilon} c_{\Gamma l}^{(i)}  - \frac{\lambda_{l^{(i)}}^{2}}{16 \pi^2 \epsilon} c_{\Gamma H} \ , \\
\delta c_{\Gamma H} = & \, - \frac{g^2 }{8 \pi^2\epsilon} \mathcal{C}_2(\bs 2) c_{\Gamma H} - \frac{g^{\prime\,2}}{8\pi^2\epsilon} y_H^2 c_{\Gamma H} + \\ & \nonumber 
- 3 \sum_i \frac{\lambda_{u^{(i)}}^{2} - \lambda_{d^{(i)}}^{2}}{16 \pi^2 \epsilon} c_{\Gamma q}^{(i)}  +  3 \sum_i\frac{\lambda_{u^{(i)}}^{2}}{16 \pi^2 \epsilon} c_{\Gamma u}^{(i)} - 3 \sum_i \frac{\lambda_{d^{(i)}}^{2} }{16 \pi^2 \epsilon} c_{\Gamma d}^{(i)}  + \\ & \nonumber
\sum_i \frac{\lambda_{l^{(i)}}^{2}}{16 \pi^2 \epsilon} c_{\Gamma l}^{(i)}  -  \sum_i \frac{\lambda_{l^{(i)}}^{2} }{16 \pi^2 \epsilon} c_{\Gamma e}^{(i)}  \ .
\end{align}
A further shift comes from the diagrams in \Fig{fig:VertexFeynmanDiagrams3}. They generate a one-loop contribution to the redundant Wilson coefficient
\be
\delta c_B = \frac{2}{3} \frac{1}{16 \pi^2 \epsilon}
\sum_{i=1}^3 \left[ 6 y_q \, c_{\Gamma q}^{(i)} + 3 y_u \, c_{\Gamma u}^{(i)} +  3 y_d \, c_{\Gamma d}^{(i)}  +
2 y_l \, c_{\Gamma l}^{(i)} +  y_e \, c_{\Gamma e}^{(i)} \right] +
 \frac{2}{3} \frac{1}{16 \pi^2 \epsilon} y_H \, c_H  \ ,
 \label{eq:deltacB}
\ee 
which becomes a shift of the independent operators upon using the equations of motion.
 
With all the loop amplitudes in hand, we can derive the RG equations. First of all we observe that the external legs and vertex corrections coming from gauge interactions have opposite divergent pieces, and therefore do not contribute to the running. The only leftover contribution from gauge interactions comes from the diagrams in \Fig{fig:VertexFeynmanDiagrams3} inducing the redundant operator. Then we make sure that every amplitude in the theory, obtained by the sum of tree-level and one-loop contributions, is finite by renormalizing the coefficients
\be
\left.  \mathcal{C}_{{\rm SM}_\chi} \right|_{\rm bare} = Z_{{\rm SM}_\chi} \left.  \mathcal{C}_{{\rm SM}_\chi} \right|_{\rm ren} =  
\left( 1 + \delta Z_\lambda + \delta Z_Y  \right) \left. \mathcal{C}\right|_{\rm ren}  \ ,
\label{eq:Wcoeffren}
\ee
where we divide the matrix $Z_{{\rm SM}_\chi}$ into identity and interacting parts, with the latter in turn given by Yukawa and hypercharge contributions. The $\delta Z$ matrices explicitly read
\be
 \!\!\!\!\!\delta Z_\lambda = \scalemath{0.56}{- \frac{1}{16\pi^2 \epsilon}  
\left(
\renewcommand{\arraystretch}{1.2}
\begin{array}{ccc:cc|ccc:cc|ccc:cc||c}
\left(\lambda_u^2 + \lambda_d^2\right) / 2  & - \lambda_u^2 / 2 & - \lambda_d^2 / 2 & 0 & 0 & 0 & 0 & 0 & 0 & 0 & 0 & 0 & 0 & 0 & 0 & \left(\lambda_u^2 + \lambda_d^2\right) / 2 \\
- \lambda_u^2 & \lambda_u^2 & 0 & 0 & 0 & 0 & 0 & 0 & 0 & 0 & 0 & 0 & 0 & 0 & 0 & \lambda_u^2 \\
- \lambda_d^2 & 0 & \lambda_d^2 & 0 & 0 & 0 & 0 & 0 & 0 & 0 & 0 & 0 & 0 & 0 & 0 & \lambda_d^2 \\ \hdashline
0 & 0 & 0 & \lambda_e^2 / 2  &  - \lambda_e^2 / 2 & 0 & 0 & 0 & 0 & 0 & 0 & 0 & 0 & 0 & 0 & \lambda_e^2 / 2 \\
0 & 0 & 0 & - \lambda_e^2 & \lambda_e^2 & 0 & 0 & 0 & 0 & 0 & 0 & 0 & 0 & 0 & 0 & \lambda_e^2 \\ \hline
0 & 0 & 0 & 0 & 0 &\left(\lambda_c^2 + \lambda_s^2\right) / 2  & - \lambda_c^2 / 2 & - \lambda_s^2 / 2 & 0 & 0 & 0 & 0 & 0 & 0 & 0 & \left(\lambda_c^2 + \lambda_s^2\right) / 2  \\
0 & 0 & 0 & 0 & 0  & - \lambda_c^2 & \lambda_c^2 & 0 & 0 & 0 & 0 & 0 & 0 & 0 & 0  &  \lambda_c^2 \\
0 & 0 & 0 & 0 & 0  & - \lambda_s^2 & 0 & - \lambda_s^2 & 0 & 0 & 0 & 0 & 0 & 0 & 0  & \lambda_s^2  \\ \hdashline
0 & 0 & 0 & 0 & 0 & 0 & 0 & 0 & \lambda_\mu^2 / 2  &  - \lambda_\mu^2 / 2 & 0 & 0 & 0 & 0 & 0 &  \lambda_\mu^2 / 2 \\
0 & 0 & 0 & 0 & 0 & 0 & 0 & 0 & - \lambda_\mu^2   &  \lambda_\mu^2  & 0 & 0 & 0 & 0 & 0 &  \lambda_\mu^2  \\ \hline
0 & 0 & 0 & 0 & 0 & 0 & 0 & 0 & 0 & 0 & \left(\lambda_t^2 + \lambda_b^2\right) / 2  & - \lambda_t^2 / 2 & - \lambda_b^2 / 2  & 0 & 0 &  \left(\lambda_t^2 + \lambda_b^2\right) / 2 \\
0 & 0 & 0 & 0 & 0 & 0 & 0 & 0 & 0 & 0 & - \lambda_t^2 & \lambda_t^2 & 0 & 0 & 0 &  \lambda_t^2 \\
0 & 0 & 0 & 0 & 0 & 0 & 0 & 0 & 0 & 0 & - \lambda_b^2 & 0 & \lambda_b^2 & 0 & 0 &  \lambda_b^2 \\ \hdashline
0 & 0 & 0 & 0 & 0 & 0 & 0 & 0 & 0 & 0 & 0 & 0 & 0 & \lambda_\tau^2 / 2  & - \lambda_\tau^2 / 2 & \lambda_\tau^2 / 2 \\
0 & 0 & 0 & 0 & 0 & 0 & 0 & 0 & 0 & 0 & 0 & 0 & 0 & - \lambda_\tau^2 & \lambda_\tau^2 & \lambda_\tau^2 \\  \hline\hline
3 \left(\lambda_u^2 - \lambda_d^2\right) &  - 3 \lambda_u^2 &  3 \lambda_d^2  &  - \lambda_e^2 & \lambda_e^2  &  
3 \left(\lambda_c^2 - \lambda_s^2\right) &  - 3 \lambda_c^2 &  3 \lambda_s^2  &  - \lambda_\mu^2 & \lambda_\mu^2 & 3 \left(\lambda_t^2 - \lambda_b^2\right) &  - 3 \lambda_t^2 &  3 \lambda_b^2 & - \lambda_\tau^2 &  \lambda_\tau^2  & 3 \sum_q \lambda_q^2 + \sum_l \lambda_l^2
\end{array}\right)}, 
\ee
and
\be
 \!\!\!\!\! \delta Z_Y = \scalemath{0.86}{   - \frac{2}{3} \frac{g^{\prime\,2}}{16 \pi^2 \epsilon}  
 \tiny  \left(
 \renewcommand{\arraystretch}{1.7}
 \begin{array}{ccc:cc|ccc:cc|ccc:cc||c}
6 y_q^2 & 3 y_q y_u  & 3 y_q y_d & 2 y_q y_l & y_q y_e & 6 y_q^2 & 3 y_q y_u  & 3 y_q y_d & 2 y_q y_l & y_q y_e & 
6 y_q^2 & 3 y_q y_u  & 3 y_q y_d & 2 y_q y_l & y_q y_e & y_q y_H \\
6 y_u y_q & 3 y^2_u  & 3 y_u  y_d & 2 y_u  y_l & y_u  y_e & 6 y_u y_q & 3 y_u^2  & 3 y_u y_d & 2 y_u  y_l & y_u y_e & 
6 y_u y_q & 3 y^2_u  & 3 y_u  y_d & 2 y_u  y_l & y_u  y_e & y_u y_H \\
6 y_d y_q & 3 y_d y_u  & 3 y^2_d & 2 y_d  y_l & y_d  y_e & 6 y_d y_q & 3 y_d y_u  & 3 y^2_d & 2 y_d  y_l & y_d  y_e & 
6 y_d y_q & 3 y_d y_u  & 3 y^2_d & 2 y_d  y_l & y_d  y_e & y_d y_H \\ \hdashline
6 y_l y_q & 3 y_l y_u  & 3 y_l y_d & 2 y^2_l & y_l  y_e & 6 y_l y_q & 3 y_l y_u  & 3 y_l y_d & 2 y^2_l & y_l  y_e & 
6 y_l y_q & 3 y_l y_u  & 3 y_l y_d & 2 y^2_l & y_l  y_e & y_l y_H \\
6 y_e y_q & 3 y_e y_u  & 3 y_e y_d & 2 y_e y_l & y^2_e & 6 y_e y_q & 3 y_e y_u  & 3 y_e y_d & 2 y_e y_l & y^2_e & 
6 y_e y_q & 3 y_e y_u  & 3 y_e y_d & 2 y_e y_l & y^2_e & y_e y_H \\ \hline
6 y_q^2 & 3 y_q y_u  & 3 y_q y_d & 2 y_q y_l & y_q y_e & 6 y_q^2 & 3 y_q y_u  & 3 y_q y_d & 2 y_q y_l & y_q y_e & 
6 y_q^2 & 3 y_q y_u  & 3 y_q y_d & 2 y_q y_l & y_q y_e & y_q y_H \\
6 y_u y_q & 3 y^2_u  & 3 y_u  y_d & 2 y_u  y_l & y_u  y_e & 6 y_u y_q & 3 y_u^2  & 3 y_u y_d & 2 y_u  y_l & y_u y_e & 
6 y_u y_q & 3 y^2_u  & 3 y_u  y_d & 2 y_u  y_l & y_u  y_e & y_u y_H \\
6 y_d y_q & 3 y_d y_u  & 3 y^2_d & 2 y_d  y_l & y_d  y_e & 6 y_d y_q & 3 y_d y_u  & 3 y^2_d & 2 y_d  y_l & y_d  y_e & 
6 y_d y_q & 3 y_d y_u  & 3 y^2_d & 2 y_d  y_l & y_d  y_e & y_d y_H \\ \hdashline
6 y_l y_q & 3 y_l y_u  & 3 y_l y_d & 2 y^2_l & y_l  y_e & 6 y_l y_q & 3 y_l y_u  & 3 y_l y_d & 2 y^2_l & y_l  y_e & 
6 y_l y_q & 3 y_l y_u  & 3 y_l y_d & 2 y^2_l & y_l  y_e & y_l y_H \\
6 y_e y_q & 3 y_e y_u  & 3 y_e y_d & 2 y_e y_l & y^2_e & 6 y_e y_q & 3 y_e y_u  & 3 y_e y_d & 2 y_e y_l & y^2_e & 
6 y_e y_q & 3 y_e y_u  & 3 y_e y_d & 2 y_e y_l & y^2_e & y_e y_H \\ \hline
6 y_q^2 & 3 y_q y_u  & 3 y_q y_d & 2 y_q y_l & y_q y_e & 6 y_q^2 & 3 y_q y_u  & 3 y_q y_d & 2 y_q y_l & y_q y_e & 
6 y_q^2 & 3 y_q y_u  & 3 y_q y_d & 2 y_q y_l & y_q y_e & y_q y_H \\
6 y_u y_q & 3 y^2_u  & 3 y_u  y_d & 2 y_u  y_l & y_u  y_e & 6 y_u y_q & 3 y_u^2  & 3 y_u y_d & 2 y_u  y_l & y_u y_e & 
6 y_u y_q & 3 y^2_u  & 3 y_u  y_d & 2 y_u  y_l & y_u  y_e & y_u y_H \\
6 y_d y_q & 3 y_d y_u  & 3 y^2_d & 2 y_d  y_l & y_d  y_e & 6 y_d y_q & 3 y_d y_u  & 3 y^2_d & 2 y_d  y_l & y_d  y_e & 
6 y_d y_q & 3 y_d y_u  & 3 y^2_d & 2 y_d  y_l & y_d  y_e & y_d y_H \\ \hdashline
6 y_l y_q & 3 y_l y_u  & 3 y_l y_d & 2 y^2_l & y_l  y_e & 6 y_l y_q & 3 y_l y_u  & 3 y_l y_d & 2 y^2_l & y_l  y_e & 
6 y_l y_q & 3 y_l y_u  & 3 y_l y_d & 2 y^2_l & y_l  y_e & y_l y_H \\
6 y_e y_q & 3 y_e y_u  & 3 y_e y_d & 2 y_e y_l & y^2_e & 6 y_e y_q & 3 y_e y_u  & 3 y_e y_d & 2 y_e y_l & y^2_e & 
6 y_e y_q & 3 y_e y_u  & 3 y_e y_d & 2 y_e y_l & y^2_e & y_e y_H \\ \hline\hline
6 y_H y_q & 3 y_H y_u  & 3 y_H y_d & 2 y_H y_l & y_H y_e & 6 y_H y_q & 3 y_H y_u  & 3 y_H y_d & 2 y_H y_l & y_H y_e &  
6 y_H y_q & 3 y_H y_u  & 3 y_H y_d & 2 y_H y_l & y_H y_e &  y^2_H
\end{array}\right)} \ .
\ee

The RG equations are derived by imposing that the bare Wilson coefficients are renormalization scale independent,
\begin{align}
\frac{d}{d \ln \mu} \left.  \mathcal{C}_{{\rm SM}_\chi} \right|_{\rm ren} =  & \, \gamma_{{\rm SM}_\chi} \left.  \mathcal{C}_{{\rm SM}_\chi} \right|_{\rm ren}  \ , \\ 
\gamma_{{\rm SM}_\chi} \equiv & \, - Z_{{\rm SM}_\chi}^{-1} \; \frac{d Z_{{\rm SM}_\chi}}{d \ln \mu} = - \frac{d}{d \ln \mu} \left( \delta Z_\lambda + \delta Z_Y \right) \ .
\label{eq:gammadef}
\end{align}
The last thing is to identify the $\mu$ dependence in the $\delta Z_i$ matrices. The key point to observe here is that we work in a mass independent scheme, namely the matrices $\delta Z_i$ do not have an explicit dependence on the mass scale but they only depend on $\mu$ through the Yukawa and gauge couplings. By defining at one loop
\begin{align}
\delta Z_\lambda \equiv & \, \frac{1}{\epsilon} \delta Z^{(1)}_\lambda(\lambda^2_i) \ , \\
\delta Z_Y \equiv & \, \frac{1}{\epsilon} \delta Z^{(1)}_Y(g^{\prime\,2}) \ ,
\end{align}
the anomalous dimension matrix reads
\be
\gamma_{{\rm SM}_\chi} \equiv \gamma_\lambda + \gamma_Y = 
\sum_i 2 \lambda_i^2 \frac{\partial \; \delta Z^{(1)}_\lambda(\lambda^2_i)}{\partial \lambda^2_i}  + 
2 g^{\prime\,2} \frac{\partial \;  \delta Z^{(1)}_Y(g^{\prime\,2})}{\partial g^{\prime\,2}}  \ .
\label{eq:gammaexplicitchi}
\ee

\subsection*{RGE in the ${\rm EMSM}_\chi$ EFT}
\label{app:EMSMchi}

Below the EWSB scale wave-function renormalization is only due to gauge interactions. The results can be derived from Eqs.(\ref{eq:Z1} - \ref{eq:Z5}) for fermions above the EWSB scale, with identical QCD contribution and electromagnetic factor obtained by replacing the hypercharges with the electric charges
\be
Q_u = \frac{2}{3} \ , \qquad \qquad Q_d = - \frac{1}{3} \ , \qquad \qquad  Q_e = - 1 \ .
\label{eq:SMelectricharges}
\ee
They all cancel again with the associated vertex corrections, therefore we do not need to further consider them.

As discussed in the paper there are two classes of vertex corrections also in this case. The first corrections are due to the SM four-fermion interactions, as shown in \Fig{fig:VertexFeynmanDiagrams4}. These diagrams do not vanish only if we start from an interaction between the DM bilinear and the {\emph {axial}} current of SM fermions. The explicit shifts read
\begin{align}
\delta c_{\Gamma Vu}^{(i)} = & \, -  \frac{G_F}{\sqrt{2}} \frac{g_{Vu}}{\pi^2 \epsilon} \, L_m  \ , \\
\delta c_{\Gamma Vd}^{(i)} = & \, -  \frac{G_F}{\sqrt{2}}  \frac{g_{Vd}}{\pi^2 \epsilon}  \, L_m  \ , \\
\delta c_{\Gamma Ve}^{(i)} = & \, -  \frac{G_F}{\sqrt{2}}  \frac{g_{Ve}}{\pi^2 \epsilon}  \, L_m  \ , \\
\delta c_{\Gamma Au}^{(i)}= & \,  -  \frac{G_F}{\sqrt{2}}  \frac{ g_{Au}}{\pi^2 \epsilon}   \, L_m \ , \\ 
\delta c_{\Gamma Ad}^{(i)}= & \, -  \frac{G_F}{\sqrt{2}}  \frac{g_{Ad}}{\pi^2 \epsilon}  \, L_m  \ , \\
\delta c_{\Gamma Ae}^{(i)} = & \, -  \frac{G_F}{\sqrt{2}}  \frac{g_{Ae} }{\pi^2 \epsilon}  \, L_m \ ,
\end{align}
where we find it convenient to isolate the common factor accounting for all the possible SM fermions in the loop of \Fig{fig:VertexFeynmanDiagrams4},
\be
L_m = 3 g_{Au} \sum_j m_{u^{(j)}}^2  c^{(j)}_{\Gamma A u}  + 
3 g_{Ad} \sum_j m_{d^{(j)}}^2  c^{(j)}_{\Gamma A d}  + g_{Ae} \sum_j m_{e^{(j)}}^2  c^{(j)}_{\Gamma A e}  \ .
\ee
 The couplings $g_{Vf}$ and $g_{Af}$ of SM fermions to the $Z$ boson are given in \Eq{eq:SMfermionstoZ}.

The second contribution is the radiative correction to the Wilson coefficient $c_F$ of the redundant operator in \Eq{eq:redundantphotonSection2}, which results in
\be
\delta c_F = \frac{4}{3} \frac{1}{16 \pi^2 \epsilon} 
\left[ 3 \sum_{i=1}^2 Q_u \, c_{\Gamma Vu}^{(i)} + 3 \sum_{i=1}^3 Q_d \,  c_{\Gamma Vd}^{(i)} +
 \sum_{i=1}^3  Q_e \, c_{\Gamma Ve}^{(i)} \right]  \ .
\ee
Using the equations of motion again leads to a shift for the independent operators.

Analogously to what we have done in \App{app:SMchi}, the RG equations are derived renormalizing the Wilson coefficients,
\be
\left.  \mathcal{C}_{{\rm EMSM}_\chi} \right|_{\rm bare} = Z_{{\rm EMSM}_\chi} \left.  \mathcal{C}_{{\rm EMSM}_\chi} \right|_{\rm ren} = \left( 1 + \delta Z_m + \delta Z_{\rm em} \right) \left.  \mathcal{C}_{{\rm EMSM}_\chi} \right|_{\rm ren} \ ,\label{eq:Wcoeffren}
\ee
where we divide again the matrix $Z_{{\rm EMSM}_\chi}$ into different contributions. They explicitly read
\be
\!\!\!\!\!\delta Z_m  =  \scalemath{0.62}{ - \frac{ G_F}{\sqrt{2} \pi^2 \epsilon}  
\left(
\renewcommand{\arraystretch}{1.7}
\begin{array}{ccccc|ccc||ccccc|ccc}
0 & 0 & 0 & 0 & 0 & 0 & 0 & 0 & 3 m_u^2 g_{Au} \, g_{Vu} & 3 m_d^2 g_{Ad} \, g_{Vu} & 3 m_c^2 g_{Au} \, g_{Vu} & 3 m_s^2 g_{Ad} \, g_{Vu} & 3 m_b^2 g_{Ad} \, g_{Vu} & m_e^2 g_{Ae} \, g_{Vu} & m_\mu^2 g_{Ae} \, g_{Vu} & m_\tau^2 g_{Ae} \, g_{Vu} \\ 
0 & 0 & 0 & 0 & 0 & 0 & 0 & 0 & 3 m_u^2 g_{Au} \, g_{Vd} & 3 m_d^2 g_{Ad} \, g_{Vd} & 3 m_c^2 g_{Au} \, g_{Vd} & 3 m_s^2 g_{Ad} \, g_{Vd} & 3 m_b^2 g_{Ad} \, g_{Vd} & m_e^2 g_{Ae} \, g_{Vd} & m_\mu^2 g_{Ae} \, g_{Vd} & m_\tau^2 g_{Ae} \, g_{Vd} \\ 
0 & 0 & 0 & 0 & 0 & 0 & 0 & 0 & 3 m_u^2 g_{Au} \, g_{Vu} & 3 m_d^2 g_{Ad} \, g_{Vu} & 3 m_c^2 g_{Au} \, g_{Vu} & 3 m_s^2 g_{Ad} \, g_{Vu} & 3 m_b^2 g_{Ad} \, g_{Vu} & m_e^2 g_{Ae} \, g_{Vu} & m_\mu^2 g_{Ae} \, g_{Vu} & m_\tau^2 g_{Ae} \, g_{Vu} \\ 
0 & 0 & 0 & 0 & 0 & 0 & 0 & 0 & 3 m_u^2 g_{Au} \, g_{Vd} & 3 m_d^2 g_{Ad} \, g_{Vd} & 3 m_c^2 g_{Au} \, g_{Vd} & 3 m_s^2 g_{Ad} \, g_{Vd} & 3 m_b^2 g_{Ad} \, g_{Vd} & m_e^2 g_{Ae} \, g_{Vd} & m_\mu^2 g_{Ae} \, g_{Vd} & m_\tau^2 g_{Ae} \, g_{Vd} \\ 
0 & 0 & 0 & 0 & 0 & 0 & 0 & 0 & 3 m_u^2 g_{Au} \, g_{Vd} & 3 m_d^2 g_{Ad} \, g_{Vd} & 3 m_c^2 g_{Au} \, g_{Vd} & 3 m_s^2 g_{Ad} \, g_{Vd} & 3 m_b^2 g_{Ad} \, g_{Vd} & m_e^2 g_{Ae} \, g_{Vd} & m_\mu^2 g_{Ae} \, g_{Vd} & m_\tau^2 g_{Ae} \, g_{Vd} \\  \hline 
0 & 0 & 0 & 0 & 0 & 0 & 0 & 0 & 3 m_u^2 g_{Au} \, g_{Ve} & 3 m_d^2 g_{Ad} \, g_{Ve} & 3 m_c^2 g_{Au} \, g_{Ve} & 3 m_s^2 g_{Ad} \, g_{Ve} & 3 m_b^2 g_{Ad} \, g_{Ve} & m_e^2 g_{Ae} \, g_{Ve} & m_\mu^2 g_{Ae} \, g_{Ve} & m_\tau^2 g_{Ae} \, g_{Ve} \\ 
0 & 0 & 0 & 0 & 0 & 0 & 0 & 0 & 3 m_u^2 g_{Au} \, g_{Ve} & 3 m_d^2 g_{Ad} \, g_{Ve} & 3 m_c^2 g_{Au} \, g_{Ve} & 3 m_s^2 g_{Ad} \, g_{Ve} & 3 m_b^2 g_{Ad} \, g_{Ve} & m_e^2 g_{Ae} \, g_{Ve} & m_\mu^2 g_{Ae} \, g_{Ve} & m_\tau^2 g_{Ae} \, g_{Ve} \\ 
0 & 0 & 0 & 0 & 0 & 0 & 0 & 0 & 3 m_u^2 g_{Au} \, g_{Ve} & 3 m_d^2 g_{Ad} \, g_{Ve} & 3 m_c^2 g_{Au} \, g_{Ve} & 3 m_s^2 g_{Ad} \, g_{Ve} & 3 m_b^2 g_{Ad} \, g_{Ve} & m_e^2 g_{Ae} \, g_{Ve} & m_\mu^2 g_{Ae} \, g_{Ve} & m_\tau^2 g_{Ae} \, g_{Ve} \\ \hline \hline
0 & 0 & 0 & 0 & 0 & 0 & 0 & 0 & 3 m_u^2 g_{Au} \, g_{Au} & 3 m_d^2 g_{Ad} \, g_{Au} & 3 m_c^2 g_{Au} \, g_{Au} & 3 m_s^2 g_{Ad} \, g_{Au} & 3 m_b^2 g_{Ad} \, g_{Au} & m_e^2 g_{Ae} \, g_{Au} & m_\mu^2 g_{Ae} \, g_{Au} & m_\tau^2 g_{Ae} \, g_{Au} \\ 
0 & 0 & 0 & 0 & 0 & 0 & 0 & 0 & 3 m_u^2 g_{Au} \, g_{Ad} & 3 m_d^2 g_{Ad} \, g_{Ad} & 3 m_c^2 g_{Au} \, g_{Ad} & 3 m_s^2 g_{Ad} \, g_{Ad} & 3 m_b^2 g_{Ad} \, g_{Ad} & m_e^2 g_{Ae} \, g_{Ad} & m_\mu^2 g_{Ae} \, g_{Ad} & m_\tau^2 g_{Ae} \, g_{Ad} \\ 
0 & 0 & 0 & 0 & 0 & 0 & 0 & 0 & 3 m_u^2 g_{Au} \, g_{Au} & 3 m_d^2 g_{Ad} \, g_{Au} & 3 m_c^2 g_{Au} \, g_{Au} & 3 m_s^2 g_{Ad} \, g_{Au} & 3 m_b^2 g_{Ad} \, g_{Au} & m_e^2 g_{Ae} \, g_{Au} & m_\mu^2 g_{Ae} \, g_{Au} & m_\tau^2 g_{Ae} \, g_{Au} \\ 
0 & 0 & 0 & 0 & 0 & 0 & 0 & 0 & 3 m_u^2 g_{Au} \, g_{Ad} & 3 m_d^2 g_{Ad} \, g_{Ad} & 3 m_c^2 g_{Au} \, g_{Ad} & 3 m_s^2 g_{Ad} \, g_{Ad} & 3 m_b^2 g_{Ad} \, g_{Ad} & m_e^2 g_{Ae} \, g_{Ad} & m_\mu^2 g_{Ae} \, g_{Ad} & m_\tau^2 g_{Ae} \, g_{Ad} \\ 
0 & 0 & 0 & 0 & 0 & 0 & 0 & 0 & 3 m_u^2 g_{Au} \, g_{Ad} & 3 m_d^2 g_{Ad} \, g_{Ad} & 3 m_c^2 g_{Au} \, g_{Ad} & 3 m_s^2 g_{Ad} \, g_{Ad} & 3 m_b^2 g_{Ad} \, g_{Ad} & m_e^2 g_{Ae} \, g_{Ad} & m_\mu^2 g_{Ae} \, g_{Ad} & m_\tau^2 g_{Ae} \, g_{Ad} \\  \hline 
0 & 0 & 0 & 0 & 0 & 0 & 0 & 0 & 3 m_u^2 g_{Au} \, g_{Ae} & 3 m_d^2 g_{Ad} \, g_{Ae} & 3 m_c^2 g_{Au} \, g_{Ae} & 3 m_s^2 g_{Ad} \, g_{Ae} & 3 m_b^2 g_{Ad} \, g_{Ae} & m_e^2 g_{Ae} \, g_{Ae} & m_\mu^2 g_{Ae} \, g_{Ae} & m_\tau^2 g_{Ae} \, g_{Ae} \\ 
0 & 0 & 0 & 0 & 0 & 0 & 0 & 0 & 3 m_u^2 g_{Au} \, g_{Ae} & 3 m_d^2 g_{Ad} \, g_{Ae} & 3 m_c^2 g_{Au} \, g_{Ae} & 3 m_s^2 g_{Ad} \, g_{Ae} & 3 m_b^2 g_{Ad} \, g_{Ae} & m_e^2 g_{Ae} \, g_{Ae} & m_\mu^2 g_{Ae} \, g_{Ae} & m_\tau^2 g_{Ae} \, g_{Ae} \\ 
0 & 0 & 0 & 0 & 0 & 0 & 0 & 0 & 3 m_u^2 g_{Au} \, g_{Ae} & 3 m_d^2 g_{Ad} \, g_{Ae} & 3 m_c^2 g_{Au} \, g_{Ae} & 3 m_s^2 g_{Ad} \, g_{Ae} & 3 m_b^2 g_{Ad} \, g_{Ae} & m_e^2 g_{Ae} \, g_{Ae} & m_\mu^2 g_{Ae} \, g_{Ae} & m_\tau^2 g_{Ae} \, g_{Ae} 
\end{array}\right)},  
\ee
and
\be
\delta Z_{\rm em} =  \scalemath{0.77}{ - \frac{4}{3} \frac{e^2}{16 \pi^2 \epsilon}
\left(
\renewcommand{\arraystretch}{1.2}
\begin{array}{ccccc|ccc||ccccc|ccc}
3 Q_u^2 & 3 Q_u Q_d & 3 Q_u^2 & 3 Q_u Q_d & 3 Q_u Q_d & Q_u Q_e & Q_u Q_e & Q_u Q_e & 0 & 0 & 0 & 0 & 0 & 0 & 0 & 0 \\
3 Q_d Q_u & 3 Q_d^2 & 3 Q_d Q_u & 3 Q_d^2 & 3 Q_d^2 & Q_d Q_e & Q_d Q_e & Q_d Q_e & 0 & 0 & 0 & 0 & 0 & 0 & 0 & 0 \\
3 Q_u^2 & 3 Q_u Q_d & 3 Q_u^2 & 3 Q_u Q_d & 3 Q_u Q_d & Q_u Q_e & Q_u Q_e & Q_u Q_e & 0 & 0 & 0 & 0 & 0 & 0 & 0 & 0 \\
3 Q_d Q_u & 3 Q_d^2 & 3 Q_d Q_u & 3 Q_d^2 & 3 Q_d^2 & Q_d Q_e & Q_d Q_e & Q_d Q_e & 0 & 0 & 0 & 0 & 0 & 0 & 0 & 0 \\
3 Q_d Q_u & 3 Q_d^2 & 3 Q_d Q_u & 3 Q_d^2 & 3 Q_d^2 & Q_d Q_e & Q_d Q_e & Q_d Q_e & 0 & 0 & 0 & 0 & 0 & 0 & 0 & 0 \\ \hline 
3 Q_e Q_u & 3 Q_e Q_d & 3 Q_e Q_u & 3 Q_e Q_d & 3 Q_e Q_d & Q_e^2 & Q_e^2 & Q_e^2 & 0 & 0 & 0 & 0 & 0 & 0 & 0 & 0 \\
3 Q_e Q_u & 3 Q_e Q_d & 3 Q_e Q_u & 3 Q_e Q_d & 3 Q_e Q_d & Q_e^2 & Q_e^2 & Q_e^2 & 0 & 0 & 0 & 0 & 0 & 0 & 0 & 0 \\
3 Q_e Q_u & 3 Q_e Q_d & 3 Q_e Q_u & 3 Q_e Q_d & 3 Q_e Q_d & Q_e^2 & Q_e^2 & Q_e^2 & 0 & 0 & 0 & 0 & 0 & 0 & 0 & 0 \\ \hline \hline
0 & 0 & 0 & 0 & 0 & 0 & 0 & 0 & 0 & 0 & 0 & 0 & 0 & 0 & 0 & 0 \\
0 & 0 & 0 & 0 & 0 & 0 & 0 & 0 & 0 & 0 & 0 & 0 & 0 & 0 & 0 & 0 \\
0 & 0 & 0 & 0 & 0 & 0 & 0 & 0 & 0 & 0 & 0 & 0 & 0 & 0 & 0 & 0 \\
0 & 0 & 0 & 0 & 0 & 0 & 0 & 0 & 0 & 0 & 0 & 0 & 0 & 0 & 0 & 0 \\
0 & 0 & 0 & 0 & 0 & 0 & 0 & 0 & 0 & 0 & 0 & 0 & 0 & 0 & 0 & 0 \\ \hline
0 & 0 & 0 & 0 & 0 & 0 & 0 & 0 & 0 & 0 & 0 & 0 & 0 & 0 & 0 & 0 \\
0 & 0 & 0 & 0 & 0 & 0 & 0 & 0 & 0 & 0 & 0 & 0 & 0 & 0 & 0 & 0 \\
0 & 0 & 0 & 0 & 0 & 0 & 0 & 0 & 0 & 0 & 0 & 0 & 0 & 0 & 0 & 0
\end{array}\right)}. 
\ee
The RG equations read
\begin{align}
\frac{d}{d \ln \mu} \left.  \mathcal{C}_{{\rm EMSM}_\chi} \right|_{\rm ren} =  & \, \gamma_{{\rm EMSM}_\chi} \left.  \mathcal{C}_{{\rm EMSM}_\chi} \right|_{\rm ren}  \ , \\ 
\gamma_{{\rm EMSM}_\chi} \equiv & \, - Z_{{\rm EMSM}_\chi}^{-1} \; \frac{d Z_{{\rm EMSM}_\chi}}{d \ln \mu} = - \frac{d}{d \ln \mu} \left( \delta Z_m + \delta Z_{\rm em} \right) \ .
\label{eq:gammadef2}
\end{align}
where the anomalous dimension can be computed analogously to what done in \Eq{eq:gammaexplicitchi}.


\section{Full Solution of the RG System}
\label{app:RGsol}

The evolution operator $U_\Lambda$ appearing in \Eq{eq:linearevolution} connects the mediator with the nuclear scale. In this Appendix we derive an expression for this operator. As described extensively in the paper, this connection requires three different steps, encoded in three different contributions
\be
U_\Lambda =  U_{{\rm EMSM}_\chi} \,U_{\rm match}\, U_{{\rm SM}_\chi}(t_\Lambda) \ .
\label{eq:evolution}
\ee
The matrix $U_{{\rm SM}_\chi}(t_\Lambda)$ evolves the Wilson coefficients from $\Lambda$ down to $m_Z$, whereas the matrix $U_{{\rm EMSM}_\chi}$ describes the evolution from $m_Z$ to $\mu \sim 1 - 2 \, {\rm GeV}$. The intermediate matching at the EWSB scale is taken care of by the matrix $U_{\rm match}$.

We start by setting up a general method to solve the RG system, which can be applied both above and below the EWSB scale. In both cases we always have to deal with a system
\be
\frac{d c(t)}{dt} = \gamma(t) c(t) \ .
\ee
The only scale dependence in $\gamma(t)$ comes from SM running couplings, namely 
\be
\gamma(t) = \sum_{j} \frac{g_j^2(t)}{16 \pi^2} \gamma_j  \ .
\ee
Here, the index $j$ runs over all SM running couplings $g_j(t)$, whereas the matrices $\gamma_j$ are constant. The running of the SM couplings can be found by solving the RG equations given in Refs.~\cite{Machacek:1983tz,Machacek:1983fi,Machacek:1984zw}. In both EFTs $t = 0$ is one boundary of our integration range, thus we rewrite the SM running couplings as follows
\be
g_j^2(t) =  g_j^2(0) + \Delta g_j^2(t) \ ,
\ee
where t = 0 at the $Z$ pole. The SM couplings at the $Z$ pole can be found in Refs.~\cite{Xing:2011aa,Buttazzo:2013uya}. The anomalous dimension matrix has then the form
\be
\gamma(t)  = \sum_{j} \frac{g_j^2(0)}{16 \pi^2} \gamma_j +
\sum_{j} \frac{\Delta g_j^2(t)}{16 \pi^2} \gamma_j \equiv \gamma_0 + \gamma_1(t) \ .
\ee
This setup is identical to a time-dependent perturbation theory problem in quantum mechanics with time replaced by the dimensionless variable $t=\ln(\mu/m_Z)$. In particular, the matrix $\gamma_1(t)$ can be treated as a time-dependent perturbation to the constant matrix $\gamma_0$, since their difference is at most a logarithmic running over two or three orders of magnitude. The procedure to set up a perturbative series is well known. We first define the ``interaction picture'' variables
\begin{align}
c_I(t) = & \, \exp[- \gamma_0 t] c(t) \ , \\
 \gamma_{1I}(t) = & \, \exp[- \gamma_0 t] \gamma_1 \exp[\gamma_0 t] \ . 
\end{align}
The Wilson coefficient vector satisfies a Schwinger-Tomonaga equation
\be
\frac{d c_I(t)}{dt} = \gamma_{1I}(t)  c_I(t) \ ,
\ee
whose formal solution is the Dyson series
\begin{align} \label{eq:Dyson}
c_I(t) = & \, \mathcal{T}\exp \left( \int_0^t d t^\prime \, \gamma_{1I}(t^\prime) \right) c_I(0) \ , \\
\mathcal{T}\exp \left( \int_0^t d t^\prime \, \gamma_{1I}(t^\prime) \right) = & \, 1 + 
  \int_0^t d t_1^\prime \, \gamma_{1I}(t_1^\prime) +  \int_0^t d t_1^\prime \int_0^{t^\prime_1} d t_2^\prime \, \gamma_{1I}(t_1^\prime) \gamma_{1I}(t_2^\prime) + \ldots \ .
\end{align}

We apply this result to derive the full evolution operator as defined in \Eq{eq:evolution}. The running from the mediator to the EWSB scale is obtained by applying the linear operator
\be
U_{{\rm SM}_\chi}(t_\Lambda) =  \left[\mathcal{T}\exp \left( - \int_0^{t_\Lambda} d t \, \gamma_{{\rm SM}_\chi 1 I}(t) \right) \right] \exp[-\gamma_{{\rm SM}_\chi 0} \, t_\Lambda] \ .
\ee
The analogous evolution from the EWSB scale down to the nuclear scale is described by
\be
U_{{\rm EMSM}_\chi} = \exp[\gamma_{{\rm EMSM}_\chi 0} t_N] \left[\mathcal{T}\exp \left(\int_0^{t_N} d t \, \gamma_{{\rm EMSM}_\chi 1 I}(t) \right) \right] \ .
\ee
Finally, the matching at the intermediate scale is achieved by using
\be
\scalemath{1}{U_{\rm match}  = \frac{1}{2}
\left(\begin{array}{ccc:cc|ccc:cc|ccc:cc||c}
1 & 1 & 0 & 0 & 0 & 0 & 0 & 0 & 0 & 0 & 0 & 0 & 0 & 0 & 0 & 2 g_{Vu} \\
1 & 0 & 1 & 0 & 0 & 0 & 0 & 0 & 0 & 0 & 0 & 0 & 0 & 0 & 0 & 2 g_{Vd} \\
0 & 0 & 0 & 0 & 0 & 1 & 1 & 0 & 0 & 0 & 0 & 0 & 0 & 0 & 0 & 2 g_{Vu} \\
0 & 0 & 0 & 0 & 0 & 1 & 0 & 1 & 0 & 0 & 0 & 0 & 0 & 0 & 0 & 2 g_{Vd} \\
0 & 0 & 0 & 0 & 0 & 0 & 0 & 0 & 0 & 0 & 1 & 0 & 1 & 0 & 0 & 2 g_{Vd} \\ \hline 
0 & 0 & 0 & 1 & 1 & 0 & 0 & 0 & 0 & 0 & 0 & 0 & 0 & 0 & 0 & 2 g_{Ve} \\
0 & 0 & 0 & 0 & 0 & 0 & 0 & 0 & 1 & 1 & 0 & 0 & 0 & 0 & 0 & 2 g_{Ve} \\
0 & 0 & 0 & 0 & 0 & 0 & 0 & 0 & 0 & 0 & 0 & 0 & 0 & 1 & 1 & 2 g_{Ve} \\ \hline \hline
- 1 & 1 & 0 & 0 & 0 & 0 & 0 & 0 & 0 & 0 & 0 & 0 & 0 & 0 & 0 & 2 g_{Au} \\
- 1 & 0 & 1 & 0 & 0 & 0 & 0 & 0 & 0 & 0 & 0 & 0 & 0 & 0 & 0 & 2 g_{Ad} \\
0 & 0 & 0 & 0 & 0 & - 1 & 1 & 0 & 0 & 0 & 0 & 0 & 0 & 0 & 0 & 2 g_{Au} \\
0 & 0 & 0 & 0 & 0 & - 1 & 0 & 1 & 0 & 0 & 0 & 0 & 0 & 0 & 0 & 2 g_{Ad} \\
0 & 0 & 0 & 0 & 0 & 0 & 0 & 0 & 0 & 0 & - 1 & 0 & 1 & 0 & 0 & 2 g_{Ad} \\ \hline 
0 & 0 & 0 & - 1 & 1 & 0 & 0 & 0 & 0 & 0 & 0 & 0 & 0 & 0 & 0 & 2 g_{Ae} \\
0 & 0 & 0 & 0 & 0 & 0 & 0 & 0 & - 1 & 1 & 0 & 0 & 0 & 0 & 0 & 2 g_{Ae} \\
0 & 0 & 0 & 0 & 0 & 0 & 0 & 0 & 0 & 0 & 0 & 0 & 0 & - 1 & 1 & 2 g_{Ae}
\end{array}\right)} \ .
\ee

\section{A Simple Recipe for an Approximate Analytical Solution}
\label{app:recipe}

In this Appendix we describe a simple analytical way to perform the RG evolution and compute direct detection cross sections based on our results. This is meant to be a useful recipe for practitioners. What is given here is the $0$-th order of the Dyson series in \Eq{eq:Dyson}, which neglects the running of the SM couplings. This approximation is quite satisfactory for our purposes since it induces errors at the level of $10\%$ in the scale $\Lambda$ if the DM couples to the top quark, and at the level of $1\%$ without this coupling.

Let us consider the situation where, starting from a UV complete model of singlet fermion WIMP, integrating out the mediators at the scale $\Lambda$ generates a subset of the dimension 6 operators in \Tab{tab:TheoryCdim6}. Then the spin-independent rate for the scattering off a target nucleus with $Z$ protons and $A$ neutrons is obtained through the following straightforward steps.
\begin{itemize}

\item[(i)] Write down the Wilson coefficients at the scale $\Lambda$ organized as in \Eq{ed:cdef}, and define
\be
 c_\Lambda \equiv \mathcal{C}_{{\rm SM}_\chi}(t_\Lambda) \ ,
\ee
where $t_\Lambda \equiv \ln \left( \Lambda / m_Z \right)$;

\item[(ii)] Evolve the vector of Wilson coefficients $c_\Lambda$ down to the nuclear scale using
\be
 c_N \equiv U_{{\rm EMSM}_\chi} U_{\rm match}  \exp[-\gamma_{{\rm SM}_\chi 0} \, t_\Lambda]  \,c_\Lambda \ ,
\ee
by applying first the exponential matrix $ \exp[-\gamma_{{\rm SM}_\chi 0} \, t_\Lambda] $ with
\be
\!\!\!\!\!\!\! \frac{\gamma_{{\rm SM}_\chi 0}}{10^{4}} \!=\! 
 \scalemath{0.72}{
\left(
\begin{array}{cccccccccccccccc}
 1.79 & 3.57 & -1.79 & -1.79 & -1.79 & 1.79 & 3.57 & -1.79 & -1.79 & -1.79 & 1.79 & 3.57 & -1.79 & -1.79 & -1.79 & 0.893 \\
 7.14 & 14.3 & -7.14 & -7.14 & -7.14 & 7.14 & 14.3 & -7.14 & -7.14 & -7.14 & 7.14 & 14.3 & -7.14 & -7.14 & -7.14 & 3.57 \\
 -3.57 & -7.14 & 3.57 & 3.57 & 3.57 & -3.57 & -7.14 & 3.57 & 3.57 & 3.57 & -3.57 & -7.14 & 3.57 & 3.57 & 3.57 & -1.79 \\
 -5.36 & -10.7 & 5.36 & 5.36 & 5.36 & -5.36 & -10.7 & 5.36 & 5.36 & 5.36 & -5.36 & -10.7 & 5.36 & 5.36 & 5.36 & -2.68 \\
 -10.7 & -21.4 & 10.7 & 10.7 & 10.7 & -10.7 & -21.4 & 10.7 & 10.7 & 10.7 & -10.7 & -21.4 & 10.7 & 10.7 & 10.7 & -5.36 \\
 1.79 & 3.57 & -1.79 & -1.79 & -1.79 & 1.79 & 3.57 & -1.79 & -1.79 & -1.79 & 1.79 & 3.57 & -1.79 & -1.79 & -1.79 & 0.894 \\
 7.14 & 14.3 & -7.14 & -7.14 & -7.14 & 7.14 & 14.3 & -7.14 & -7.14 & -7.14 & 7.14 & 14.3 & -7.14 & -7.14 & -7.14 & 3.57 \\
 -3.57 & -7.14 & 3.57 & 3.57 & 3.57 & -3.57 & -7.14 & 3.57 & 3.57 & 3.57 & -3.57 & -7.14 & 3.57 & 3.57 & 3.57 & -1.79 \\
 -5.36 & -10.7 & 5.36 & 5.36 & 5.36 & -5.36 & -10.7 & 5.36 & 5.36 & 5.36 & -5.36 & -10.7 & 5.36 & 5.36 & 5.36 & -2.68 \\
 -10.7 & -21.4 & 10.7 & 10.7 & 10.7 & -10.7 & -21.4 & 10.7 & 10.7 & 10.7 & -10.7 & -21.4 & 10.7 & 10.7 & 10.7 & -5.36 \\
 1.79 & 3.57 & -1.79 & -1.79 & -1.79 & 1.79 & 3.57 & -1.79 & -1.79 & -1.79 & 61.8 & -56.4 & -1.81 & -1.79 & -1.79 & 60.9 \\
 7.14 & 14.3 & -7.14 & -7.14 & -7.14 & 7.14 & 14.3 & -7.14 & -7.14 & -7.14 & -113. & 134. & -7.14 & -7.14 & -7.14 & 124 \\
 -3.57 & -7.14 & 3.57 & 3.57 & 3.57 & -3.57 & -7.14 & 3.57 & 3.57 & 3.57 & -3.61 & -7.14 & 3.61 & 3.57 & 3.57 & -1.74 \\
 -5.36 & -10.7 & 5.36 & 5.36 & 5.36 & -5.36 & -10.7 & 5.36 & 5.36 & 5.36 & -5.36 & -10.7 & 5.36 & 5.36 & 5.35 & -2.67 \\
 -10.7 & -21.4 & 10.7 & 10.7 & 10.7 & -10.7 & -21.4 & 10.7 & 10.7 & 10.7 & -10.7 & -21.4 & 10.7 & 10.7 & 10.7 & -5.34 \\
 5.36 & 10.7 & -5.36 & -5.36 & -5.36 & 5.36 & 10.7 & -5.36 & -5.36 & -5.36 & 365. & -349. & -5.23 & -5.37 & -5.34 & 363 \\
\end{array}
\right)},
\ee

then the matching matrix
\be
\!\!\!\!\!\!U_{\rm match} \! = \!\scalemath{0.72}{\left(
\begin{array}{cccccccccccccccc}
 0.5 & 0.5 & 0 & 0 & 0 & 0 & 0 & 0 & 0 & 0 & 0 & 0 & 0 & 0 & 0 & 0.19 \\
 0.5 & 0 & 0.5 & 0 & 0 & 0 & 0 & 0 & 0 & 0 & 0 & 0 & 0 & 0 & 0 & -0.35 \\
 0 & 0 & 0 & 0 & 0 & 0.5 & 0.5 & 0 & 0 & 0 & 0 & 0 & 0 & 0 & 0 & 0.19 \\
 0 & 0 & 0 & 0 & 0 & 0.5 & 0 & 0.5 & 0 & 0 & 0 & 0 & 0 & 0 & 0 & -0.35 \\
 0 & 0 & 0 & 0 & 0 & 0 & 0 & 0 & 0 & 0 & 0.5 & 0 & 0.5 & 0 & 0 & -0.35 \\
 0 & 0 & 0 & 0.5 & 0.5 & 0 & 0 & 0 & 0 & 0 & 0 & 0 & 0 & 0 & 0 & -0.038 \\
 0 & 0 & 0 & 0 & 0 & 0 & 0 & 0 & 0.5 & 0.5 & 0 & 0 & 0 & 0 & 0 & -0.038 \\
 0 & 0 & 0 & 0 & 0 & 0 & 0 & 0 & 0 & 0 & 0 & 0 & 0 & 0.5 & 0.5 & -0.038 \\
 -0.5 & 0.5 & 0 & 0 & 0 & 0 & 0 & 0 & 0 & 0 & 0 & 0 & 0 & 0 & 0 & -0.5 \\
 -0.5 & 0 & 0.5 & 0 & 0 & 0 & 0 & 0 & 0 & 0 & 0 & 0 & 0 & 0 & 0 & 0.5 \\
 0 & 0 & 0 & 0 & 0 & -0.5 & 0.5 & 0 & 0 & 0 & 0 & 0 & 0 & 0 & 0 & -0.5 \\
 0 & 0 & 0 & 0 & 0 & -0.5 & 0 & 0.5 & 0 & 0 & 0 & 0 & 0 & 0 & 0 & 0.5 \\
 0 & 0 & 0 & 0 & 0 & 0 & 0 & 0 & 0 & 0 & -0.5 & 0 & 0.5 & 0 & 0 & 0.5 \\
 0 & 0 & 0 & -0.5 & 0.5 & 0 & 0 & 0 & 0 & 0 & 0 & 0 & 0 & 0 & 0 & 0.5 \\
 0 & 0 & 0 & 0 & 0 & 0 & 0 & 0 & -0.5 & 0.5 & 0 & 0 & 0 & 0 & 0 & 0.5 \\
 0 & 0 & 0 & 0 & 0 & 0 & 0 & 0 & 0 & 0 & 0 & 0 & 0 & -0.5 & 0.5 & 0.5 \\
\end{array}
\right)} ,
\ee
and finally the matrix
\be
\!\!\!\!\!\!U_{{\rm EMSM}_\chi} \!=\! 
 \scalemath{0.72}{
 \renewcommand{\arraystretch}{1.2}
\left(
\begin{array}{cccccccccccccccc}
 0.99 & 0.0048 & -0.0097 & 0.0048 & 0.0048 & 0.0048 & 0.0048 & 0.0048 & 0 & 0 & 0 & 0 & 0 & 0 & 0 & 0 \\
 0.0048 & 1 & 0.0048 & -0.0024 & -0.0024 & -0.0024 & -0.0024 & -0.0024 & 0 & 0 & 0 & 0 & 0 & 0 & 0 & 0 \\
  -0.0097 & 0.0048 & 0.99 & 0.0048 & 0.0048 & 0.0048 & 0.0048 & 0.0048 & 0 & 0 & 0 & 0 & 0 & 0 & 0 & 0 \\
  0.0048 & -0.0024 & 0.0048 & 1 & -0.0024 & -0.0024 & -0.0024 & -0.0024 & 0 & 0 & 0 & 0 & 0 & 0 & 0 & 0 \\
0.0048 & -0.0024 & 0.0048 & -0.0024 & 1 & -0.0024 & -0.0024 & -0.0024 & 0 & 0 & 0 & 0 & 0 & 0 & 0 & 0 \\
 0.015 & -0.0073 & 0.015 & -0.0073 & -0.0073 & 0.99 & -0.0073 & -0.0073 & 0 & 0 & 0 & 0 & 0 & 0 & 0 & 0 \\
  0.015 & -0.0073 & 0.015 & -0.0073 & -0.0073 & -0.0073 & 0.99 & -0.0073 & 0 & 0 & 0 & 0 & 0 & 0 & 0 & 0 \\
   0.015 & -0.0073 & 0.015 & -0.0073 & -0.0073 & -0.0073 & -0.0073 & 0.99 & 0 & 0 & 0 & 0 & 0 & 0 & 0 & 0 \\
    0 & 0 & 0 & 0 & 0 & 0 & 0 & 0 & 1 & 0 & 0 & 0 & 0 & 0 & 0 & 0 \\
 0 & 0 & 0 & 0 & 0 & 0 & 0 & 0 & 0 & 1 & 0 & 0 & 0 & 0 & 0 & 0 \\
 0 & 0 & 0 & 0 & 0 & 0 & 0 & 0 & 0 & 0 & 1 & 0 & 0 & 0 & 0 & 0 \\
 0 & 0 & 0 & 0 & 0 & 0 & 0 & 0 & 0 & 0 & 0 & 1 & 0 & 0 & 0 & 0 \\
 0 & 0 & 0 & 0 & 0 & 0 & 0 & 0 & 0 & 0 & 0 & 0 & 1 & 0 & 0 & 0 \\
 0 & 0 & 0 & 0 & 0 & 0 & 0 & 0 & 0 & 0 & 0 & 0 & 0 & 1 & 0 & 0 \\
 0 & 0& 0& 0& 0 & 0 & 0 & 0 & 0 & 0 & 0 & 0 & 0 & 0 & 1 & 0 \\
 0 & 0 & 0 & 0 & 0 & 0 & 0 & 0 & 0 & 0 & 0 & 0 & 0 & 0 & 0 & 1 \\
\end{array}
\right)}.
\ee

\item[(iii)] After obtaining $c_N$, its first two components
\begin{align}
\left(c_N\right)_{1}& \equiv   c_{VV u}^{(1)} \, ,\;\;\;\;\;\;
\left(c_N\right)_{2} \equiv c_{VV d}^{(1)} 
\end{align}
directly appear in the spin-independent WIMP-nucleus cross section which can be computed according to
 \be
\sigma_{\mathcal{N}}^{\rm SI} = \frac{m_\chi^2\, m_\mathcal{N}^2}{(m_\chi + m_\mathcal{N})^2\,\pi\,\Lambda^4}\,
\left| c_{VV u}^{(1)}  (A+Z) + c_{VV d}^{(1)} (2 A - Z)  \right|^2 \ .
 \ee
where $m_\mathcal{N}$ is the mass of the target nucleus.

\end{itemize}

\bibliographystyle{JHEP}
\bibliography{DMdim6}

\end{document}